\documentclass[11pt,a4paper]{article}

\usepackage[margin=2.3cm]{geometry}
\usepackage{color}
\usepackage[colorlinks=true, pdfstartview=FitV, linkcolor=blue, citecolor=blue, urlcolor=blue]{hyperref}
\usepackage[sort&compress,numbers,colon,merge]{natbib}

\usepackage[utf8]{inputenc}
\usepackage{url}
\usepackage{amsmath}
\usepackage{amssymb}
\usepackage{graphicx}
\usepackage[english]{babel}
\usepackage[per-mode=fraction,separate-uncertainty=true]{siunitx}
\usepackage{multirow, makecell, hhline}
\usepackage{enumitem}
\usepackage{bbm}
\usepackage[compat=1.1.0]{tikz-feynman}
\tikzfeynmanset{warn luatex=false}
\usepackage{slashed}
\usepackage{nicematrix}
\usepackage{orcidlink}

\usepackage{physics}
    \renewcommand{\qcomma}{,\qquad}
	\newcommand{\qcand}{,\qand}
\usepackage{ulem}
\usepackage{subcaption}
\usepackage{bm}

\usepackage[retainorgcmds]{IEEEtrantools}
\newenvironment{eqns}
{\begin{IEEEeqnarray}{+rCl"s+x*}}
{\end{IEEEeqnarray}}
\newenvironment{eqns*}
{\begin{IEEEeqnarray*}{+rCl"s+x*}}
{\end{IEEEeqnarray*}}

\newcommand{\eq}[1]{\text{Eq.}~\eqref{eq:#1}}

\newcommand{\lr}[1]{\overset{\leftrightarrow}{#1}}
\newcommand{\DI}{{\,\raisebox{2mm}{${}^\leftrightarrow$}\hspace{-4mm} D_\mu^{\,a}}}

\renewcommand{\vev}{v_T}

\makeatletter
\g@addto@macro\bfseries\boldmath
\makeatother

\begin{document}

\begin{titlepage}
\vspace*{-1cm}
\phantom{hep-ph/2211.15954}
\flushright
\hfil{CPPC-2022-12}

\vskip 1.5cm
\begin{center}
\mathversion{bold}
{\LARGE\bf
Riding the Seesaw: \\ What Higgsstrahlung May Reveal \\ about Massive Neutrinos
}\\[3mm]
\mathversion{normal}
\vskip 0.3cm
\end{center}
\vskip 0.5cm
\begin{center}
{\large Tobias Felkl \orcidlink{0000-0003-0738-0264}},
{\large Adam Lackner \orcidlink{0009-0009-7386-7274}},
{\large Michael A.~Schmidt \orcidlink{0000-0002-8792-5537}}
\\
\vskip 0.7cm
{\footnotesize
Sydney Consortium for Particle Physics and Cosmology, School of Physics, The University of New South Wales, Sydney, NSW 2052, Australia\\[0.3cm]

\vskip 0.5cm
\begin{minipage}[l]{.9\textwidth}
\begin{center}
\textit{E-mail:}
\tt{t.felkl@unsw.edu.au}, \tt{a.lackner@unsw.edu.au}, \tt{m.schmidt@unsw.edu.au}
\end{center}
\end{minipage}
}
\end{center}
\vskip 1cm
\begin{abstract}
    We investigate if the projected high-precision measurements of the cross section of the Higgsstrahlung process $e^+ e^- \to Zh$ at a future electron-positron collider can be utilised to indirectly probe the fermionic Seesaw models. We consider the two centre-of-mass energies $\sqrt{s}=240$ GeV and $365$ GeV, and compare the collider reaches to constraints from electroweak observables, probes of lepton flavour universality and the existing and prospective bounds from searches for lepton flavour violation. For the analysis we assume the limit of an exactly conserved lepton-number symmetry. We find that while any appreciable correction to the Higgsstrahlung cross section is already strictly constrained in the Type-I Seesaw model,  effects of up to $\mathcal{O}(10\%)$ are possible within Type-III Seesaw.
\end{abstract}
\end{titlepage}
\pagenumbering{arabic}
\tableofcontents
\newpage

\section{Introduction}

Upon the discovery of the Higgs boson in 2012~\cite{ATLAS:2012yve,CMS:2012qbp}, the particle content as proposed within the Standard Model (SM) was confirmed to exist. Precise determinations of the properties of the Higgs boson, as well as in other sectors of physics, are a major component of the proposed program of next-generation lepton colliders. There are five proposals for so-called electron-positron ``Higgs factories"~\cite{Dawson:2022zbb}: the Circular Electron Positron Collider (CEPC)~\cite{CEPC-SPPCStudyGroup:2015csa,CEPCStudyGroup:2018ghi,CEPCPhysicsStudyGroup:2022uwl}, the International Linear Collider (ILC)~\cite{Baer:2013cma,Asner:2013psa,Bambade:2019fyw,ILCInternationalDevelopmentTeam:2022izu}, the Future Circular Collider (FCC-ee)~\cite{TLEPDesignStudyWorkingGroup:2013myl,10150/633402,Agapov:2022bhm,Bernardi:2022hny}, the Compact Linear Collider (CLIC)~\cite{Lebrun:2012hj,CLIC:2018fvx,CLICdp:2018cto} and the Cool Copper Collider (C$^3$)~\cite{Bai:2021rdg,Nanni:2022oha,Dasu:2022nux}.
The designation ``Higgs factory" mainly refers to the stage of a future $e^+e^-$ collider being run at a centre-of-mass energy of roughly $\sqrt{s} =$ 240--\SI{250}{\GeV}, where the integrated cross section of the Higgsstrahlung process peaks and so dominates over all Higgs production mechanisms. Other stages involve the operation as a ``$Z$ factory" at the $Z$ pole for $\sqrt{s} = m_Z$, close to the $W^+W^-$ production threshold at $\sqrt{s} = 2m_W$, and close to or at the $t\bar t$ threshold for $\sqrt{s} =$ 350--\SI{370}{\GeV}, as well as potential upgrades for runs at even higher centre-of-mass energies~\cite{Aslanides:2019gew}.

The incompleteness of the SM has been corroborated upon the measurement of neutrino oscillations by the Super-Kamiokande~\cite{Super-Kamiokande:1998kpq} and SNO collaborations~\cite{SNO:2001kpb,SNO:2002tuh}, which evidence the need to introduce new physics in order to account for neutrino masses. A particularly straightforward way to do so is via postulating the existence of \emph{sterile neutrinos}, several searches for which at future lepton colliders have been proposed. Most of the sensitivity studies focus on direct production of sterile neutrinos in electroweak processes and Higgs production~\cite{delAguila:2005ssc,delAguila:2005pin,Antusch:2015mia,Antusch:2015gjw,Antusch:2016vyf,Zhang:2018rtr,Das:2018usr,Barducci:2020icf,Gao:2021one}, see also Refs.~\cite{Deppisch:2015qwa,Cai:2017mow,Das:2018hph,Abdullahi:2022jlv} for reviews on heavy neutral leptons at colliders. Similarly, electroweak triplet fermions at colliders have been studied in, e.g., Refs.~\cite{Das:2020uer,Arguelles:2022tki}. The anticipated high precision attainable at future lepton colliders also serves as a motivation to consider virtual corrections, see e.g.~Ref.~\cite{Ge:2016zro}. In Refs.~\cite{Baglio:2016bop,Baglio:2016ijw}, the contribution of sterile neutrinos to the triple-Higgs coupling was studied.

In this work, we investigate the potential of electron-positron colliders to test the fermionic Seesaw scenarios~\cite{Minkowski:1977sc, Yanagida:1979as, GellMann:1980vs, Glashow1979,Mohapatra:1979ia,Foot:1988aq}, given the sub-percent precision to which these colliders are projected to measure the cross section of the Higgsstrahlung process $e^+e^-\to Zh$. Achieving this precision is possible due to the so-called ``recoil method" that refers to the selection of Higgsstrahlung events solely through measuring the four-momenta of the decay products of the $Z$ boson which recoils against the Higgs boson. This method is in principle applicable for any Higgs decay mode and thus allows for a model-independent reconstruction of the Higgs-boson mass, see for instance Ref.~\cite{Dawson:2022zbb}. 
The Higgsstrahlung process is also well-understood in the SM. Recently, two-loop electroweak corrections to the SM cross section were calculated in Refs.~\cite{Freitas:2022hyp,Chen:2022mre}; see also Refs.~\cite{FLEISCHER1983469,Kniehl1991RadiativeCF,Denner:1992bc,Bondarenko:2018sgg,Gong:2016jys,Sun:2016bel,Chen:2018xau}.

We consider the process at two benchmark centre-of-mass energies, $\sqrt{s} = \SI{240}{\GeV}$ and \SI{365}{\GeV}. As argued for in Ref.~\cite{Abramowicz:2016zbo}, the smaller cross section at larger $s$ can be partly compensated for by a higher instantaneous luminosity which scales approximately linearly with $s$. Moreover, the additional boost of the $Z$ and $h$ bosons allows for a better separation of the respective jets and therefore a more precise measurement of $\sigma_{Zh}\times \text{BR}(H\to X)$. Further advantages mentioned are the immediate access to top-pair production as well as the $e^+e^-\to H\nu_e\overline{\nu_e}$ process via $W^+W^-$ fusion which enables a precise determination of the Higgs-boson width, and the possibility to measure the Higgs-boson mass with a precision similar to the one at $\sqrt{s} = \SI{240}{\GeV}$.

In the fermionic Seesaw models, the smallness of the absolute mass scale of the light neutrinos generically requires the neutrino Yukawa couplings to be tiny for TeV-scale sterile states, and consequently lepton-number conserving processes are suppressed~\cite{Kersten:2007vk,Drewes:2019byd,Abada:2022wvh} and beyond the reach of the proposed lepton colliders. In contrast, \emph{symmetry-protected Seesaw models}~\cite{Wyler:1982dd,Bernabeu:1987gr,Branco:1988ex,Tommasini:1995ii,Pilaftsis:2004xx,Pilaftsis:2005rv,Shaposhnikov:2006nn,Kersten:2007vk,Abada:2007ux,Gavela:2009cd,Eboli:2011ia,Fernandez-Martinez:2015hxa} offer the possibility to entertain relatively light singlets or triplets around the \si{\tera\electronvolt} scale without the need to assume tiny Yukawa couplings only. Essentially, the new interaction states are assigned units of lepton number such that it is (approximately) conserved, which implies that some elements of the Dirac and Majorana mass matrices can be sizeable, whereas the remaining ones must be comparatively suppressed. The smallness of active neutrino masses is then guaranteed via the proportionality to these small entries, and does not only rely on an overall suppression through the large mass scale. In addition, no fine-tuned cancellation between (a priori unrelated) elements of the mass matrices are needed. 

The remaining paper is organised as follows. In Section~\ref{sec:theory} we introduce the relevant theoretical framework. We discuss Higgsstrahlung and other relevant processes which are sensitive to the same parameters in Section~\ref{sec:pheno}. Our results are summarised and conclusions are drawn in Section~\ref{sec:summary}. In the Appendices, more technical details and useful formulae are collected.


\section{Theory Framework}
\label{sec:theory}

Firstly, to set our conventions for notation, signs, and normalisation, we state the electroweak (EW) and leptonic parts of the SM Lagrangian:
\begin{eqns}
    \mathcal{L}_{\text{EW}} &=& -\frac{1}{4} W_{\mu\nu}^a W^{a\mu\nu} - \frac{1}{4} B_{\mu\nu} B^{\mu\nu}
    + (D_\mu H)^\dagger (D^\mu H) + \mu^2 H^\dagger H - \lambda (H^\dagger H)^2
    \\\nonumber &&
    \qquad {} + \overline{L_i} i\slashed{D} L_i + \overline{e_{Ri}} i\slashed{D} e_{Ri} 
    - Y^e_{ij} (\overline{L_i} e_{Rj} H + H^\dagger \overline{e_{Rj}} L_i) 
\end{eqns}
with the gauge-covariant derivative
\begin{equation}
    D_\mu = \partial_\mu - ig_2 W_\mu^a T^a - ig_1 Y B_\mu, 
    \qquad\qquad
    T^a = \begin{cases} 
        \dfrac{\sigma^a}{2} & \text{EW doublet}\\
        0 & \text{EW singlet}\\
    \end{cases},
\end{equation}
in terms of the Pauli matrices $\sigma^a$ with $a = 1,2,3$.


\subsection{Effective Field Theory}

\subsubsection{SMEFT}
\label{sec:SMEFT}

In the Standard Model Effective Field Theory (SMEFT) we extend the SM Lagrangian with a set of dimension-five and -six operators $\mathcal{O}_i$:
\begin{equation}
    \mathcal{L} = \mathcal{L}_{\text{SM}} + \sum_i C_i \mathcal{O}_i.
\end{equation}
In our convention the Wilson coefficients $C_i$ are dimensional; for later use we also define the dimensionless variants $\hat C_i \equiv C_i \times \mathrm{TeV}^2$ of the dimension-six coefficients.

The conventional basis of operators is the so-called \emph{Warsaw basis}~\cite{Grzadkowski:2010es}. For the convenience of the reader, we reproduce the most relevant operators for the fermionic Seesaw models (additional operators relevant to Higgsstrahlung are also listed in Table~\ref{table:warsaw basis}):
\begin{eqns}
    \mathcal{L} & \supset &
    \left[C_{5,ij}
    \,
    (\widetilde H^\dagger L_i)^T C (\widetilde H^\dagger L_j) + \mathrm{h.c.}\right]
    +
    \left[C_{eB,ij} \, (\overline{L_i} \sigma^{\mu\nu} e_{Rj}) H B_{\mu\nu} + \mathrm{h.c.}\right] 
    \\\nonumber & & {} 
    +
    \left[C_{eW,ij} \, (\overline{L_i} \sigma^{\mu\nu} \sigma^a e_{Rj}) H W^a_{\mu\nu} + \mathrm{h.c.}\right]
    +
    \left[C_{eH,ij} \,(H^\dagger H) (\overline{L_i} e_{Rj} H) + \mathrm{h.c.}\right]
    \\\nonumber & & {}
    +
    C_{HL,ij}^{(1)} \,(H^\dagger i \lr{D}_\mu H)(\overline{L_i} \gamma^\mu L_j)
    + 
    C_{HL,ij}^{(3)} \, (H^\dagger i \DI H)(\overline{L_i} \sigma^a \gamma^\mu L_j),
\end{eqns}%
with $\widetilde H = i\sigma_2 H^*$, and $C$ the charge conjugation matrix satisfying $C^\dagger \gamma^\mu C = -(\gamma^\mu)^T$.

In addition to directly contributing to physical processes, the SMEFT operators also modify the relations between observed quantities and SM parameters, as well as the relations between different parameters. The result of these effects is that the SM parameters will be numerically different in SMEFT compared to the SM.
For example, the measured value of the Fermi constant, $\hat{G}_F$, in muon decay will differ from its standard expression $G_F = 1/(\sqrt{2} \vev^2)$, $\vev$ being the Higgs VEV in SMEFT (see e.g.~Ref.~\cite{Berthier:2015oma}), by
\begin{equation}\label{eq:shift_GF}
    \var \hat G_F = \hat{G}_F - G_F = \frac{1}{\sqrt{2}} \left(C_{HL,11}^{(3)} + C_{HL,22}^{(3)}\right) - \frac{1}{2\sqrt{2}}(C_{LL,1221} + C_{LL,2112}).
\end{equation}
When computing the shift in a quantity such as a cross section from its SM value, one must therefore take care to include the contributions from these parameter shifts. Concretely, if $\sigma_{\text{SM}}(g_i)$ is a cross section computed in the SM written as a function of the parameters $g_i$, then
\begin{equation}
    \Delta \sigma = \sigma_{\text{SMEFT}} - \sigma_{\text{SM}} = \Delta \sigma_{\text{Direct}} + \sum_i \pdv{\sigma_{\text{SM}}}{g_i} \var g_i,
\end{equation}
where $\Delta \sigma_{\text{Direct}}$ captures direct contributions of new operators, and $\var g_i$ is the shift in the parameter $g_i$. The values of the parameter shifts critically depend on one's choice of input parameters; in this work we use $(\alpha, m_Z, G_F)$ as our electroweak inputs.
A thorough review of such shifts may be found for instance in Ref.~\cite{Alonso:2013hga}, and we provide additional exposition in Appendix~\ref{app:SMEFT}.


\subsubsection{LEFT}

To describe physics taking place at scales below the electroweak scale we utilise Low-Energy Effective Field Theory (LEFT). The relevant part of the Lagrangian for purely leptonic transitions reads in the Jenkins-Manohar-Stoffer (JMS) basis~\cite{Jenkins:2017jig}
\begin{eqns}
    \mathcal{L} & \supset &
    C_{ee,ijkl}^{VLL} (\overline{\ell_i} \gamma^\mu P_L \ell_j)(\overline{\ell_k} \gamma_\mu P_L \ell_l)
    +
    C_{ee,ijkl}^{VRR} (\overline{\ell_i} \gamma^\mu P_R \ell_j)(\overline{\ell_k} \gamma_\mu P_R \ell_l)
    \\\nonumber & & {}
    +
    C_{ee,ijkl}^{VLR} (\overline{\ell_i} \gamma^\mu P_L \ell_j)(\overline{\ell_k} \gamma_\mu P_R \ell_l)
    +
    \bqty{C_{e\gamma,ij}(\overline{\ell_i} \sigma^{\mu\nu} P_R \ell_j) F_{\mu\nu} + \mathrm{h.c.}}. \nonumber
\end{eqns}%
We define the covariant derivative in QED as in $D_\mu=\partial_\mu + i Q e A_\mu$, following~Ref.~\cite{Jenkins:2017jig}. For semi-leptonic neutral-current transitions, the relevant part of the Lagrangian is
\begin{eqns}
    \mathcal{L} & \supset & 
    C_{eq}^{VLL} (\overline \ell \gamma^\mu P_L \ell)  (\overline q \gamma_\mu P_L q)
    +
    C_{eq}^{VRR} (\overline \ell \gamma^\mu P_R \ell) (\overline q \gamma_\mu P_R q)
    \\\nonumber & & {}
    +
    C_{eq}^{VLR} (\overline \ell \gamma^\mu P_L \ell)  (\overline q \gamma_\mu P_R q)
    +
    C_{qe}^{VLR}  (\overline q \gamma^\mu P_L q) (\overline \ell \gamma_\mu P_R \ell).
\end{eqns}%
To obtain the LEFT Wilson coefficients for the Seesaw models introduced below, we utilise the software package \texttt{DsixTools}~\cite{Celis:2017hod}
to (i) compute the renormalisation group (RG) running of the SMEFT coefficients between the Seesaw scale and the electroweak scale, $\mu = m_Z$, (ii) match the SMEFT and LEFT coefficients, and (iii) run the LEFT coefficients to the low scale $\mu = \SI{5}{\GeV}$.
As there are no sizeable contributions to quark-field operators in the Seesaw models, we assume that this procedure captures the main contributions from RG running in LEFT, and further effects at lower scales do not appreciably change the results.
See Appendix~\ref{sec:approx_matching} for more details.


\subsection{Seesaw Models}

\subsubsection{Type-I}

\begin{table*}[tbp]
    \centering
    \def\arraystretch{1.3}
    \begin{tabular}{|c||c|c|}\hline
         Coefficient & Type-I & Type-III
        \\ \hline
         $C_{5,ij}$ & $\frac{1}{2} \left(Y^{\nu*} (M^{\nu\dagger})^{-1} Y^{\nu\dagger}\right)_{ij}$ & $\frac{1}{2} (Y^{\Sigma*} (M^{\Sigma\dagger})^{-1} Y^{\Sigma\dagger})_{ij}$
        \\ 
         $C_{HL,ij}^{(1)}$ & $\frac{1}{4} \left(Y^\nu (M^{\nu\dagger}M^\nu)^{-1} Y^{\nu\dagger}\right)_{ij}$ & $\frac{3}{4} \left(Y^\Sigma (M^{\Sigma\dagger}M^\Sigma)^{-1} Y^{\Sigma\dagger}\right)_{ij}$
        \\ 
         $C_{HL,ij}^{(3)}$ & $-\frac{1}{4} \left(Y^\nu (M^{\nu\dagger}M^\nu)^{-1} Y^{\nu\dagger}\right)_{ij}$ & $\frac{1}{4} \left(Y^\Sigma (M^{\Sigma\dagger}M^\Sigma)^{-1} Y^{\Sigma\dagger}\right)_{ij}$
        \\ 
         $C_{eH,ij}$ & 0 & $\left(Y^\Sigma (M^{\Sigma\dagger}M^\Sigma)^{-1} Y^{\Sigma\dagger} Y^e\right)_{ij}$
        \\\hline 
         $C_{eB,ij}$ & $\frac{1}{16\pi^2} \frac{g_1}{24}\left(Y^\nu (M^{\nu\dagger}M^\nu)^{-1} Y^{\nu\dagger} Y^e\right)_{ij}$ & $\frac{1}{16\pi^2} \frac{g_1}{8} \left(Y^\Sigma (M^{\Sigma\dagger}M^\Sigma)^{-1} Y^{\Sigma\dagger} Y^e\right)_{ij}$
        \\ 
         $C_{eW,ij}$ & $\frac{1}{16\pi^2} \frac{5g_2}{24}\left(Y^\nu (M^{\nu\dagger}M^\nu)^{-1} Y^{\nu\dagger} Y^e\right)_{ij}$ & $\frac{1}{16\pi^2} \frac{3g_2}{8} \left(Y^\Sigma (M^{\Sigma\dagger}M^\Sigma)^{-1} Y^{\Sigma\dagger} Y^e\right)_{ij}$
        \\ \hline
    \end{tabular}
    
    \caption{SMEFT Wilson coefficients obtained from matching the Type-I and Type-III Seesaw models at the scale $\mu = M^X$~\cite{Du:2022vso, Zhang:2021jdf, Coy:2021hyr}. 
    In order to properly account for the stringent bounds from the non-observation of lepton flavour violation, the electroweak dipole operators $C_{eB}$ and $C_{eW}$ (last two rows) are matched to one-loop order, while all other operators are matched at tree level.}
    \label{table:SMEFT matching}
\end{table*}

In the Type-I Seesaw model~\cite{Minkowski:1977sc, Yanagida:1979as, GellMann:1980vs, Glashow1979,Mohapatra:1979ia} the SM Lagrangian is extended by adding $n_\nu$ right-handed sterile neutrinos $\{ \nu_{Ri} \}_{i=1}^{n_\nu}$ (so that there are a total of $3 + n_\nu$ neutrinos), accompanied by a new Yukawa interaction to generate Dirac neutrino masses, as well as Majorana mass terms for the $\nu_R$:
\begin{equation}
    \mathcal{L}_{\nu_R} = \sum_{i=1}^{n_\nu} \overline{\nu_{Ri}} i\slashed{\partial} \nu_{Ri}
    - \frac{1}{2} \sum_{i,j=1}^{n_\nu}\left( M^\nu_{ij} \overline{\nu_{Ri}^c} \nu_{Rj} + \text{h.c.}\right)
    - \sum_{i=1}^3 \sum_{j=1}^{n_\nu} \left(Y^\nu_{ij} \overline{L_i} \nu_{Rj} \widetilde{H} + \mathrm{h.c.}\right),
\end{equation}
where the conjugate fields $\nu^c$ are defined as $\nu^c \equiv \gamma^0 C \nu^*$ \cite{Pal2010}.
After electroweak symmetry breaking we are able to express the combined mass terms as the matrix equation
\begin{equation}
    \mathcal{L}_{\nu \text{ mass}} = - \frac{1}{2} \mqty( \overline{\nu_L} & \overline{\nu_R^c} ) \mqty( 0 & m^\nu \\ m^{\nu T} & M^\nu ) \mqty( \nu_L^c \\ \nu_R ) + \text{h.c.},
\end{equation}
where we refer to $m^\nu_{ij} \equiv Y^\nu_{ij} v/\sqrt{2}$ as the Dirac mass matrix, and $M^\nu$ is the Majorana mass matrix. Matching this theory onto SMEFT at the scale $\mu = M^\nu$ yields the effective operators collected on the left side of Table~\ref{table:SMEFT matching}.


\subsubsection{Type-III}

In the Type-III Seesaw model \cite{Foot:1988aq} the SM Lagrangian is extended by adding $n_\Sigma$ right-handed weak fermion triplets $\{ \Sigma_{Ri}^a \}_{i=1}^{n_\Sigma}$ with vanishing hypercharge, a new Yukawa interaction to generate Dirac neutrino masses, and Majorana mass terms for the $\Sigma_R$~\cite{Abada:2007ux,Biggio:2019eeo}:
\begin{equation}
    \mathcal{L}_{\Sigma_R} =
    \sum_{i=1}^{n_\Sigma} \overline{\Sigma_{Ri}} i\slashed{D} \Sigma_{Ri}
    - \frac{1}{2} \sum_{i,j=1}^{n_\Sigma} \left(M_{ij}^\Sigma\,\overline{\Sigma_{Ri}^{c}} \Sigma_{Rj} + \text{h.c.}\right)
    - \sum_{i=1}^3 \sum_{j=1}^{n_\Sigma} \left(Y_{ij}^\Sigma \overline{L_i}\sigma^a\Sigma_{Rj}^a \widetilde{H} + \mathrm{h.c.}\right),
\end{equation}
where $a = 1,2,3$ is elided from all but the Yukawa term.
For a fixed triplet generation $i$, the eigenstates of electric charge are given by the combinations~\cite{Abada:2007ux}
\begin{equation}
    \Sigma_{Ri}^\pm \equiv \frac{\Sigma_{Ri}^1 \mp i\Sigma_{Ri}^2}{\sqrt{2}}, \quad \Sigma_{Ri}^0 \equiv \Sigma_{Ri}^3.
\end{equation}
In a manner completely analogous to the Type-I Seesaw model, after electroweak symmetry breaking we obtain the neutrino mass matrix
\begin{equation}
    \mathcal{L}_{\nu\text{ mass}} = - \frac{1}{2} \mqty( \overline{\nu_L} & \overline{\Sigma^{0c}_R} ) \mqty( 0 & m^\Sigma \\ m^{\Sigma T} & M^\Sigma ) \mqty( \nu_L^c \\ \Sigma^0_R ) + \text{h.c.},
\end{equation}
where we refer to $m^\Sigma_{ij} \equiv Y^\Sigma_{ij} v/\sqrt{2}$ as the Dirac mass matrix, and $M^\Sigma$ is the Majorana mass matrix.
The states $\Sigma_{Ri}^\pm$ instead mix into the charged leptons.
Matching this theory onto SMEFT at the scale $\mu = M^\Sigma$ yields the effective operators collected on the right side of Table~\ref{table:SMEFT matching}.


\subsection{Conserved Lepton-Number Symmetry}

In this work, we study symmetry-protected versions of the fermionic Seesaw models, wherein a lepton-number (LN) symmetry decouples the physics of neutrino masses from the phenomenology associated with the conservation of LN~\cite{Kersten:2007vk,Abada:2007ux,Gavela:2009cd,Eboli:2011ia,Fernandez-Martinez:2015hxa}. Without loss of generality, we fix $n_\nu = n_\Sigma = 2$, that is, we focus on the case of two heavy fermion singlet or triplet interaction states which is consistent with at least two massive active neutrinos, as is dictated by neutrino oscillation data. The heavy fermion states are assigned 1 and $-1$ unit of LN, respectively. After electroweak symmetry breaking, the Dirac mass matrix is given by
\begin{equation}
    m^X_{ij} = \frac{v}{\sqrt{2}}\mqty(Y^X & \epsilon Y^{X'})_{ij} = \frac{v}{\sqrt{2}}\mqty(Y^{X_e} & \epsilon Y^{X'_e} \\ Y^{X_\mu} & \epsilon Y^{X'_\mu} \\ Y^{X_\tau} & \epsilon Y^{X'_\tau})_{ij},
\end{equation}
and the Majorana mass matrix reads
\begin{equation}
    M^X_{ij} = \mqty(\mu_1M^X & M^X \\ M^X & \mu_2M^X)_{ij},
\end{equation}
where $\epsilon$ and $\mu_{1,2}$ are dimensionless parameters. We parametrise the mixing of the SM neutrino $\nu_i$ with the fermion singlet $\nu_{Ri}$, or the neutral component $\Sigma_{Ri}^0$ of the fermion triplet  in terms of the dimensionless ratios
\begin{equation}\label{eq:mixing}
    \theta_i = \frac{m^X_{i1}}{M^X} =  \frac{Y^{X_i}}{\sqrt{2}}\frac{v}{M^X},
\end{equation}
which are equal to the active-sterile \emph{mixing angles} in the small-mixing approximation, that is, if $\mathcal{O}\big((v/M^X)^3\big)$ effects are neglected. For simplicity, we will refer to $\theta_e$ also as the ``electron(-flavour) mixing angle", and to $|\theta_e|$ as ``electron(-flavour) mixing", and equivalently for the other flavours. Light neutrino masses are then proportional to $\epsilon$ and $\mu_2$ which break LN:
\begin{equation}
    m_v = \frac{v^2}{2} \Big[\mu_2  Y^X(M^X)^{-1}Y^{X^T} - \epsilon \left(Y^{X'}(M^X)^{-1}Y^{X^T} + Y^X(M^X)^{-1}Y^{X'^T}\right) \Big].
\end{equation}
The limit $\mu_2\neq 0$ and $\mu_1 = \epsilon = 0$ is referred to as \emph{inverse Seesaw}~\cite{Wyler:1982dd,Mohapatra:1986aw,Mohapatra:1986bd}, and $\epsilon\neq 0$ and $\mu_{1,2} = 0$ is commonly known as \emph{linear Seesaw}~\cite{Akhmedov:1995ip,Akhmedov:1995vm}.\footnote{Note that with the help of Table~\ref{table:SMEFT matching} one can check that---as long as the contributions of the Weinberg operator to the dimension-six operators via RG running can be neglected---the only change resulting from
$\mu_{1,2} \neq 0$
compared to the LN-conserving limit is
$\theta_i^2 \to \frac{1 + \mu_2^2}{(1 + \mu_1 \mu_2)^2} \theta_i^2$
for each mixing angle in all of our formulae.
}

We adopt the LN-conserving limit $\epsilon = \mu_{1,2} = 0$ with non-zero $M^X$ in this work, which results in massless active neutrinos and a heavy Dirac neutrino of mass $M^X$, and hence assume the textures
\begin{eqns}
    Y_{ij}^X = 
    \mqty( Y^{X_e} & 0 \\ Y^{X_\mu} & 0 \\ Y^{X_\tau} & 0 ) \qand M_{ij}^X = \mqty(0 & M^X \\ M^X & 0)
\end{eqns}%
for both $X = \nu$ and $\Sigma$. In this way, we neglect the phenomenological implications of LN violation, and instead focus on LN-conserving effects.\footnote{Note that this also means that the parameters $\theta_i$ can be treated as independent. If data on lepton mixing and the hierarchy of neutrino masses is to be properly accommodated, the $\theta_i$ exhibit non-trivial correlations as is discussed for instance in Refs.~\cite{Donini:2012tt,Biggio:2019eeo}.} Note that one may add a further singlet or triplet with vanishing LN such that
\begin{equation}
    Y_{ij}^X = \mqty( Y^{X_e} & 0 & 0 \\ Y^{X_\mu} & 0 & 0 \\ Y^{X_\tau} & 0 & 0 ) \qand
    M_{ij}^X = \mqty(0 & M^X & 0 \\ M^X & 0 & 0 \\ 0 & 0 & M^{X'}),
\end{equation}
which supports three massive active neutrinos if one departs from the LN-conserving limit. Still, the additional state trivially decouples from the phenomenology.\footnote{As is discussed in Ref.~\cite{Abada:2007ux}, the determinant of the full neutrino mass matrix also vanishes in the (LN-violating) case $\mu_1 \neq 0$ and $\epsilon = \mu_2 = 0$, which however generally guarantees only one massless active neutrino. One may also consider a vanishing Majorana mass matrix and a completely general Dirac mass matrix instead of the textures assumed herein; nonetheless, this scenario yields light Dirac neutrinos and is not suitable for a study in SMEFT.}


\section{Phenomenology}
\label{sec:pheno}

We choose the benchmark value $M^X = \SI{1}{\TeV}$ for the masses of the new interaction states in our analysis. To our knowledge, this is consistent with all performed direct searches for heavy neutral leptons at colliders, see for instance Ref.~\cite{Abdullahi:2022jlv} for a recent overview. In Ref.~\cite{CMS:2018iaf}, for sterile neutrinos of a mass $M^{\nu} \approx \SI{1}{\TeV}$ the constraint $|\theta_e|\sim|\theta_\mu|\lesssim\mathcal{O}(1)$ was derived via a search for the signature of three charged leptons with any combination of electron and muon flavours. Ref.~\cite{CMS:2022rqc} reports the constraint $|\theta_\mu|^2\lesssim\mathcal{O}(0.1)$ for \si{\TeV}-scale Majorana neutrinos, based on a search for same-sign dimuon final states, see also Ref.~\cite{Abada:2022wvh}. The bound $M^{\Sigma_0} \geq \SI{910}{\GeV}$ was derived in a recent study~\cite{ATLAS:2022yhd} which focuses on leptonic final states and takes into account earlier ATLAS results.

In our phenomenological discussion we consider the following observables:
\begin{enumerate}[label=\textbullet]
    \item the relative shift $\Delta\sigma/\sigma_0$ in the Higgsstrahlung cross section from its SM prediction, 
    \item the effective leptonic weak mixing angle $\sin^2(\theta^{\text{lept}}_{w,\text{eff}})$ and the $W$-boson mass $m_W$, 
    \item the ratios $g^X_{\mu/e}$ and $g^X_{\tau/\mu}$ of leptonic gauge couplings as probes of lepton flavour universality (LFU), and the ratios $R(K_{\ell3})$ and $R(V_{us})$, and
    \item the branching ratios of the LFV processes $\mu\to e\gamma$, $\mu\to3e$, $\tau\to e\gamma$, $\tau\to3e$, and the ratios of the $\mu-e$ conversion rates over the muon capture rate in different target nuclei.
\end{enumerate}
In Tables~\ref{table:shifts_higgsstrahlung}, \ref{table:electroweak_shifts}, \ref{table:LFU} and~\ref{table:LFV}, the theoretical expressions for these observables in the fermionic Seesaw models are listed as functions of the mixing angles $\theta_e$, $\theta_\mu$, $\theta_\tau$, as defined in Eq.~\eqref{eq:mixing}, for a matching scale $\mu = M^X = \SI{1}{\tera\electronvolt}$. While our expressions hold for complex Yukawa couplings, we do not consider CP violation in the analysis and treat the mixings as real numbers which are only constrained from existing bounds on LN-conserving processes.
If effects from RG running in SMEFT above the matching scale are neglected, one may na\"ively interpret the results also for a larger mass and appropriately rescale the couplings. Nonetheless, we also computed the relevant expressions for $\mu = M^X = \SI{10}{\tera\electronvolt}$, which is commented on in Section~\ref{sec:larger_seesaw_scale}.

The following discussions are supported by plots in the $\theta_e$--$\theta_\mu$ plane, as well as plots in the $\theta_e$--$\theta_\tau$ plane for the LFU and LFV observables.\footnote{We do not include plots in the $\theta_\mu$--$\theta_\tau$ plane, as the major objective of our work is to investigate how the sensitivity of the fermionic Seesaw models to the Higgsstrahlung process compares to other observables, and we are thus primarily concerned with phenomenological effects related to electron flavour.} The third mixing angle is fixed to the benchmark values $\theta_\tau = \num{e-2}$ and $\theta_\mu = \num{e-6}$; these choices are most transparently justified (at least for Type-III) by Figure~\ref{fig:summary}, which depicts the most competitive constraints for both models. We have explicitly checked that there are no appreciable changes in the resulting phenomenology if these values are tuned smaller or even zero, apart from the fact that the LFV bounds become weaker and eventually vanish. In each of these plots we produce exclusion regions which either reflect the current bounds at $2\sigma$ for the electroweak and LFU observables, or the upper limits on the LFV processes at 90\% C.L.


\subsection{Higgsstrahlung}
\label{sec:higgsstrahlung}

\subsubsection{SM Tree-Level Contribution}

The tree-level differential cross section in the SM is well-known, and is given by~\cite{Kniehl:1994vk, Craig:2014una}
\begin{equation}
    \dv{\sigma_0}{\cos\theta} = \frac{\sqrt{\lambda}}{32\pi s^2} \overline{\abs{\mathcal{M}_t}^2}, \label{eq:SM differential cross section}
\end{equation}
where unpolarised beams are assumed, with
\begin{equation}
    \overline{\abs{\mathcal{M}_t}^2} = \frac{s}{2} (g_L^2 + g_R^2) \pqty{ \frac{g_{ZZh}}{s-m_Z^2} }^2 \pqty{ 1 + \frac{\lambda \sin^2\theta}{8sm_Z^2} }
\end{equation}
the spin-averaged matrix element. Here
\begin{equation}
    g_{ZZh} = \frac{g_2 m_Z}{\cos\theta_w} \qcomma
    g_L = g_Z (-\tfrac{1}{2} + \sin^2\theta_w) \qcomma
    g_R = g_Z \sin^2\theta_w \qcomma
    g_Z = \frac{g_2}{\cos\theta_w},
\end{equation}
$\theta$ is the angle between the incoming electron and outgoing $Z$ boson, and
\begin{equation}
    \lambda = (s - m_Z^2 - m_h^2)^2 - 4 m_Z^2 m_h^2
\end{equation}
is the relevant K\"all\`en function.
\begin{figure}[tbp]
    \centering
    \includegraphics[width=0.7\linewidth]{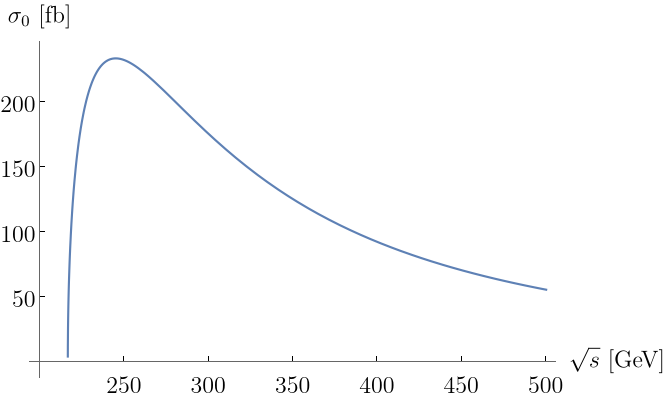}
    \caption{The tree-level cross section for the Higgsstrahlung process in the SM as a function of the centre-of-mass energy $\sqrt{s}$.}
    \label{fig:TreeLevelHiggsstrahlung}
\end{figure}
The corresponding integrated cross section is
\begin{equation}
    \sigma_0 = \frac{\sqrt{\lambda}}{32\pi s} (g_L^2 + g_R^2) \pqty{ \frac{g_{ZZh}}{s-m_Z^2} }^2 \pqty{ 1 + \frac{\lambda}{12sm_Z^2} }.
\end{equation}
The dependence of $\sigma_0$ on $\sqrt{s}$ is depicted in Figure~\ref{fig:TreeLevelHiggsstrahlung}. It peaks around a centre-of-mass energy of $\sqrt{s} \approx \SI{245}{\GeV}$.


\subsubsection{Corrections in SMEFT}

Including corrections from new physics, we write
\begin{equation}
    \dv{\sigma}{\cos\theta} = \frac{\sqrt{\lambda}}{32\pi s^2} \pqty{ \overline{\abs{\mathcal{M}_t}^2} + \var \overline{\abs{\mathcal{M}_t}^2} + 2\Re \overline{\mathcal{M}_t^* \mathcal{M}_c} },
\end{equation}
where $\var \overline{\abs{\mathcal{M}_t}^2}$ denotes the effect of parameter shifts in SMEFT to the tree-level cross section, as discussed in Section \ref{sec:SMEFT}, and $2\Re \overline{\mathcal{M}_t^* \mathcal{M}_c}$ is the interference term of the tree-level amplitude with corrections from new operators. The explicit result reads\footnote{See Ref.~\cite{Kniehl:1994vk} for a calculation of the Higgsstrahlung cross section in the Type-I Seesaw model which does not rely on effective field theory.}
\begin{equation}
    \Delta \dv{\sigma}{\cos\theta} = \frac{\sqrt{\lambda}}{32\pi s^2} \bqty{ 2\pqty{ \frac{\var g_{ZZh}}{g_{ZZh}} + \frac{g_L \var g_L + g_R \var g_R}{g_L^2 + g_R^2} } \overline{|\mathcal{M}_t|^2} + \frac{g_{ZZh} v}{2} \sum_{i=2}^5 d_i F_i },
\end{equation}
where the parameter shifts and coefficients $d_i$ are presented below, and the form factors $F_i$ may be found in Ref.~\cite{Craig:2014una}.

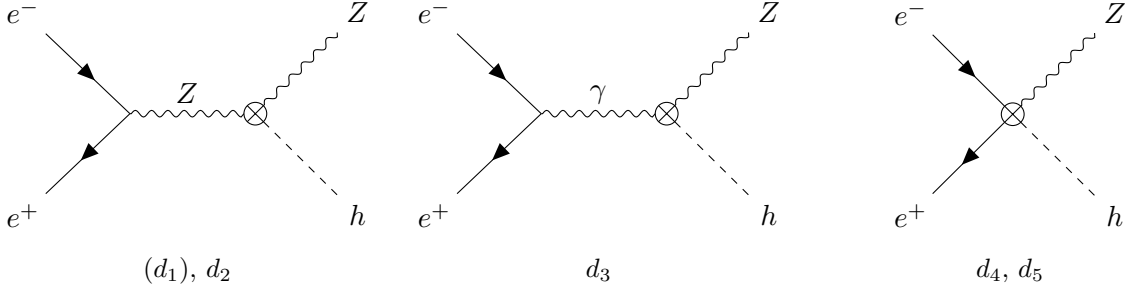
\begin{figure}[tbp]
    \begin{subfigure}[t]{.33\textwidth}
        \centering
        \begin{tikzpicture}
        \begin{feynman}
            \vertex (a);
    		\vertex [above left=1.5cm of a] (i1) {$e^-$};
    		\vertex [below left=1.5cm of a] (i2) {$e^+$};
    		\vertex [right=1.5cm of a, crossed dot] (b) {};
    		\vertex [above right=1.9cm of b] (f1) {$Z$};
    		\vertex [below right=1.9cm of b] (f2) {$h$};
    		
    		\diagram* {
    			(i1) -- [fermion] (a) -- [fermion] (i2),
    			(a) -- [boson, edge label=$Z$] (b),
    			(b) -- [boson] (f1),
    			(b) -- [scalar] (f2),
    		};
        \end{feynman}
        \end{tikzpicture}
        \caption*{($d_1$), $d_2$}
    \end{subfigure}%
    \begin{subfigure}[t]{.33\linewidth}
        \centering
        \begin{tikzpicture}
        \begin{feynman}
            \vertex (a);
    		\vertex [above left=1.5cm of a] (i1) {$e^-$};
    		\vertex [below left=1.5cm of a] (i2) {$e^+$};
    		\vertex [right=1.5cm of a, crossed dot] (b) {};
    		\vertex [above right=1.9cm of b] (f1) {$Z$};
    		\vertex [below right=1.9cm of b] (f2) {$h$};
    		
    		\diagram* {
    			(i1) -- [fermion] (a) -- [fermion] (i2),
    			(a) -- [boson, edge label=$\gamma$] (b),
    			(b) -- [boson] (f1),
    			(b) -- [scalar] (f2),
    		};
        \end{feynman}
        \end{tikzpicture}
        \caption*{$d_3$}
    \end{subfigure}%
    \begin{subfigure}[t]{.33\textwidth}
        \centering
        \begin{tikzpicture}
        \begin{feynman}
            \vertex [crossed dot] (a) {};
    		\vertex [above left=1.9cm of a] (i1) {$e^-$};
    		\vertex [below left=1.9cm of a] (i2) {$e^+$};
    		\vertex [above right=1.9cm of a] (f1) {$Z$};
    		\vertex [below right=1.9cm of a] (f2) {$h$};
    		
    		\diagram* {
    			(i1) -- [fermion] (a) -- [fermion] (i2),
    			(a) -- [boson] (f1),
    			(a) -- [scalar] (f2),
    		};
        \end{feynman}
        \end{tikzpicture}
        \caption*{$d_4$, $d_5$}
    \end{subfigure}
    \caption{Diagrams with effective vertices contributing to \eq{full_cross_section_shift}.}
    \label{fig:higgsstrahlung feynman}
\end{figure}

The full resulting fractional shift for the cross section is
\begin{equation}\label{eq:full_cross_section_shift}
    \frac{\Delta \sigma}{\sigma_0} = 2 \pqty{ \frac{\var g_{ZZh}}{g_{ZZh}} + \frac{g_L \var g_L + g_R \var g_R}{g_L^2 + g_R^2} } + \frac{v}{g_{ZZh}} \sum_{i=2}^5 d_i f_i.
\end{equation}
Here the parameter shifts in the $(\alpha, m_Z, G_F)$ input scheme are
\begin{eqns}
    \frac{\var g_{ZZh}}{g_{ZZh}} &=& \vev^2 \pqty{ C_{H\square} + \frac{1}{4} C_{HD} - \frac{1}{\sqrt{2}} \var \hat G_F },
    \IEEEyesnumber \label{eq:Higgsstrahlung_ParameterShift1} \IEEEyessubnumber
    \\
    \frac{\var g_L}{g_Z} &=& \frac{1}{8(c_w^2 - s_w^2)} \vev^2 \Big( 8s_w c_w C_{HWB} + C_{HD} + 2\sqrt{2} \var \hat G_F \Big) - \frac{1}{2} \vev^2 \pqty{ C_{HL,11}^{(1)} + C_{HL,11}^{(3)} } \qcand
    \IEEEyessubnumber
    \\
    \frac{\var g_R}{g_Z} &=& \frac{s_w^2}{4(c_w^2 - s_w^2)} \vev^2 \bigg( 4\frac{c_w}{s_w} C_{HWB} + C_{HD} + 2\sqrt{2} \var \hat G_F \bigg) - \frac{1}{2} \vev^2 C_{He,11},
    \IEEEyessubnumber
\end{eqns}%
where we adopt the shorthands $s_w \equiv \sin\theta_w$ and $c_w \equiv \cos\theta_w$, and $\var \hat G_F$ is defined in \eq{shift_GF}. The integrated form factors $f_i$ are
\begin{eqns}
    f_2 &=& 12m_Z^2 \frac{s(s + m_Z^2 - m_h^2)}{12sm_Z^2 + \lambda},
    \IEEEyesnumber\IEEEyessubnumber
    \\
    f_3 &=& -12 em_Z^2 \frac{g_L + g_R}{g_L^2 + g_R^2} \frac{(s-m_Z^2)(s + m_Z^2 - m_h^2)}{12s m_Z^2 + \lambda},
    \IEEEyessubnumber
    \\
    f_4 &=& \frac{2g_L}{g_L^2 + g_R^2} (s-m_Z^2) \qcand
    \IEEEyessubnumber
    \\
    f_5 &=& \frac{2g_R}{g_L^2 + g_R^2} (s-m_Z^2),
    \IEEEyessubnumber
\end{eqns}%
and their corresponding coefficients $d_i$ are
\begin{eqns}
    d_2 &=& 4(s_w^2 C_{HB} + s_w c_w C_{HWB} + c_w^2 C_{HW}),
    \IEEEyesnumber\IEEEyessubnumber
    \\
    d_3 &=& -4 s_w c_w C_{HB} - 2(c_w^2 - s_w^2) C_{HWB} + 4s_w c_w C_{HW}
    \IEEEyessubnumber
    \\
    d_4 &=& -g_Z (C_{HL,11}^{(1)} + C_{HL,11}^{(3)}) \qcand
    \IEEEyessubnumber
    \label{eq:direct_contrbs_higgsstrahlung_d4}
    \\
    d_5 &=& -g_Z C_{He,11}.
    \IEEEyessubnumber
\end{eqns}%
The diagrams giving rise to the $d_i$'s are depicted in Figure~\ref{fig:higgsstrahlung feynman}. Note that $d_1$ and $f_1$ are absent here, as we have elected instead to absorb their contribution into $\var g_{ZZh}$. The dimension-six SMEFT operators which constitute these corrections are listed in Table~\ref{table:warsaw basis}. We comment that the $d_i$'s are zero at tree level in the Seesaw models, apart from $d_4$ in Type-III.

\begin{table}[tbp]
    \centering
    \def\arraystretch{1.3}
    \begin{tabular}{|r|l||r|l||r|l|}
        \hhline{|-|-||-|-||-|-|}
        $\mathcal{O}_{HW}$ & $H^\dagger H W_{\mu\nu}^a W^{a\mu\nu}$
        & $\mathcal{O}_{H\square}$ & $(H^\dagger H) \square (H^\dagger H)$
        & $\mathcal{O}_{HL,ij}^{(1)}$ & $(H^\dagger i \lr{D}_\mu H)(\overline{L_i} \gamma^\mu L_j)$ \\
        $\mathcal{O}_{HB}$ & $H^\dagger H B_{\mu\nu} B^{\mu\nu}$
        & $\mathcal{O}_{HD}$ & $(H^\dagger D_\mu H)^* (H^\dagger D^\mu H)$
        & $\mathcal{O}_{HL,ij}^{(3)}$ & $(H^\dagger i \DI H)(\overline{L_i} \sigma^a \gamma^\mu L_j)$ \\
        $\mathcal{O}_{HWB}$ & $H^\dagger \sigma^a H W_{\mu\nu}^a B^{\mu\nu}$
        & $\mathcal{O}_{LL,ijkl}$ & $(\overline{L_i} \gamma^\mu L_j)(\overline{L_k} \gamma_\mu L_l)$
        & $\mathcal{O}_{He,ij}$ & $(H^\dagger i \DI H)(\overline{e_{Ri}} \gamma^\mu e_{Rj})$
        \\
        \hhline{|-|-||-|-||-|-|}
    \end{tabular}
    \caption{Dimension-6 SMEFT operators in the Warsaw basis~\cite{Grzadkowski:2010es} which enter the correction to the Higgsstrahlung cross section. Here $H^\dagger i\DI H \equiv iH^\dagger \sigma^a D_\mu H - i(D_\mu H)^\dagger \sigma^a H$.}
    \label{table:warsaw basis}
\end{table}


\subsubsection{Discussion}

An essential part of the program of most of the proposed next-generation lepton colliders is to be run as ``Higgs factories" at a centre-of-mass energy of $\sqrt{s} = \SI{240}{\GeV}$ (CEPC, FCC-ee) or \SI{250}{\GeV} (ILC), and to perform a scan of the $t\bar t$ production threshold in the range $\sqrt{s} = 350$--\SI{380}{\GeV}. Measurements of the Higgsstrahlung cross section are foreseen at both stages for all collider proposals considered herein, apart from CLIC which is envisioned to directly run at $\sqrt{s} = \SI{380}{\GeV}$ in its initial stage. Therefore, we evaluate the relative shift $\Delta\sigma/\sigma_0$ of the Higgsstrahlung cross section, computed in Eq.~(\ref{eq:full_cross_section_shift}), at $\sqrt{s} = \SI{240}{\GeV}$ and $\SI{365}{\GeV}$. Regarding the precision of the measurement, we assume the two benchmark values of 0.5\% and 1.0\% for $\sqrt{s} = \SI{240}{\GeV}$, and 1.0\% for $\sqrt{s} = \SI{365}{\GeV}$. This is representative of the results of several conducted studies of the attainable precision, which are collected in Table~\ref{tab:forecast_precision_Higgsstrahlung}.

\begin{table}[tb]
    \centering
    \def\arraystretch{1.3}
    \begin{tabular}{|c|c|c|c|c|}
        \hhline{-----} Collider & $L_{\text{int}}\,[\text{ab}^{-1}]$ & $Z$-decay final states & $\sqrt{s}\,[\si{\GeV}]$ & Precision
        \\ \hhline{=====} \multirow{2}{*}{CEPC} & 20 & $\ell^+\ell^-$, $q\bar q$, $\nu\bar\nu$ & 240 & 0.26\%~\cite{CEPCPhysicsStudyGroup:2022uwl}
        \\
        & 1 & $\ell^+\ell^-$, $q\bar q$, $\nu\bar\nu$ & 360 & 1.4\%~\cite{CEPCPhysicsStudyGroup:2022uwl}
        \\
        \hline
        \multirow{2}{*}{FCC-ee} & 5 & $\ell^+\ell^-$ & 240 & 0.5\%~\cite{Bernardi:2022hny}
        \\
        & 1.5 & $\ell^+\ell^-$, $q\bar q$, $\nu\bar\nu$ & 365 &
        0.9\%~\cite{Bernardi:2022hny}
        \\
        \hline
        \multirow{3}{*}{ILC} & 1.35 & $\ell^+\ell^-$ & 250 &
        1.1\%~\cite{Yan:2016xyx}
        \\
        & 0.115\,(0.5) & $\ell^+\ell^-$ ($q\bar q$) & 350 & 5\%\,(1.63\%)~\cite{Yan:2016xyx,Thomson:2015jda}
        \\
        &
        1.6\,(0.5)
        & $\ell^+\ell^-$ ($q\bar q$) & 500 & 2.9\%\,(3.9\%)~\cite{Yan:2016xyx,Miyamoto:2013zva}
        \\
        \hline
        CLIC & 0.5 & $\ell^+\ell^-$, $q\bar q$ & 350 & 1.65\%~\cite{Abramowicz:2016zbo}
        \\ \hhline{-----}
    \end{tabular}
    \caption{Forecast (statistical) precision of measurements of the Higgsstrahlung cross section at different proposed next-generation colliders. In the third column, the $Z$-decay final states taken into account in the respective analysis are given; $\ell^+\ell^-$ always implies both $Z\to e^+e^-$ and $Z\to \mu^+\mu^-$. For the results from Ref.~\cite{Yan:2016xyx} a polarisation $(P_{e^-},P_{e^+}) = (-80\%,+30\%)$ is assumed, still, this changes the expected Higgsstrahlung event rate by maximally 50\% compared to unpolarised beams~\cite{Thomson:2015jda,Abramowicz:2016zbo}. No polarisation is assumed for the ILC precision for $\sqrt{s} = \SI{350}{\GeV}$ and hadronic $Z$-boson decays in Ref.~\cite{Thomson:2015jda}, wherein an attainable precision of 1.76\% for polarised beams and an integrated luminosity of $0.35\;\text{ab}^{-1}$ is reported as well.}
    \label{tab:forecast_precision_Higgsstrahlung}
\end{table}

\begin{table}[tb]
    \centering
    \def\arraystretch{1.3}
    \begin{tabular}{|r||c|c|}
        \hline
        & Type-I & Type-III \\
        \hline\hline
        $\Delta\sigma/\sigma_0$ (\SI{240}{\GeV}) & $0.95\,|\theta_e|^2 + 1.10\,|\theta_\mu|^2 + 0.02\,|\theta_\tau|^2$ & $27.59\, |\theta_e|^2 - 1.08\, |\theta_\mu|^2 - 0.01\, |\theta_\tau|^2$
        \\
        $\Delta\sigma/\sigma_0$ (\SI{365}{\GeV}) & $0.87\, |\theta_e|^2 + 1.12\, |\theta_\mu|^2 + 0.04\, |\theta_\tau|^2$ & $66.15\, |\theta_e|^2 - 1.09\, |\theta_\mu|^2 - 0.01\, |\theta_\tau|^2$
        \\
        $\Delta\sigma/\sigma_0$ (\SI{500}{\GeV}) & $0.80\,|\theta_e|^2 + 1.14\,|\theta_\mu|^2 + 0.05\,|\theta_\tau|^2$ & $126.39\, |\theta_e|^2 - 1.10\, |\theta_\mu|^2 - 0.01\, |\theta_\tau|^2$ \\
        \hline
    \end{tabular}
    \caption{Shifts of the Higgsstrahlung cross section at different centre-of-mass energies in terms of the mixing angles, when $M^X = \SI{1}{\TeV}$. For the sake of comparison, we have also computed them for the larger centre-of-mass energy $\sqrt{s} = \SI{500}{\GeV}$.}
    \label{table:shifts_higgsstrahlung}
\end{table}

The shifts in the cross section as functions of the mixing angles are listed for both Seesaw models in Table~\ref{table:shifts_higgsstrahlung}.
To aid in the following discussion, we moreover find for $\sqrt{s}=\SI{240}{\GeV}$
\begin{equation}
    \frac{\Delta \sigma}{\sigma_0} \approx 0.90\,\hat{C}_{HL,11}^{(1)} + 0.77\,\hat{C}_{HL,11}^{(3)} - 0.13\,\hat{C}_{HL,22}^{(3)},
\end{equation}
and for $\sqrt{s} = \SI{365}{\GeV}$
\begin{equation}
    \frac{\Delta \sigma}{\sigma_0} \approx 2.09\,\hat{C}_{HL,11}^{(1)} + 1.96\,\hat{C}_{HL,11}^{(3)} - 0.13\,\hat{C}_{HL,22}^{(3)},
\end{equation}
where the Wilson coefficients are evaluated at $\mu = \sqrt{s}$, respectively, and as a reminder, we define the dimensionless Wilson coefficients by $\hat{C} = C\times\si{\TeV\squared}$.
These approximate expressions deviate from the exact results, presented in Table~\ref{table:shifts_higgsstrahlung}, by maximally 5\% in either model.

As the couplings of the $Z$ boson to charged leptons are not directly altered at tree level in the Type-I Seesaw model, $\sigma(e^+ e^- \to Zh)$ is predominantly modified via the shift in the Fermi constant, \eq{shift_GF}, which enters through the shifts $\var g_{ZZh}$, $\var g_L$, and $\var g_R$ in Eq.~\eqref{eq:Higgsstrahlung_ParameterShift1}.
While the contributions of electron and muon mixing are of fairly similar magnitudes, $\Delta\sigma/\sigma_0$ turns out to be slightly more sensitive to the latter. This is due to a partial cancellation of the coefficient $d_4$ in Eq.~(\ref{eq:direct_contrbs_higgsstrahlung_d4}) (which acquires a nonzero value from RG running) against the dominant contribution from the Fermi constant for electron mixing. As the corresponding form factor $f_4$ scales with $s$, the resulting sensitivity to electron mixing shrinks even further at higher energies.
Consequently, if sterile neutrinos are to be searched for via precision Higgs measurements, we do not expect running a next-generation lepton collider at higher centre-of-mass energies to reveal much for Type-I.

Contrariwise, in Type-III the couplings of the $Z$ boson to left-handed charged leptons are modified at tree level which results in a sizeable contribution to $d_4$. That is, in an EFT language, the model induces the effective four-point interaction $\overline e\gamma^\mu P_LeZ_\mu h$ depicted on the right in Figure~\ref{fig:higgsstrahlung feynman} which interferes with the tree-level contribution to Higgsstrahlung in the SM. As the latter is suppressed by $s$ due to the $Z$-boson propagator, this results in a very pronounced sensitivity of the ratio $\Delta\sigma/\sigma_0$ to electron mixing which approximately scales with $s$. Consequently, if enough luminosity can be attained to compensate for smaller statistics, fermion triplets may well be searched for in Higgsstrahlung measurements at larger centre-of-mass energies. Note that the contributions from electron mixing could in principle be (partly) cancelled by large muon mixing, however, we will find that this scenario is tightly constrained by existing phenomenological bounds. From Table~\ref{table:shifts_higgsstrahlung} one can immediately deduce that if only electron mixing is sizeable, a minimal shift of $\Delta\sigma/\sigma_0 \ge 1\%$ for $\sqrt{s} = \SI{240}{\GeV}$ requires $|\theta_e|\gtrsim 0.019$, whereas $|\theta_e|\gtrsim 0.013$ is sufficient if the relative cross-section shift can be as small as 0.5\%, or if $\sqrt{s} = \SI{365}{\GeV}$ is considered instead.


\subsection{Electroweak Sector}
\label{sec:electroweak sector}

In this section, we introduce the shifts in the weak mixing angle and the mass of the $W$ boson. The expressions obtained for the fermionic Seesaw models in terms of the mixing angles are listed in Table~\ref{table:electroweak_shifts}.

\begin{table*}[tbp]
    \centering\def\arraystretch{1.3}
    \begin{tabular}{|c|cl|cl|}
    \hline
    \multicolumn{5}{|c|}{\textsc{ Electroweak Sector}}\\
        \hline
        Observable & \multicolumn{2}{|c|}{SM prediction} & \multicolumn{2}{|c|}{Measurement} \\
        \hline	$\sin^2(\theta^\text{lept}_\text{eff})$ & $0.231534\pm 0.000030$ & \cite{deBlas:2021wap} & $0.23153\pm 0.00026$ & \cite{ALEPH:2005ab}
        \\
        $m_W\,[\si{\GeV}]$ & $80.356\pm 0.006$ & \cite{Workman:2022ynf} & $80.377\pm 0.012$ & \cite{Workman:2022ynf} \\
        \hline
    \end{tabular}
    \vspace{0.25cm}
    \\
    \begin{tabular}{|r||c|c|}
        \hline Shift
        & Type-I & Type-III \\
        \hline\hline
        $\var s_w^2$ & $- 0.157(|\theta_e|^2 + |\theta_\mu|^2) + 0.003\,|\theta_\tau|^2$ & $0.017(|\theta_e|^2 + |\theta_\mu|^2) - 0.143\,|\theta_\tau|^2$ \\
        $\var m_W/\si{\GeV}$ & $8.24(|\theta_e|^2 + |\theta_\mu|^2) - 0.13\,|\theta_\tau|^2$ & $- 8.51(|\theta_e|^2 + |\theta_\mu|^2) - 0.13\,|\theta_\tau|^2$ \\
        \hline
    \end{tabular}
    \caption{SM predictions for and current measurements of the electroweak observables considered in this work, together with approximate expressions for their shifts in terms of the mixing angles. The $W$-boson mass listed in the PDG~\cite{Workman:2022ynf} corresponds to the mass parameter in a Breit-Wigner distribution with a mass-dependent width. The SM prediction for the effective leptonic weak mixing angle is taken from Table II in Ref.~\cite{deBlas:2021wap}. The model predictions are obtained from matching onto SMEFT at the Seesaw scale $\mu = M^X = \SI{1}{\tera\electronvolt}$ and running to the electroweak scale $\mu = m_Z$.}
    \label{table:electroweak_shifts}
\end{table*}


\subsubsection{Weak Mixing Angle}
\label{sec:weak_mixing_angle}

In SMEFT the weak mixing angle is modified in accordance with (see e.g.~Ref.~\cite{Berthier:2015oma})
\begin{equation}\label{eq:weak_mixing_angle_SMEFT}
    \var s_w^2 = \frac{c_w s_w}{c_w^2 - s_w^2} \vev^2 \bigg( \frac{1}{2} c_w s_w C_{HD} +  C_{HWB} + \sqrt{2} c_w s_w \var \hat G_F \bigg),
\end{equation}
where the shift in the Fermi constant $\var \hat G_F$ is defined in Eq.~\eqref{eq:shift_GF}. There are numerous ways to extract the weak mixing angle from data; the most precise determination is that of the \emph{effective leptonic weak mixing angle} $s^2_{w,\text{eff}}\equiv\sin^2(\theta^\text{lept}_{w,\text{eff}})$ at LEP~\cite{ALEPH:2005ab}, achieved via measurements of the left-right asymmetry factor
\begin{equation}
    \mathcal{A}_f = \frac{g_L^2 - g_R^2}{g_L^2 + g_R^2} = \frac{2(1-4s_w^2)}{1+(1-4s_w^2)^2}.
\end{equation}
Apart from the general shift in Eq.~(\ref{eq:weak_mixing_angle_SMEFT}), we must also take into account the fact that a modification of the $Z$ couplings to charged leptons will directly affect the extraction of $s^2_{w,\text{eff}}$ from $\mathcal{A}_f$. Incorporating the ``direct'' shifts to these couplings,
\begin{equation}
    \var g_{L,ij}^{\text{direct}} = -\frac{1}{2} g_Z \vev^2 \left(C_{HL,ij}^{(1)} + C_{HL,ij}^{(3)}\right) 
    \qand
    \var g_{R,ij}^{\text{direct}} = -\frac{1}{2} g_Z \vev^2 C_{He,ij},
\end{equation}%
we find
\begin{eqns}\label{eq:weak_mixing_angle}
    s_{w,\text{eff}}^2 &=& s_{w,\text{SM}}^2 + \delta s_w^2 + \frac{1}{3} \dv{s_w^2}{\mathcal{A_\ell}} \pqty{ \pdv{\mathcal{A_\ell}}{g_L}\sum_{i=1}^3\delta g_{L,ii}^{\text{direct}} + \pdv{\mathcal{A_\ell}}{g_R}\sum_{i=1}^3\delta g_{R,ii}^{\text{direct}} }
    \\\nonumber
    &\approx& s_{w,\text{SM}}^2 + 0.020\left(\hat{C}_{HL,11}^{(3)} + \hat{C}_{HL,22}^{(3)}\right) - 0.005\sum_{i=1}^3\left(\hat{C}_{HL,ii}^{(1)} + \hat{C}_{HL,ii}^{(3)}\right),
\end{eqns}%
where the right-hand side is evaluated at the scale $\mu = m_Z$. Contributions from $C_{He}$ are not sourced at tree level in either fermionic Seesaw model and thus we neglect them in the approximate expression above.


\subsubsection{\texorpdfstring{$W$}{W}-Boson Mass}

The shift incurred in SMEFT is \cite{Berthier:2015oma}
\begin{equation}
    \frac{\var m_W^2}{m_W^2} = -\frac{1}{2(c_w^2 - s_w^2)} \vev^2 \pqty{ 4c_w s_w C_{HWB} + c_w^2 C_{HD} + 2\sqrt{2} s_w^2 \var \hat G_F },
\end{equation}
which approximately evaluates to
\begin{equation}\label{eq:Wmass_approx}
    m_W \approx m_{W,\text{SM}} - \num{1.05}\left(\hat{C}_{HL,11}^{(3)} + \hat{C}_{HL,22}^{(3)}\right)\si{\GeV}
\end{equation}
at the scale $\mu = m_Z$. Evidently, the predicted $W$-boson mass is mainly sensitive to $\mathcal{O}^{(3)}_{HL}$ as the major correction is induced via the modified extraction of the Fermi constant in the fermionic Seesaw models.


\subsubsection{Discussion}

\begin{figure}[tbp]
    \centering
    \includegraphics[width=\linewidth]{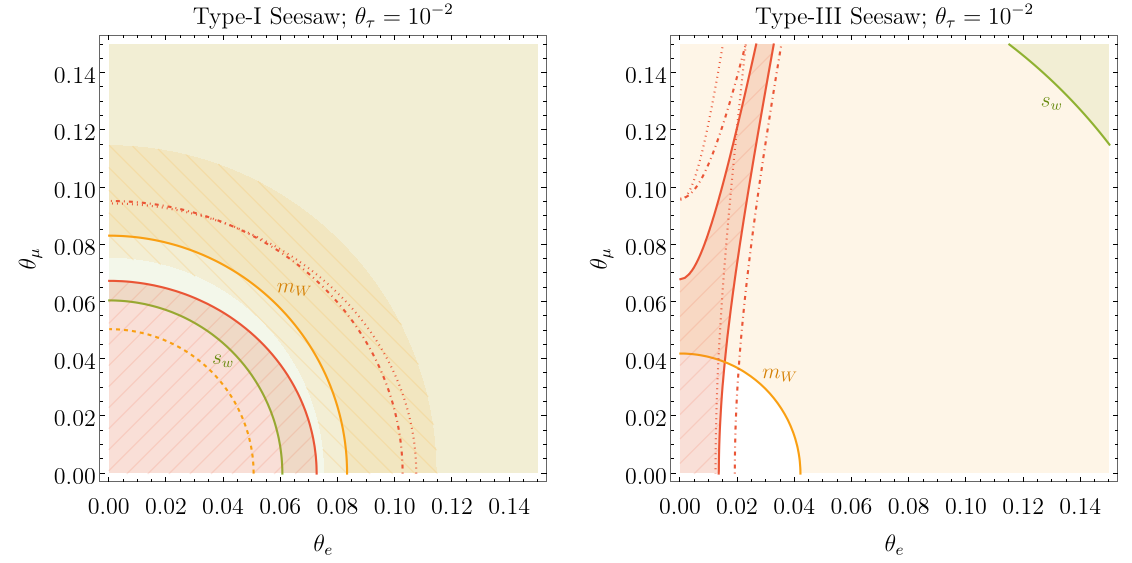}
    \caption{Current constraints from electroweak observables at $2\sigma$, in comparison with projected sensitivities of precision Higgsstrahlung measurements. The red-ruled regions indicate $|\Delta\sigma/\sigma| < 0.5\%$ at $\sqrt{s} = \SI{240}{\GeV}$. The dot-dashed and dotted red lines are the corresponding 1\% contours at $\sqrt{s} = \SI{240}{\GeV}$ and \SI{365}{\GeV}, respectively. 
    The ivory regions for large $\theta_{e,\mu}$ in the upper-right of the plots are excluded by measurements of both $s_w$ and $m_W$.
    For Type-I Seesaw, the orange dashed line marks where the current experimental world average for $m_W$ is exactly accommodated, and in the orange-ruled region the CDF measurement~\cite{CDF:2022hxs} is explained at $2\sigma$.}
    \label{fig:electroweak}
\end{figure}

The constraints arising from the electroweak observables $s_{w,\text{eff}}^2$ and $m_W$ are illustrated in Figure \ref{fig:electroweak}.

For Type-I Seesaw one immediately notices that observing a deviation in the Higgsstrahlung cross section is already in direct conflict with the determination of $s_{w,\text{eff}}^2$, with a 0.5\% shift at $\sqrt{s} = \SI{240}{\GeV}$ suffering from a $2.9\sigma$ tension, and a 1\% shift excluded at $\sim 6\sigma$. While this tension can be reduced by turning up $\theta_\tau$ (see the expression in Table~\ref{table:electroweak_shifts}), LFU constraints discussed in Section~\ref{sec:LFU} below preclude this from occurring.
Thus, the Type-I Seesaw model is unlikely to be a viable minimal SM extension that can be probed in precision Higgs measurements, unless a significant reduction of the statistical uncertainty of these measurements can be attained.

In the Type-III Seesaw model, the effects of the triplets contributing to the extraction of the Fermi constant and from direct contributions to the leptonic gauge couplings largely cancel out, which equally applies to electron and muon flavour, see Eq.~(\ref{eq:weak_mixing_angle}). Thus, $s_{w,\text{eff}}^2$ acts as a rather weak constraint on the Type-III Seesaw model, and is in fact most relevant for tau-flavour mixing, implying $|\theta_\tau| \lesssim 0.06$ at $2\sigma$.

The existing tension between the SM prediction for the $W$-boson mass $m_W$ and the larger experimental world average is however exacerbated in Type-III, leading to a much stronger constraint. In contrast, as $C_{HL}^{(3)}$ is induced with equal magnitude, but opposite signs via matching at the Seesaw scale in the two models under consideration, the tension can be alleviated in Type-I; if uncertainties are ignored, the current world average is reproduced for $\sqrt{|\theta_e|^2 + |\theta_\mu|^2} \approx 0.051$, and the CDF measurement~\cite{CDF:2022hxs} for $\sqrt{|\theta_e|^2 + |\theta_\mu|^2} \approx 0.097$. A detectable shift in $\sigma(e^+ e^- \to Zh)$ could in principle only be induced in the latter case.

Testing the Type-III Seesaw model via Higgsstrahlung measurements is generally compatible with the current constraint arising from $m_W$. Still, it disfavours detectable $\Delta\sigma/\sigma_0$ induced via muon mixing, and clearly prefers contributions from electron mixing. The bound from $m_W$ can be expected to become more competitive when the CDF measurement will be included in the experimental average in the future.


\subsection{Lepton Flavour Universality}
\label{sec:LFU}

In SMEFT the $W\ell\nu$ coupling is directly altered due to $\mathcal{O}_{HL}^{(3)}$:
\begin{equation}\label{eq:W coupling}
    \mathcal{L} \supset -i \frac{g_2}{\sqrt{2}} \pqty{ \delta_{ij} + \vev^2 C_{HL,ij}^{(3)} } \overline{\ell_i}\gamma^\mu P_L\nu_jW_\mu.
\end{equation}
This modifies (semi-)leptonic decays mediated by $W$ bosons, which, among other effects, drives the predictions for LFU ratios away from 1.
In the following we consider ratios constituted by the leptonic decays $\pi \to \ell  \nu$, $K \to \ell \nu$ and $\tau\to \ell \nu \bar \nu$. Furthermore, we utilise the extraction of the Cabibbo-Kobayashi-Maskawa (CKM)-matrix element $V_{us}$ from the semi-leptonic decay $K_{\ell3}: K\to \pi\ell\nu$ and from nuclear beta decays, together with the assumption of unitarity of the CKM matrix.


\subsubsection{Ratios of Leptonic Gauge Couplings}

\begin{table}[tbp]
    \centering
    \def\arraystretch{1.5}
    \begin{tabular}{|c|cl|c|}
        \hline
        \multicolumn{4}{|c|}{\textsc{Lepton Flavour Universality}}\\
        \hline
        Observable & \multicolumn{2}{c|}{Measurement} & Model prediction
        \\
        \hline
        $g^\pi_{\mu/e}$
        & $1.0010\pm0.0009$ & \cite{Pich:2020qna} & \multirow{4}{*}{$ 1 \pm 0.48 (|\theta_e|^2 - |\theta_\mu|^2)$}
        \\
        $g^\ell_{\mu/e}$ & $1.0017\pm0.0016$ & \cite{Pich:2020qna} &
        \\
        $g^K_{\mu/e}$ & $0.9978\pm0.0018$ & \cite{Pich:2020qna} &
        \\
        $R(K_{\ell3})$ & $1.001295\pm 0.002891$ & \cite{Seng:2022wcw} &
        \\
        \hline
        $g^\pi_{\tau/\mu}$ & $0.9965\pm0.0026$ & \cite{Pich:2020qna} & \multirow{2}{*}{$ 1 \pm 0.48 (|\theta_\mu|^2 - |\theta_\tau|^2)$}
        \\
        $g^\ell_{\tau/\mu}$ & $1.0011\pm0.0014$ & \cite{Pich:2020qna} &
        \\
        \hline
        $R(V_{us})$ & $0.98898\pm 0.00606$ & \cite{Workman:2022ynf} &
        $1 \pm 0.47 |\theta_e|^2 \pm 8.80 |\theta_\mu|^2 \mp 0.04 |\theta_\tau|^2$
        \\
        \hline
    \end{tabular}
    \caption{Current constraints on and model predictions for the LFU ratios taken into account in this work. For $R(V_{us})$, the given experimental value refers to the case of $N_f = 2 + 1 + 1$ dynamical quark flavours in the lattice simulations from which the relevant decay constants are extracted. The case of $N_f = 2 + 1$ quark flavours gives rise to a less competitive bound~\cite{Workman:2022ynf}. The model predictions are obtained from matching onto SMEFT at the Seesaw scale $\mu = M^X = \SI{1}{\tera\electronvolt}$ and running to the electroweak scale $\mu = m_Z$. In the right column, the upper sign refers to Type-I Seesaw, whereas the lower sign stands for Type-III Seesaw.}
    \label{table:LFU}
\end{table}

We consider the ratios of leptonic gauge couplings collected in Table~\ref{table:LFU} which are probes of LFU. For the first three listed therein, \cite{Crivellin:2020lzu,Bryman:2021teu}
\begin{equation}
    g^X_{\mu/e} \equiv \left(\frac{g_\mu}{g_e}\right)^X
    \approx 1 + \vev^2 C_{HL,22}^{(3)} - \vev^2 C_{HL,11}^{(3)}
    \approx 1 + 0.06\left(\hat{C}_{HL,22}^{(3)} - \hat{C}_{HL,11}^{(3)}\right),
\end{equation}
where $(g_\mu/g_e)^X$ is the ratio of leptonic gauge couplings extracted from the ratio $\Gamma(X\to\mu\nu)/\Gamma(X\to e\nu)$ for $X=\pi,K$, 
and $\Gamma(\tau\to\mu \nu\bar\nu)/\Gamma(\tau\to e\nu\bar\nu)$ for $X=\ell$.\footnote{The flavour of the neutrinos produced in all these decays is assumed to coincide with the one of the respective associated charged lepton, which allows for interference with the SM and can thus be expected to represent the channel dominantly affected by BSM physics.}
Leptonic $W$-boson decays can also be used to derive constraints on the ratios of leptonic gauge couplings as in $g^W_{\mu/e}\equiv (g_\mu/g_e)^W$ and $g^W_{\tau/\mu}\equiv(g_\tau/g_\mu)^W$; still, we do not list them in the table, as the corresponding bounds are merely weaker versions of other constraints. We similarly have
\begin{equation}
    g^X_{\tau/\mu} \equiv \left(\frac{g_\tau}{g_\mu}\right)^X
    \approx 1 + 0.06\left(\hat{C}_{HL,33}^{(3)} - \hat{C}_{HL,22}^{(3)}\right),
\end{equation}
where $(g_\tau/g_\mu)^X$ is the ratio of leptonic gauge couplings extracted from the ratio $\Gamma(\tau\to e \nu\bar\nu)/\Gamma(\mu\to e\nu\bar\nu)$ for $X=\ell$, and $\Gamma(\tau\to\pi \nu)/\Gamma(\pi\to\mu\nu)$ for $X=\pi$.

In our Seesaw models the predicted deviation of the considered LFU ratios from 1 is at leading order proportional to $\pm(|\theta_\mu|^2 - |\theta_j|^2)$ where $j = e$ or $\tau$, and the sign depends on $j$ as well as which Seesaw model is considered, see Table~\ref{table:LFU}. The derived constraints therefore give rise to hyperbolic contours in the figures presented in this section. Clearly, if the data favours a ratio to be, say, smaller than 1, predicting it to be larger than 1 will lead to a tighter constraint. This then translates into one of the mixing angles being slightly more stringently bounded than another one, with the roles reversed if the other Seesaw model is considered. Consequently, if the contribution from a specific mixing angle accommodates the data well in one model, the other model will necessarily increase the tension with the SM.


\subsubsection{Light Quark Mixing}

In Table~\ref{table:LFU} we furthermore consider ratios of the CKM-matrix element $V_{us}$ extracted from the semi-leptonic kaon decays $K_{\mu3}$ and $K_{e3}$, the leptonic kaon decay $K_{\mu 2}$, and nuclear beta decay -- these are
\begin{equation}
    R(K_{\ell3}) \equiv \frac{V_{us}^{K_{\mu3}}}{V_{us}^{K_{e3}}} \qand R(V_{us}) \equiv \frac{V_{us}^{K_{\mu2}}}{V_{us}^\beta}.
\end{equation}
The dependence of $R(K_{\ell3})$ on new physics is identical to that of $g^\pi_{\mu/e}$ and $g^K_{\mu/e}$, of which both can similarly be viewed as CKM ratios, with the latter being commonly denoted by $R(K_{\ell2})$. In the following we illustrate why $R(V_{us})$ is a special case, and refer the reader to Ref.~\cite{Crivellin:2020lzu}, wherein the ratio was originally proposed, for further information.

The extracted value of $V_{us}$ from the leptonic kaon decay $K_{\mu 2}$ is
\begin{equation}
    V_{us}^{K_{\mu 2}} = V_{us} \left(1 + \vev^2 C_{HL,22}^{(3)} - \frac{\var \hat G_F}{G_F}\right),
\end{equation}
where $V_{us}$ is a Lagrangian parameter, the $C_{HL,22}^{(3)}$ is due to the new direct contribution, \eq{W coupling}, and $\var \hat G_F/G_F$ parameterises the shift in the Fermi constant. Similarly, beta decay results in 
\begin{equation}
    V_{ud}^\beta = V_{ud} \left( 1 + \vev^2 C_{HL,11}^{(3)} - \frac{\var \hat G_F}{G_F}\right),
\end{equation}
which, as per Ref.~\cite{Crivellin:2020lzu}, is translated to $V_{us}^\beta$ using CKM unitarity of the Lagrangian parameters:
\begin{equation}
    V_{us}^\beta = \sqrt{1-|V_{ud}^\beta|^2 - |V_{ub}|^2}
    \approx V_{us} \bqty{1 - \pqty{ \frac{V_{ud}}{V_{us}} }^2 \left(\vev^2 C_{HL,11}^{(3)} - \frac{\var \hat G_F}{G_F}\right) }.
\end{equation}
From this we derive the ratio
\begin{eqns}
    R(V_{us}) &=& \frac{V_{us}^{K_{\mu2}}}{V_{us}^\beta}
    = 1 + \vev^2 C_{HL,22}^{(3)} - \frac{\var \hat G_F}{G_F} + \pqty{\frac{V_{ud}}{V_{us}}}^2 \pqty{ \vev^2 C_{HL,11}^{(3)} - \frac{\var \hat G_F}{G_F} }
    \\\nonumber
    &\approx& 1 - \vev^2 \bqty{ \pqty{ \frac{V_{ud}}{ V_{us}} }^2 C_{HL,22}^{(3)} + C_{HL,11}^{(3)} }.
\end{eqns}%
As is explained in Ref.~\cite{Crivellin:2020lzu}, a crucial feature of $R(V_{us})$ is the enhanced sensitivity to new physics due to $(V_{ud}/V_{us})^2 \approx 20$. Experimental data favours the ratio to be smaller than 1 with a significance between $1\sigma$ and $2\sigma$, depending on the number of quark flavours assumed for the calculation of the relevant decay constant, see also Table~\ref{table:LFU}.


\subsubsection{Discussion}

\begin{figure}[h!]
    \centering
    \begin{subfigure}[t]{\textwidth}
        \centering
        \includegraphics[width=\linewidth]{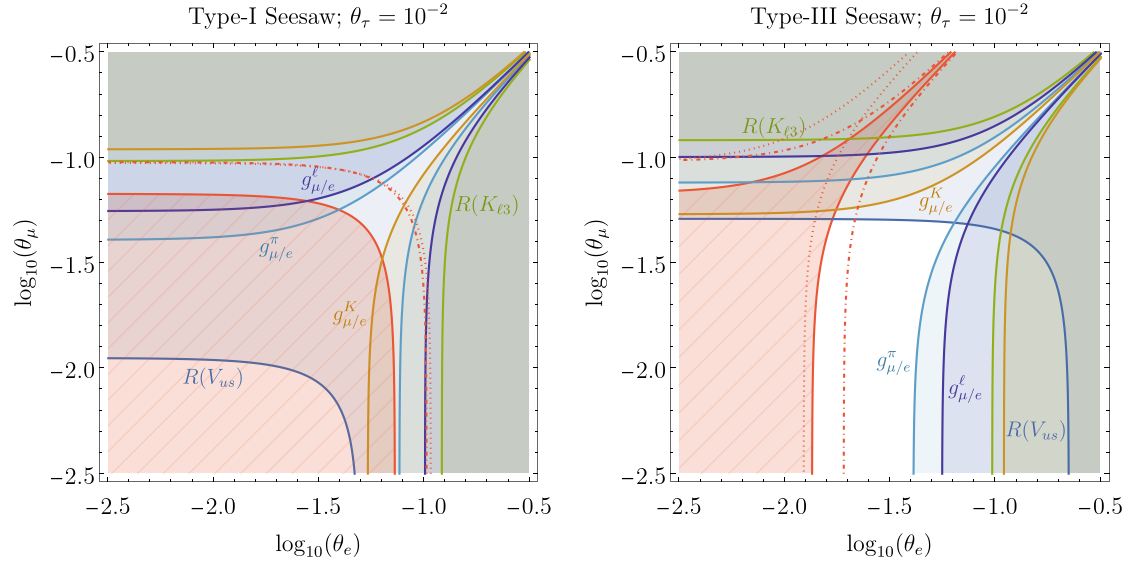}
        \caption{LFU constraints in the $\theta_e$--$\theta_\mu$ plane.}
        \label{fig:LFU_e_mu}
    \end{subfigure}
    \begin{subfigure}[t]{\textwidth}
        \centering
        \includegraphics[width=\linewidth]{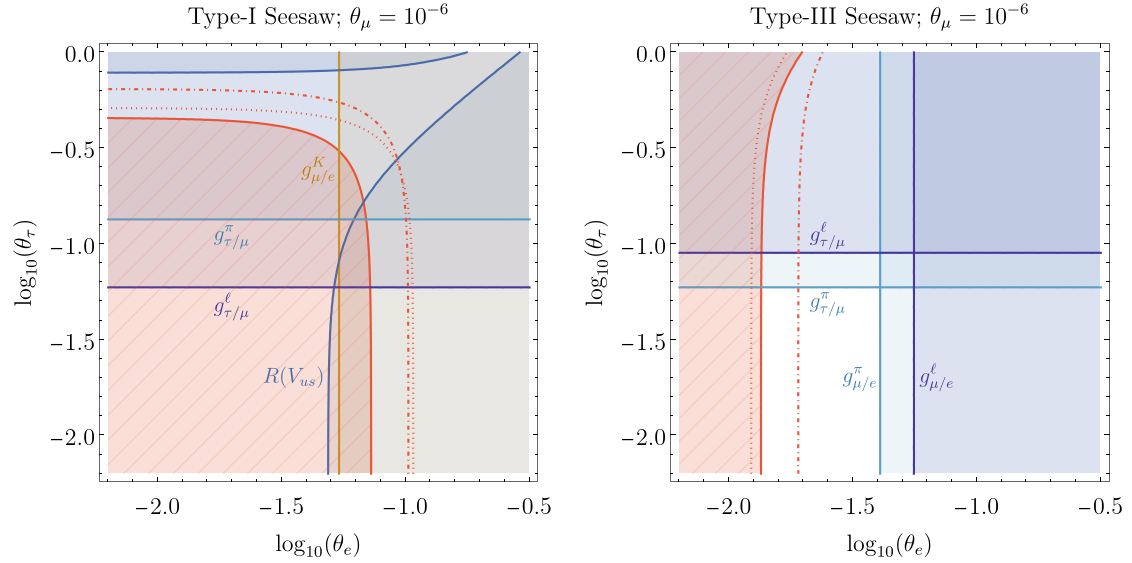}
        \caption{LFU constraints in the $\theta_e$--$\theta_\tau$ plane. To avoid clutter, subdominant constraints like the one from $R(K_{\ell3})$ are not drawn, but their locations may be inferred from the above plots.}
        \label{fig:LFU_e_tau}
    \end{subfigure}
    \caption{Current constraints arising from LFU ratios at $2\sigma$ in comparison with projected sensitivities of precision Higgsstrahlung measurements. The red-ruled regions indicate $|\Delta\sigma/\sigma| < 0.5\%$ at $\sqrt{s} = \SI{240}{\GeV}$. The dot-dashed and dotted red lines are the corresponding 1\% contours at $\sqrt{s} = \SI{240}{\GeV}$ and \SI{365}{\GeV}, respectively.}
    \label{fig:LFU}
\end{figure}

In Figure \ref{fig:LFU}, the constraints arising from LFU ratios are illustrated.
Among those, $R(V_{us})$ is very sensitive to muon mixing and highly relevant for the fermionic Seesaw models.
For Type-I Seesaw, the ratio is enhanced and thus driven further away from the data. In fact, $R(V_{us})$ is in conflict with any visible effect in Higgsstrahlung induced via electron or muon mixing, even more so than $s_w$. The next-to-most competitive bounds on electron and muon mixing arise from $g^K_{\mu/e}$ and $g^\pi_{\mu/e}$, respectively.

In the case of Type-III Seesaw, the most important constraints stem from $g^\pi_{\mu/e}$ and $R(V_{us})$. Since requiring a discernible deviation in the Higgsstrahlung cross section mainly translates into a lower bound on the electron mixing angle in this model and muon mixing may thus be tuned arbitrarily small, the obtained upper limits on $|\theta_e|$ are more relevant in this context. $g^\pi_{\mu/e}$ demands $|\theta_e|\lesssim\num{0.04}$ at $2\sigma$, unless one allows for larger muon mixing and thus a cancellation, whereupon the maximal electron mixing angle allowed by LFU bounds can be rendered up to 50\% larger.
$R(V_{us})$ implies $|\theta_\mu|\lesssim\num{0.05}$ at $2\sigma$, which holds largely independently of the value chosen for $|\theta_e|$.
Note that in a vein similar to $m_W$, the bounds from $g^K_{\mu/e}$ and $R(V_{us})$ both constrain muon mixing efficiently enough so that no appreciable cancellations of the contributions to Higgsstrahlung from electron mixing can occur for Type-III Seesaw.

As can be seen in Figure~\ref{fig:LFU_e_tau}, LFU data constrains tau mixing to $|\theta_\tau| \lesssim 0.06$ for either Seesaw model, which arises from $g^\ell_{\tau/\mu}$ in Type-I, and $g^\pi_{\tau/\mu}$ for Type-III. These constraints can in principle be weakened if $|\theta_\mu|$ is more sizeable; still, large changes are only observed for~$\order{0.1}$ muon mixing, a scenario which is nonetheless excluded by other observables. While the contributions to $R(V_{us})$ from electron and tau mixing may cancel, this does not open up parameter space in the Type-I Seesaw model, as tau mixing itself is too constrained.

Further LFU ratios not contained in Table~\ref{table:LFU} deviate from the SM prediction by (close to) $2\sigma$ and thus present moderate anomalies in themselves, see Ref.~\cite{Pich:2020qna}. Explicitly, $g^K_{\tau/\mu}$ is measured to be smaller than 1, while $g^\ell_{\tau/e}$ and $g^W_{\tau/e}$ exceed the SM expectation. If the models under consideration ought to accommodate the data on $g^K_{\tau/\mu}$, one would require large tau mixing in comparison with muon mixing for Type-I Seesaw, with the flavours swapped for Type-III Seesaw. The latter is unlikely to be realised for scenarios which are testable via Higgsstrahlung measurements, as large $\Delta\sigma/\sigma_0$ are likely induced via electron mixing in this model, which then demands muon mixing to be very small due to the bounds arising from LFV, see the following section. Similarly, for Type-III Seesaw, the other two ratios necessitate tau mixing to substantially exceed electron mixing in magnitude, which is not a promising scenario either to be tested in the given context, in particular in light of the bound on BR($\tau\to 3e$).


\subsection{Lepton Flavour Violation}
\label{sec:LFV}

\begin{table}[tbp]
    \centering
    \def\arraystretch{1.3}
    \begin{NiceTabular}{|c|cl|cl|c|c|}
        \hline
        \multicolumn{7}{|c|}{\textsc{Lepton Flavour Violation}}\\
        \hline
        \multirow{2}{*}{Observable}
        & \multicolumn{4}{c|}{Experiment} & \multicolumn{2}{c|}{Model predictions} \\
        \cline{2-7}
        & \multicolumn{2}{c|}{Current bound} & \multicolumn{2}{c|}{Future reach} & Type-I & Type-III \\
        \hline
        $\mathrm{BR}(\mu\to e\gamma)$ & \num{4.2E-13}
        & \cite{MEG:2016leq} & \num{6E-14}
        & \cite{MEGII:2021fah} & $\num{0.82e-3} \,|\theta_e \theta_\mu|^2$ & $\num{1.27e-3} \,|\theta_e \theta_\mu|^2$
        \\
        $\mathrm{BR}(\mu\to 3e)$ & \num{1E-12}
        & \cite{SINDRUM:1987nra} &
        \num{1E-16}
        & \cite{Blondel:2013ia} & $\num{0.14e-3}\,|\theta_e \theta_\mu|^2$ & $\num{0.72}\,|\theta_e \theta_\mu|^2$
        \\
        CR($\mu - e;\;$Au) & \num{7E-13}
        & \cite{SINDRUMII:2006dvw} & --- & & $\num{0.04e-3}\,|\theta_e \theta_\mu|^2$ & $27.1\,|\theta_e \theta_\mu|^2$
        \\
        CR($\mu - e;\;$Al) & --- & &
        \makecell{\num{2.6E-17} \\ \num{8E-17}}
        & \makecell{\cite{COMET:2018auw} \\ \cite{Hedges:2022tnh}}
        & $\num{0.15e-3}\,|\theta_e \theta_\mu|^2$ & $6.7\,|\theta_e \theta_\mu|^2$
        \\
        CR($\mu - e;\;$Ti) & \num{6.1e-13} & \cite{Wintz:1998rp} & $\order{10^{-18}}$ & \cite{Kuno:2005mm} & $\num{0.18e-3}\,|\theta_e \theta_\mu|^2$ & $13.5\,|\theta_e \theta_\mu|^2$
        \\
        CR($\mu - e;\;$Pb) & \num{4.6e-11} & \cite{SINDRUMII:1996fti} & --- & & $\num{0.02e-3}\,|\theta_e \theta_\mu|^2$ & $20.3\,|\theta_e \theta_\mu|^2$
        \\
        CR($\mu - e;\;$S) & $\num{7e-11}$ & \cite{Badertscher:1980bt} & --- & & $\num{0.21e-3}\,|\theta_e \theta_\mu|^2$ & $6.4\,|\theta_e \theta_\mu|^2$
        \\
        $\mathrm{BR}(\tau\to e\gamma)$ & \num{3.3E-8}
        & \cite{BaBar:2009hkt} & \num{9E-9} & \cite{Banerjee:2022xuw} & $\num{0.15e-3} \,|\theta_e \theta_\tau|^2$ & $\num{0.23e-3} \,|\theta_e \theta_\tau|^2$
        \\
        $\mathrm{BR}(\tau\to 3e)$ & \num{2.7E-8}
        & \cite{Hayasaka:2010np} & \num{4.7E-10} & \cite{Banerjee:2022xuw} & $\num{0.02e-3}\,|\theta_e \theta_\tau|^2$ & $\num{0.13}\,|\theta_e \theta_\tau|^2$
        \\
        \hline
    \end{NiceTabular}
    \caption{Current and projected constraints on the LFV observables taken into account in this work, together with the pertinent model predictions. The current bounds hold at 90\% C.L. The future reach listed for BR$(\mu\to3e)$ refers to Phase II of the Mu3e experiment; an initial sensitivity of BR$(\mu\to3e)\lesssim\num{2E-15}$ is expected after Phase I. The upper (lower) value listed for the future reach of CR($\mu - e;\;$Al) refers to COMET (Mu2e). The model predictions are obtained from matching onto SMEFT at the Seesaw scale $\mu = M^X = \SI{1}{\tera\electronvolt}$, running to the electroweak scale $\mu = m_Z$, matching onto LEFT and running to the low scale $\mu = \SI{5}{\GeV}$.}
    \label{table:LFV}
\end{table}

As is generically the case for models of neutrino mass generation, the Seesaw models predict sizeable rates for flavour-violating decays of charged leptons. These processes have not been observed to date and thus impart stringent bounds on the parameter space, which will likely get refined in the near future due to several ongoing or upcoming experiments, see Table~\ref{table:LFV}. The scales relevant for these LFV decays are within the realm of LEFT and thus it is instrumental to discuss them in terms of the contributions to LEFT operators. Since we focus on the comparison with the sensitivities to the Higgsstrahlung process at colliders, we restrict ourselves to observables involving electron-flavoured transitions. A comprehensive investigation of LFV effects in the symmetry-protected Type-I Seesaw model can for instance be found in Ref.~\cite{Crivellin:2022cve}. We will relegate the explicit matching conditions used in this section to Appendix~\ref{sec:approx_matching}.


\subsubsection{Radiative Charged-Lepton Decays}

The branching ratios for radiative flavour-violating charged-lepton decays read~\cite{Kuno:1999jp} 
\begin{equation}
    \mathrm{BR}(\ell_i\to \ell_j\gamma) = \frac{m_{\ell_i}^3}{4\pi \Gamma_{\ell_i}} \left(|C_{e\gamma,ij}|^2 + |C_{e\gamma,ji}|^2\right),
\end{equation}
with the full decay width $\Gamma_{\ell_i}$. We approximately find
\begin{eqns}
    \mathrm{BR}(\mu\to e\gamma) &\approx& \num{7.117E6}\vqty{ \hat{C}_{eB,12} - 0.55\,\hat{C}_{eW,12} + \left(1.77\,\hat{C}_{HL,12}^{(3)} - 0.48\,\hat{C}_{HL,12}^{(1)}\right)\num{E-6} }^2 \text{ and}
    \\
    \mathrm{BR}(\tau\to e\gamma) &\approx& \num{0.004E6}\vqty{ \hat{C}_{eB,13} - 0.55\,\hat{C}_{eW,13} + \left(29.69\,\hat{C}_{HL,13}^{(3)} - 8.12\,\hat{C}_{HL,13}^{(1)}\right)\num{E-6} }^2,
\end{eqns}%
where the SMEFT Wilson coefficients on the right are evaluated at the electroweak scale $\mu = m_Z$. In both Seesaw models the one-loop matching contributions to the electromagnetic dipole operator $\mathcal{O}_{e\gamma}$ from the electroweak dipole operators $\mathcal{O}_{eB}$ and $\mathcal{O}_{eW}$ are of the same order of magnitude as the contributions from $\mathcal{O}_{HL}^{(1)}$ and $\mathcal{O}_{HL}^{(3)}$ which originate from RG running, see also Eq.~(\ref{eq:dipole_matching}) in Appendix~\ref{sec:approx_matching}.


\subsubsection{Trilepton Decays}

The branching ratio for trilepton decays with identical flavours in the final state is given by~\cite{Calibbi:2021pyh}\footnote{See also \cite{Kuno:1999jp,Brignole:2004ah} for earlier work. In the case of $\tau$ decays, we do not expect significantly more stringent constraints if some of the final-state electrons are swapped for muons.}
\begin{eqns}
    \mathrm{BR}(\ell_i\to \ell_j\ell_j \overline{\ell_j}) &=&  \frac{m_{\ell_i}^5}{3(16\pi)^3 \Gamma_{\ell_i}}
    \Bigg[64\left|C_{ee,jijj}^{VLL}\right|^2 + 64\left|C_{ee,jijj}^{VRR}\right|^2 + 8 \left|C_{ee,jijj}^{VLR}\right|^2 + 8 \left|C_{ee,jjji}^{VLR}\right|^2 
    \\\nonumber &&
    \quad {} + \frac{256 e^2}{m_{\ell_i}^2} \left(\ln \frac{m_{\ell_i}^2}{m_{\ell_j}^2} -\frac{11}{4}\right)\left(\left|C_{e\gamma}^{ij}\right|^2 + \left|C_{e\gamma}^{ji}\right|^2\right)
    \\\nonumber &&
    \quad {} - \frac{64 e}{m_{\ell_i}} \Re\Big[\left(4\, C_{ee,jijj}^{VLL}+C_{ee,jijj}^{VLR}\right) C_{e\gamma}^{ji*} + \left(4\, C_{ee,jijj}^{VRR} + C_{ee,jjji}^{VLR}\right)C_{e\gamma}^{ij}\Big]\Bigg].
\end{eqns}%
In the Type-III Seesaw model, these decays are dominated by the vector operators $\mathcal{O}^{VLL}_{ee}$ and $\mathcal{O}^{VLR}_{ee}$, with the flavour change occurring in the left-handed lepton bilinear. These operators receive large contributions from tree-level matching of the Type-III Seesaw model onto SMEFT, and then onto LEFT (see Appendix~\ref{sec:approx_matching}). By neglecting all Wilson coefficients apart from $C^{VLX}_{ee,jijj}$ with $X=L,R$, we thus find
\begin{eqns}
    \mathrm{BR}(\mu\to 3e) &\approx& \num{1.2E-4} \Big[ 64\left|0.27\Big(\hat{C}_{HL,12}^{(1)} + \hat{C}_{HL,12}^{(3)}\Big)\right|^2 + 8 \left|0.49\Big(\hat{C}_{HL,12}^{(1)} + \hat{C}_{HL,12}^{(3)}\Big)\right|^2 \Big] \text{ and}
    \\
    \mathrm{BR}(\tau\to 3e) &\approx& \num{0.2E-4} \Big[ 64\left|0.27
    \Big(\hat{C}_{HL,13}^{(1)} + \hat{C}_{HL,13}^{(3)}\Big)\right|^2 + 8 \left|0.49\Big(\hat{C}_{HL,13}^{(1)} + \hat{C}_{HL,13}^{(3)}\Big)\right|^2 \Big],
\end{eqns}%
where the SMEFT Wilson coefficients on the right are evaluated at the electroweak scale $\mu = m_Z$. In the Type-I Seesaw model, all Wilson coefficients entering the branching ratios for trilepton decays
receive contributions from matching onto SMEFT only at loop level.
In this case, the branching ratios are relatively more sensitive to the contributions from the electromagnetic dipole operator $\mathcal{O}_{e\gamma}$, and the above approximations are only accurate to about 20\%.


\subsubsection{\texorpdfstring{$\mu-e$}{mu-e} Conversion in Nuclei}

As the scalar and gluon operators are suppressed in the fermionic Seesaw models, the $\mu-e$ conversion rate takes the simple form~\cite{Kitano:2002mt,Cirigliano:2009bz}
\begin{equation}\label{eq:mueconvrate}
    \omega_{\rm conv} =
    \left|
    -\frac{C_{e\gamma,12}}{2 m_\mu} D
    + \tilde g_{LV}^{(p)} V^{(p)}
    + \tilde g_{LV}^{(n)} V^{(n)}
    \right|^2 
    + 
    \left| 
    -\frac{C_{e\gamma,21}^*}{2m_\mu} D 
    + \tilde g_{RV}^{(p)} V^{(p)} 
    + \tilde g_{RV}^{(n)} V^{(n)}
    \right|^2,
\end{equation}%
where the overlap integrals $D$, $V^{(p)}$ and $V^{(n)}$, and muon capture rates $\omega_{\rm capt}$ can be found in Refs.~\cite{Kitano:2002mt,PhysRevC.35.2212}, and the effective coupling constants are
\begin{eqns}
    \tilde g_{LV}^{(p)} &=& 2\left(C_{eu,1211}^{VLL} + C_{eu,1211}^{VLR}\right) +\left(C_{ed,1211}^{VLL} + C_{ed,1211}^{VLR}\right),
    \IEEEyesnumber\IEEEyessubnumber
    \\
    \tilde g_{RV}^{(p)} &=& 2 \left(C_{eu,1211}^{VRR} + C_{ue,1112}^{VLR}\right) + \left(C_{ed,1211}^{VRR} + C_{de,1112}^{VLR}\right),
    \IEEEyessubnumber
    \\
    \tilde g_{LV}^{(n)} &=& \left(C_{eu,1211}^{VLL} + C_{eu,1211}^{VLR}\right) + 2\left(C_{ed,1211}^{VLL} + C_{ed,1211}^{VLR}\right) \qcand
    \IEEEyessubnumber
    \\
    \tilde g_{RV}^{(n)} &=& \left(C_{eu,1211}^{VRR} + C_{ue,1112}^{VLR}\right) + 2 \left(C_{ed,1211}^{VRR} + C_{de,1112}^{VLR}\right);
    \IEEEyessubnumber
\end{eqns}%
see Appendix~\ref{sec:approx_matching} for approximate matching expressions.

We are interested in the conversion ratio $\mathrm{CR}(\mu \to e)$, defined as the ratio of the $\mu-e$ conversion rate $\omega_{\rm conv}$ over the muon capture rate $\omega_{\rm capt}$. For the Type-I and Type-III Seesaw models it approximates to
\begin{eqns}
    \mathrm{CR}(\mu \to e) &\approx&
    \left( \frac{m_\mu}{\si{\GeV}} \right)^5
    \left\{
    \begin{array}{c}
         5.87 \\
         1.16 \\
         21.54
    \end{array}
    \right\}
    \times\num{E5}
    \\\nonumber &&
    \Bigg|\Big[2\left(\hat{C}_{eu,1211}^{VLL} +\hat{C}_{eu,1211}^{VLR}\right) + \left(\hat{C}_{ed,1211}^{VLL} +\hat{C}_{ed,1211}^{VLR}\right)  \Big] 
    \left\{
    \begin{array}{c}
         0.0396 \\
         0.0974 \\
         0.0161
    \end{array}
    \right\}
    \\\nonumber && \quad 
    {} + \Big[\left(\hat{C}_{eu,1211}^{VLL} + \hat{C}_{eu,1211}^{VLR}\right) + 2\left(\hat{C}_{ed,1211}^{VLL} + \hat{C}_{ed,1211}^{VLR}\right)\Big]
    \left\{
    \begin{array}{c}
         0.0468 \\
         0.146 \\
         0.0173
    \end{array}
    \right\}
    \Bigg|^2,
\end{eqns}%
where the Wilson coefficients on the right are evaluated at the low scale $\mu = \SI{5}{\GeV}$, and where the upper, middle and lower entry in the brackets refers to a titanium (Ti), gold (Au) and aluminium (Al) target, respectively. We also include the predictions for the conversion ratios for lead (Pb) and sulfur (S) in Table~\ref{table:LFV}, from which one can infer that the respective current bounds do not impose relevant constraints. As is reflected by the above approximation, $\mu-e$ conversion is dominated by contributions from left-handed vector operators in both Seesaw models. This is evident in Type-III where these contributions are sourced at tree level, but also holds in Type-I. The electroweak dipole operator $\mathcal{O}_{e\gamma,12}$ plays a subdominant role in both models.


\subsubsection{Discussion}

\begin{figure}[h!]
    \begin{subfigure}[t]{\textwidth}
        \centering
        \includegraphics[width=\linewidth]{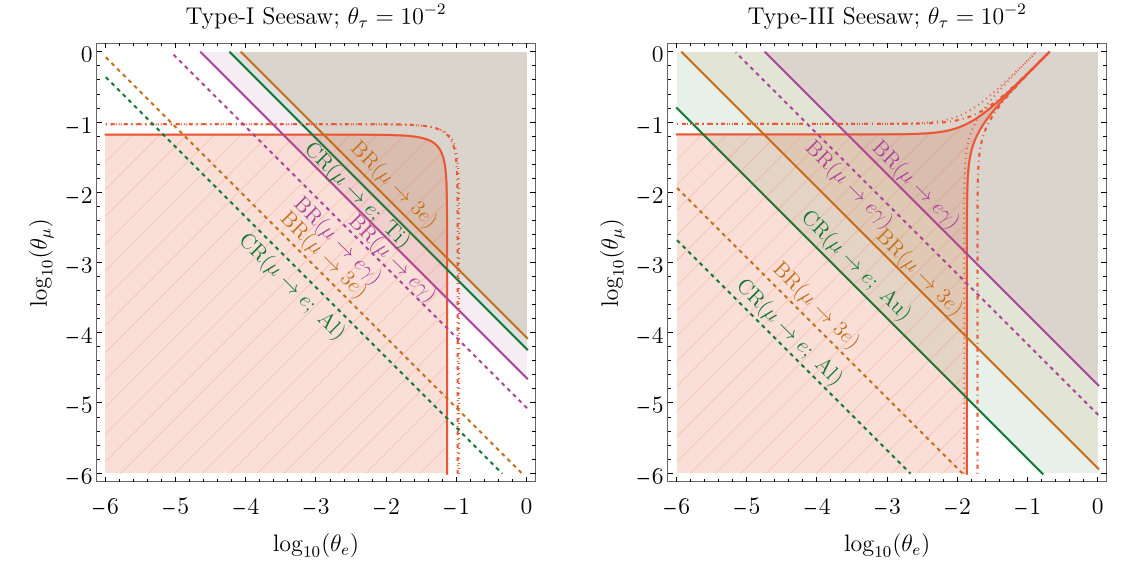}
        \caption{LFV constraints in the $\theta_e$--$\theta_\mu$ plane. Though we checked the current bounds arising from $\mu-e$ conversion for different target materials, here we display only the most competitive ones for each model, respectively. The PRISM/PRIME~\cite{Kuno:2005mm} proposal may even further extend the new-physics reach of $\mu-e$ conversion.}
        \label{fig:LFVa}
    \end{subfigure}
    \begin{subfigure}[t]{\textwidth}
        \centering
        \includegraphics[width=\linewidth]{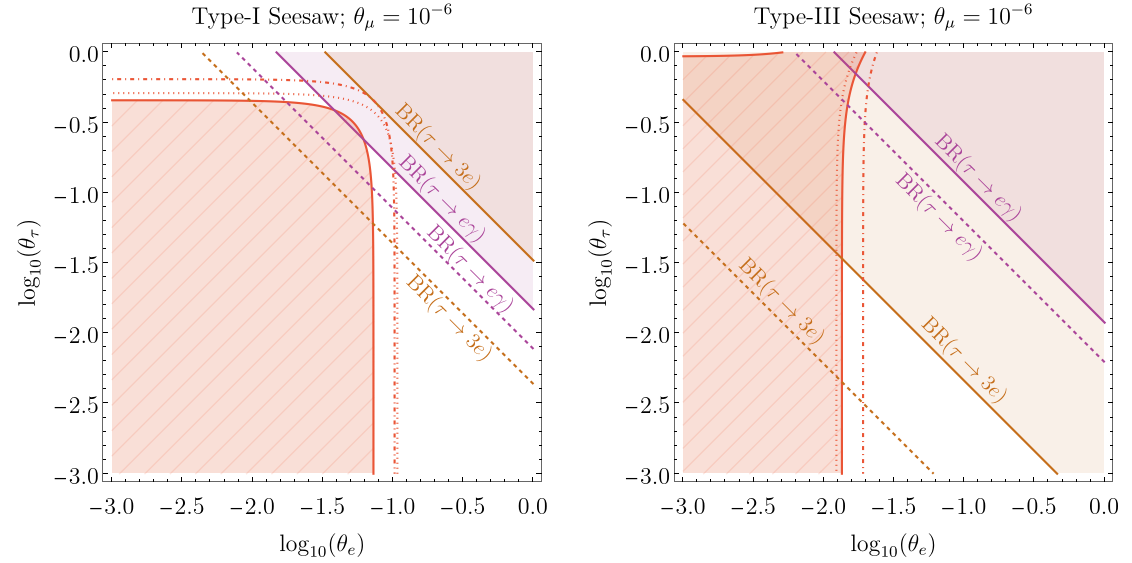}
        \caption{LFV constraints in the $\theta_e$--$\theta_\tau$ plane.}
        \label{fig:LFVb}
    \end{subfigure}
    \caption{Current constraints from LFV observables depicted at 90\% C.L., together with their prospective future reaches, in comparison with projected sensitivities of precision Higgsstrahlung measurements. The red-ruled regions indicate $|\Delta\sigma/\sigma| < 0.5\%$ at $\sqrt{s} = \SI{240}{\GeV}$. The dot-dashed and dotted red lines are the corresponding 1\% contours at $\sqrt{s} = \SI{240}{\GeV}$ and \SI{365}{\GeV}, respectively.
    We do not depict the bounds from tau-flavoured processes in the $\theta_e$--$\theta_\mu$ plots, and vice versa. If included, they would appear as vertical lines with positioning highly dependent on the choice of the third mixing angle, which limits their relevance.}
    \label{fig:LFV}
\end{figure}

We find that for the Type-I Seesaw model, as per the absence of tree-level contributions to trilepton decays and $\mu-e$ conversion, the most competitive bounds currently arise from the non-observation of $\mu\to e\gamma$ and $\tau\to e\gamma$, see Figure~\ref{fig:LFV}. Therefore, a detectable shift in $\sigma(e^+ e^- \to Zh)$ would enforce either $|\theta_e|\gtrsim 0.1$ and $|\theta_\mu|\lesssim 10^{-4}$, or vice versa. Still, in the $\mu-e$ sector, we expect the limits from BR($\mu\to3e$) and CR($\mu-e$) to become more stringent in the future, and further improve this bound by up to two orders of magnitude. The relatively loose current bound from CR($\mu-e;\,\text{Au}$) is due to cancellations between the effective vector couplings to protons and neutrons. These generically occur for all target materials in both fermionic Seesaw models, but are insignificant for Type-III. In the case of Type-I, matching at a larger scale reduces the cancellations for CR($\mu-e;\,\text{Au}$), cf.~Section~\ref{sec:larger_seesaw_scale}. See also for instance Ref.~\cite{Alonso:2012ji} for a pertinent discussion.
Currently, the non-observation of $\tau\to e\gamma$ enforces $|\theta_\tau|\lesssim0.1$ for $|\theta_e|\gtrsim0.1$, and $|\theta_e|\lesssim0.02$ for $|\theta_\tau|\gtrsim0.6$. These limits can be expected to become slightly more stringent in light of the future reaches of BR($\tau\to e\gamma$) and BR($\tau\to3e$), still, the improvements are expected to be less than an order of magnitude.

In Type-III Seesaw, due to tree-level contributions to the respective pertinent operators, the bounds from BR($\mu\to 3e$) and CR($\mu-e$) are stricter than that of BR($\mu\to e\gamma$), and $\tau\to3e$ is also more competitive than $\tau\to e\gamma$. Note that the projected sensitivity to BR($\mu\to e\gamma$) at MEG II cannot even be expected to supersede the current bound on BR($\mu\to 3e$). Moreover, the existing bound on $\mu-e$ conversion in gold effectively enforces the muon mixing angle to be smaller than $|\theta_\mu|\lesssim\num{E-5}$ for $|\theta_e|>\num{E-2}$ which is required for an observable deviation of the Higgsstrahlung cross section at a future lepton collider. Thus, the measurements of the cross section are only sensitive to a rather pronounced hierarchy $|\theta_\mu/\theta_e|\lesssim\num{E-3}$.
A hierarchy $|\theta_e/\theta_\mu|\lesssim\num{e-5}$, corresponding to the region visible in the top-left of Figure~\ref{fig:LFVa}, is in principle also compatible with the LFV bounds, but still disfavoured by the constraints from electroweak and LFU observables, see the relevant sections above.

Similarly, the non-observation of $\tau\to3e$ presently constrains the tau mixing angle to $|\theta_\tau|\lesssim\num{0.03}$ if a non-SM signature in $\sigma(e^+ e^- \to Zh)$ is to be attainable, that is, tau mixing should not substantially exceed electron mixing in magnitude. This limit on $|\theta_\tau|$ would be strengthened by an order of magnitude if no decay $\tau\to3e$ is observed at Belle II, which will then also necessitate a hierarchy $|\theta_\tau/\theta_e|\lesssim\num{0.1}$ if a chance of detecting $\Delta\sigma/\sigma_0$ is to be retained. In a similar vein, if no signals in $\mu\to 3e$ or $\mu-e$ conversion in aluminium are observed in the future, the ratio of the relevant mixing angles will be constrained to be even as small as $|\theta_\mu/\theta_e|\lesssim\num{E-4}$ or $\num{E-5}$, respectively.


\subsection{Larger Seesaw Scale}
\label{sec:larger_seesaw_scale}

\begin{table}[tbp]
    \centering
    \begin{tabular}{|r||c|c|}
        \hline
        & Type-I & Type-III \\
        \hline\hline
        $\Delta\sigma/\sigma_0$ (\SI{240}{\GeV}) & $0.74\,|\theta_e|^2 + 1.11\,|\theta_\mu|^2 + 0.06\,|\theta_\tau|^2$ & $26.30\, |\theta_e|^2 - 1.07\, |\theta_\mu|^2 - 0.01\, |\theta_\tau|^2$ \\
        $\Delta\sigma/\sigma_0$ (\SI{365}{\GeV}) & $0.41\,|\theta_e|^2 + 1.18\,|\theta_\mu|^2 + 0.12\,|\theta_\tau|^2$ & $63.12\, |\theta_e|^2 - 1.09\, |\theta_\mu|^2 - 0.03\, |\theta_\tau|^2$
        \\
        $\Delta\sigma/\sigma_0$ (\SI{500}{\GeV}) & $-0.05\,|\theta_e|^2 + 1.26\,|\theta_\mu|^2 + 0.19\,|\theta_\tau|^2$ & $120.66\, |\theta_e|^2 - 1.11\, |\theta_\mu|^2 - 0.05\, |\theta_\tau|^2$ \\
        \hline
        $\var s_w^2$ & $- 0.151(|\theta_e|^2 + |\theta_\mu|^2) + 0.005\,|\theta_\tau|^2$ & $0.021(|\theta_e|^2 + |\theta_\mu|^2) - 0.135\,|\theta_\tau|^2$ \\
        $\var m_W/\si{\GeV}$ & $7.95(|\theta_e|^2 + |\theta_\mu|^2) - 0.24\,|\theta_\tau|^2$ & $- 8.44(|\theta_e|^2 + |\theta_\mu|^2) - 0.24\,|\theta_\tau|^2$ \\
        \hline
        $g^X_{\mu/e}$ & $1 + 0.47(|\theta_e|^2 - |\theta_\mu|^2)$ & $1 - 0.47(|\theta_e|^2 - |\theta_\mu|^2)$ \\ $g^X_{\tau/\mu}$ & $1 + 0.47(|\theta_\mu|^2 - |\theta_\tau|^2)$ & $1 - 0.47(|\theta_\mu|^2 - |\theta_\tau|^2)$ \\
        $R(V_{us})$ & $1 + 0.44\,|\theta_e|^2 + 8.58\,|\theta_\mu|^2 - 0.07\,|\theta_\tau|^2$ & $1 - 0.44\,|\theta_e|^2 - 8.58\,|\theta_\mu|^2 + 0.07\,|\theta_\tau|^2$ \\
        \hline
        $\mathrm{BR}(\mu \to e\gamma)$ & $\num{0.80e-3} \,|\theta_e \theta_\mu|^2$ & $\num{1.21e-3} \,|\theta_e \theta_\mu|^2$ \\
        $\mathrm{BR}(\mu \to 3e)$ & $\num{0.34e-3}\,|\theta_e \theta_\mu|^2$ & $\num{0.66}\,|\theta_e \theta_\mu|^2$ \\
        $\mathrm{CR}(\mu \to e; \mathrm{Au})$ & $\num{0.94e-3}\,|\theta_e \theta_\mu|^2$ & $24.7\,|\theta_e \theta_\mu|^2$ \\
        $\mathrm{CR}(\mu \to e; \mathrm{Al})$ & $\num{0.03e-3}\,|\theta_e \theta_\mu|^2$ & $6.1\,|\theta_e \theta_\mu|^2$
        \\
        $\mathrm{CR}(\mu \to e; \mathrm{Ti})$ & $\num{0.15e-3}\,|\theta_e \theta_\mu|^2$ & $\num{12.3}\,|\theta_e \theta_\mu|^2$
        \\
        $\mathrm{CR}(\mu \to e; \mathrm{Pb})$ & $\num{0.76e-3}\,|\theta_e \theta_\mu|^2$ & $\num{18.5}\,|\theta_e \theta_\mu|^2$
        \\
        $\mathrm{CR}(\mu \to e; \mathrm{S})$ & $\num{0.01e-3}\,|\theta_e \theta_\mu|^2$ & $5.8\,|\theta_e \theta_\mu|^2$
        \\
        $\mathrm{BR}(\tau \to e\gamma)$ & $\num{0.15e-3} \,|\theta_e \theta_\tau|^2$ & $\num{0.23e-3} \,|\theta_e \theta_\tau|^2$ \\
        $\mathrm{BR}(\tau \to 3e)$ & $\num{0.06e-3}\,|\theta_e \theta_\tau|^2$ & $\num{0.12}\,|\theta_e \theta_\tau|^2$ \\
        \hline
    \end{tabular}
    \caption{Contributions to the observables considered in the phenomenology study in this work, but obtained from matching onto SMEFT at a Seesaw scale $\mu = M^X = \SI{10}{\TeV}$.}
    \label{table:approximate expressions 10 TeV}
\end{table}

Lastly, we will comment on the scenario with a larger matching scale. In Table~\ref{table:approximate expressions 10 TeV} we have collected our results for the shifts in the observables considered in our work for the Seesaw scale $\mu = M^X = \SI{10}{\TeV}$.

In the case of the Type-III Seesaw model, we find that the results discussed in this section seem to be fairly robust with respect to raising the triplet mass to $\mathcal{O}(\SI{10}{\TeV})$ at least. That is, the numerical coefficients entering the expressions for the considered observables typically change by less than 20\%. The only notable differences lie in the sensitivity of $\delta m_W$ and $\Delta\sigma/\sigma_0$ to tau-flavour mixing at larger centre-of-mass energies, where the respective coefficients grow by a factor of 2--5. Still, it is a subleading effect, as these observables remain much more sensitive to electron- and muon-flavour mixing. Therefore, the results for the Type-III Seesaw model discussed in Section~\ref{sec:pheno} so far, for which $M^\Sigma = \SI{1}{\tera\electronvolt}$ is assumed, will also approximately hold for (moderately) larger masses.

In contrast, the observable phenomenology of Type-I Seesaw morphs somewhat nontrivially upon raising the Seesaw scale to $M^\nu = \mathcal{O}(\SI{10}{\TeV})$. Most profoundly, the Higgsstrahlung shift $\Delta\sigma/\sigma_0$ now experiences a crossing near $\sqrt{s} = \SI{500}{\GeV}$, whereupon the dependence on the squared electron mixing angle $|\theta_e|^2$ reverses from positive to negative. Additionally, the trilepton decay rates receive a relative numerical boost of 200\%, and the $\mu-e$ conversion rates are, in general, significantly altered. Specifically, $\mu-e$ conversion in gold and lead increase substantially due to the fact that the effective left-handed vector couplings to neutrons increase by a factor larger than 2, while the proton couplings and the dipole operator remain largely unchanged and thus the cancellations are much less efficient. On the contrary, $\mu-e$ conversion in aluminium and titanium experience a suppression. This implies in particular that the current bound arising from $\mu-e$ conversion in gold is clearly stronger than the one from titanium, unlike the scenario with $M^\nu = \SI{1}{\tera\electronvolt}$.


\section{Summary}
\label{sec:summary}

\begin{figure}[h!]
    \centering
    \begin{subfigure}[t]{\textwidth}
        \centering
        \includegraphics[width=\linewidth]{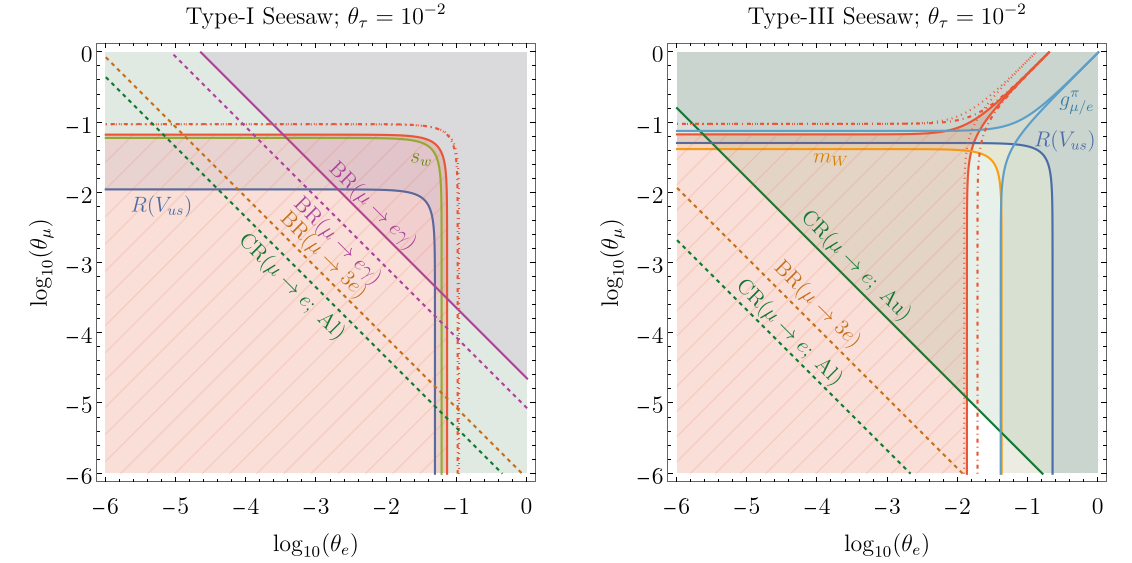}
        \caption{Summary plots in the $\theta_e$--$\theta_\mu$ plane.}
        \label{fig:summary_e_mu}
    \end{subfigure}
    \begin{subfigure}[t]{\textwidth}
        \centering
        \includegraphics[width=\linewidth]{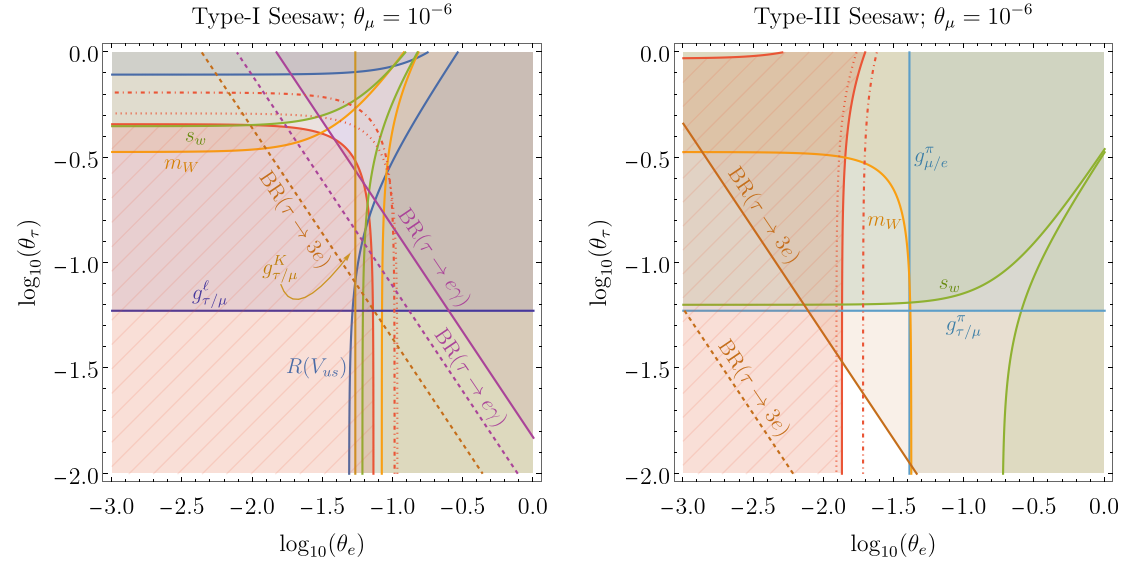}
        \caption{Summary plots in the $\theta_e$--$\theta_\tau$ plane.}
        \label{fig:summary_e_tau}
    \end{subfigure}
    \caption{Summary plots featuring the most competitive current constraints as well as future reaches considered in this work in comparison with projected sensitivities of precision Higgsstrahlung measurements. The red-ruled regions indicate $|\Delta\sigma/\sigma| < 0.5\%$ at $\sqrt{s} = \SI{240}{\GeV}$. The dot-dashed and dotted red lines are the corresponding 1\% contours at $\sqrt{s} = \SI{240}{\GeV}$ and \SI{365}{\GeV}, respectively.}
    \label{fig:summary}
\end{figure}

We have computed the correction to the tree-level cross section of the Higgsstrahlung process $e^+e^-\to Zh$ in the LN-conserving limit of the Type-I and Type-III Seesaw models, and compared several benchmark sensitivities of next-generation lepton colliders to existing and prospective constraints from electroweak precision measurements, and LFU and LFV probes. Summary plots in the $\theta_e$--$\theta_\mu$ and $\theta_e$--$\theta_\tau$ planes are presented in Figure~\ref{fig:summary}.

\begin{figure}[tbp]
    \centering
    \includegraphics[width=\linewidth]{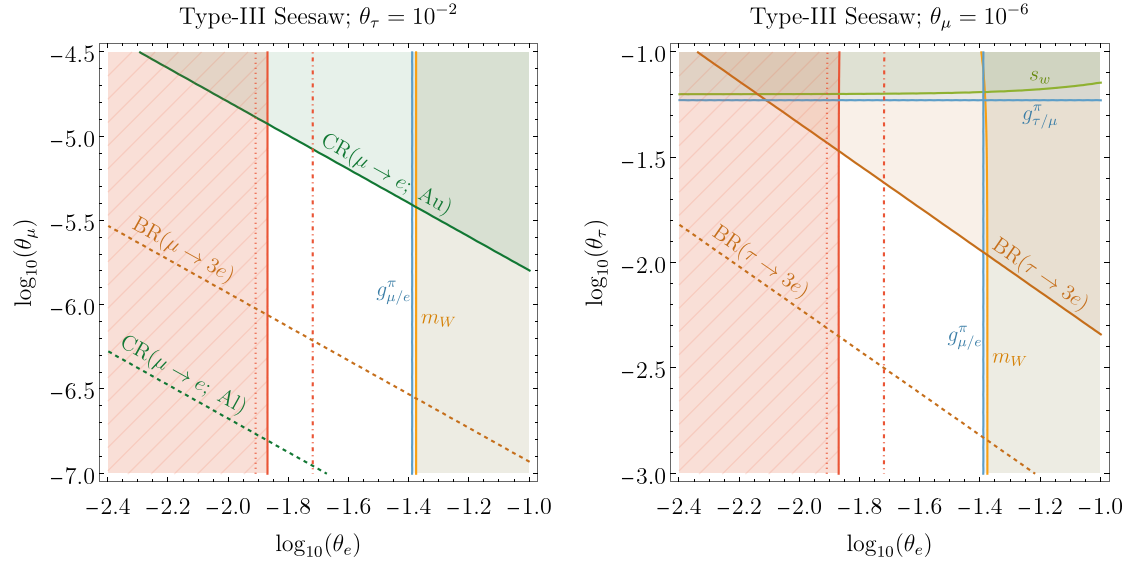}
    \caption{Plots zoomed in on the viable parameter regions in the Type-III Seesaw model. Only the most constraining observables are depicted. The red-ruled regions indicate $|\Delta\sigma/\sigma| < 0.5\%$ at $\sqrt{s} = \SI{240}{\GeV}$. The dot-dashed and dotted red lines are the corresponding 1\% contours at $\sqrt{s} = \SI{240}{\GeV}$ and \SI{365}{\GeV}, respectively.}
    \label{fig:typeIII_zoomed_in}
\end{figure}

As a major result, we found that existing data on the effective leptonic weak mixing angle and LFU observables preclude substantial corrections to the Higgsstrahlung cross section for Type-I Seesaw. The most likely signature of this model at a future lepton collider is therefore the \emph{absence} of a detectable deviation from the SM prediction, at least if no further new physics modifying the electroweak and LFU sectors is introduced.
For Type-III Seesaw, the current constraints (at $2\sigma$) leave genuinely viable parameter space that can be probed at an $e^+e^-$ Higgs factory. Figure~\ref{fig:typeIII_zoomed_in} provides a magnified view of this region. Concretely, for a centre-of-mass energy $\sqrt{s} = \SI{240}{\GeV}$ the largest permitted shift in the Higgsstrahlung cross section is $\sim$5\%; at $\sqrt{s} = \SI{365}{\GeV}$ it is $\sim$12\%.

The viable region in Type-III is isolated by three main considerations.
Firstly, the non-observation of LFV tightly constrains any scenario with sizeable mixing of heavy fermion singlets or triplets with two lepton flavours. These constraints are particularly strong for Type-III Seesaw, which induces tree-level contributions to trilepton decays and $\mu-e$ conversion. Indeed, a detectable deviation in $\sigma(e^+ e^- \to Zh)$ already necessitates a sizeable hierarchy between $\theta_e$ and $\theta_\mu$ which will become more pronounced if signals of LFV remain elusive in the future. (The situation is similar for Type-I Seesaw, where in the absence of contributions to LFV at tree level, the radiative decays $\mu\to e\gamma$ and $\tau\to e\gamma$ are more important.)
Secondly (and thirdly), both the W-boson mass $m_W$ and LFU data currently disfavour detectable corrections induced via muon mixing at the level of $2\sigma$, but leave room for visible effects due to electron-flavoured couplings, which together with the LFV constraints enforces a hierarchy $|\theta_\mu/\theta_e| \lesssim \num{e-3}$.

Focusing on the viable region in Figure~\ref{fig:typeIII_zoomed_in}, the constraints arising from the LFU ratio $g^\pi_{\mu/e}$ and $m_W$ are similarly competitive and provide the most stringent constraint on electron mixing in the Type-III Seesaw model, with $|\theta_e|\lesssim0.04$ at $2\sigma$.
Note also that in the region of parameter space where $\Delta\sigma/\sigma_0$ is detectable, tau-flavour mixing is more strictly constrained by the current bound on BR($\tau\to3e$) than by measurements of the weak mixing angle or pion decays. As is expected from Section~\ref{sec:LFV}, the most competitive upper limit on muon mixing in the given context currently arises from the non-observation of $\mu-e$ conversion in gold, and will be further constrained by Mu3e as well as the searches for $\mu-e$ conversion in aluminium at COMET and Mu2e.
If an observation of these transitions remains elusive in the future, the currently viable parameter space will retreat to $|\theta_\mu| \lesssim 10^{-7.5}$. Similarly, if $\tau\to3e$ is not observed at Belle II, tau mixing would need to be smaller than $\theta_\tau = 10^{-2}$.

Since fermion triplets induce a tree-level contribution to $e^+e^-\to Zh$ which is not mediated via the $s$ channel as in the SM, deviations from the cross section induced via electron-flavour mixing grow approximately with $s$. In that sense, if the drop in statistics can be compensated by higher luminosity, the Type-III Seesaw model motivates precision measurements of the Higgsstrahlung process at higher centre-of-mass energies as well, whereas this is not indicated for Type-I Seesaw.

Overall, we have corroborated the expectation that a rich interplay of neutrino, Higgs, electroweak and flavour physics is to be expected for Seesaw models at low energies, and demonstrated the benefit of measuring the Higgsstrahlung cross section at multiple centre-of-mass energies for the Type-III Seesaw model. One may extend the research conducted in this work along two major avenues. Firstly, although the list of processes which we consider in the analysis captures a wide range of phenomenology of the fermionic Seesaw models, it is not exhaustive. In particular, taking into account observables sensitive to angular distributions for Higgs physics~\cite{Craig:2015wwr} as well as a comprehensive global fit in the electroweak sector should help to further differentiate between the low-energy signatures of Seesaw models. Secondly, since we relied on the assumption of an exactly conserved LN symmetry on the Lagrangian level, existing data on lepton mixing and the mass hierarchies in the neutrino sector was \emph{per definition} not incorporated. While we expect the implications of explicit breakings of lepton number for the induced low-energy phenomenology to be small in general, any viable model of neutrino mass generation eventually needs to be tested against them. Lastly, we leave similar studies for different models of neutrino mass generation for future work.


\section*{Acknowledgements}

We acknowledge support by the Australian Research Council through the ARC Discovery Project DP200101470. AL acknowledges support from an Australian Government Research Training Program Scholarship. TF has also been supported by a UNSW Science PhD Writing Scholarship. The Feynman diagrams in this article were generated using TikZ-Feynman~\cite{Ellis:2016jkw}.


\appendix

\section{A Brief Overview of Parameter Shifts in SMEFT}
\label{app:SMEFT}

Here we supplement this work with a somewhat pedagogical overview aimed at explaining how to arrive at the parameter shifts presented in the main text. This overview is not entirely self-contained, and we recommend the reader refer to Section 5 of Ref.~\cite{Alonso:2013hga}, which provides an excellent guided tour of much upon which this exposition relies. We will match their notation with the exception of the gauge fields and their couplings, as their redefinition due to $\mathcal{O}_{HW}$ and $\mathcal{O}_{HB}$ is just a matter of bookkeeping. (Practically, this means we will continue to write e.g.~$g_2$ and $Z_\mu$ instead of $\bar{g}_2$ and $\mathcal{Z}_\mu$ respectively.) Herein we work to order $(v/\Lambda)^2$, where $\Lambda$ is the cutoff scale of the effective theory, equal to $M^X$ in the Seesaw models.

\paragraph{Part 1: Effective Parameters}

We denote with a bar, e.g.~$\bar{g}$, an effective parameter that appears in place of its unbarred form in the SMEFT Lagrangian after all effective operators have been expanded out and the dust settles. For example, the $Z$-boson part of the gauge-covariant derivative, which in the SM reads
\begin{equation}
    D_\mu \supset -ig_Z (T^3 - s_w^2 Q) Z_\mu,
\end{equation}
becomes in SMEFT
\begin{equation}
    D_\mu \supset -i\bar{g}_Z (T^3 - \bar{s}_w^2 Q) Z_\mu,
\end{equation}
where \cite{Alonso:2013hga}
\begin{eqns}
    \bar{g}_Z &=& \sqrt{g_1^2 + g_2^2} \pqty{ 1 + \frac{g_1 g_2}{g_1^2 + g_2^2} v_T^2 C_{HWB} } \qcand \IEEEyesnumber\IEEEyessubnumber
    \\
    \bar{s}_w^2 &=& \frac{g_1^2}{g_1^2 + g_2^2} \pqty{ 1 - \frac{g_2}{g_1} \frac{g_1^2 - g_2^2}{g_1^2 + g_2^2} v_T^2 C_{HWB} }. \IEEEyessubnumber
\end{eqns}%
To denote the shifts of these parameters from their SM expressions we write
\begin{IEEEeqnarray}{rCl"t"rCl}
    \bar{g}_Z &=& g_Z + \var \bar{g}_Z & where & \frac{\var \bar{g}_Z}{g_Z} &=& \frac{g_1 g_2}{g_1^2 + g_2^2} v_T^2 C_{HWB} \qcand \IEEEyesnumber\IEEEyessubnumber
    \\
    \bar{s}_w^2 &=& s_w^2 + \var \bar{s}_w^2 & where & \frac{\var \bar{s}_w^2}{s_w^2} &=& - \frac{g_2}{g_1} \frac{g_1^2 - g_2^2}{g_1^2 + g_2^2} v_T^2 C_{HWB}. \IEEEyessubnumber
\end{IEEEeqnarray}%
Note that $\var \bar{g}_Z/g_Z = \var \bar{g}_Z/\bar{g}_Z$ at order $(v/\Lambda)^2$ (and similarly so for every other parameter), so we will consistently opt to write the former, which uses less ink.

To reiterate the notation and have a referenceable equation, the shift for a general parameter $g$ is written
\begin{equation}
    \bar{g} = g + \var \bar{g}. \label{eq:g bar}
\end{equation}
There is one exception to this notational rule in $v_T$, which should be written $\bar{v}$ for consistency, but isn't by convention. In this case we have $v_T = v + \var\bar{v}$.

\paragraph{Part 2: Input Parameters}

\begin{figure*}[tbp]
    \begin{subfigure}[t]{.33\linewidth}
        \centering
        \begin{tikzpicture}
        \begin{feynman}
            \vertex (i) {$\mu^-$};
            \vertex [right=1.5cm of i] (a);
            \vertex [above right=0.7cm and 1.1cm of a] (f1) {$\nu_\mu$};
            \vertex [below right=0.7cm and 1.0cm of a, crossed dot] (b) {};
            \vertex [above right=0.9cm and 1.3cm of b] (f2) {$e^-$};
            \vertex [below right=0.9cm and 1.3cm of b] (f3) {$\overline{\nu}_e$};
    		
    		\diagram* {
    			(i) -- [fermion] (a) -- [fermion] (f1),
    			(a) -- [boson, edge label'=$W$] (b),
    			(f3) -- [fermion] (b) -- [fermion] (f2),
    		};
        \end{feynman}
        \end{tikzpicture}
        \caption*{$C_{HL,11}^{(3)}$}
    \end{subfigure}%
    \begin{subfigure}[t]{.33\textwidth}
        \centering
        \begin{tikzpicture}
        \begin{feynman}
            \vertex (i) {$\mu^-$};
            \vertex [right=1.7cm of i, crossed dot] (a) {};
            \vertex [above right=0.95cm and 1.35cm of a] (f1) {$\nu_\mu$};
            \vertex [below right=0.7cm and 1.0cm of a] (b);
            \vertex [above right=0.7cm and 1.1cm of b] (f2) {$e^-$};
            \vertex [below right=0.7cm and 1.1cm of b] (f3) {$\overline{\nu}_e$};
    		
    		\diagram* {
    			(i) -- [fermion] (a) -- [fermion] (f1),
    			(a) -- [boson, edge label'=$W$] (b),
    			(f3) -- [fermion] (b) -- [fermion] (f2),
    		};
        \end{feynman}
        \end{tikzpicture}
        \caption*{$C_{HL,22}^{(3)}$}
    \end{subfigure}%
    \begin{subfigure}[t]{.33\textwidth}
        \centering
        \begin{tikzpicture}
        \begin{feynman}
            \vertex [crossed dot] (a) {};
    		\vertex [left=1.8cm of a] (i) {$\mu^-$};
    		\vertex [above right=1.9cm of a] (f1) {$\nu_\mu$};
    		\vertex [right=1.7cm of a] (f2) {$e^-$};
    		\vertex [below right=1.9cm of a] (f3) {$\overline{\nu}_e$};
    		
    		\diagram* {
    			(i) -- [fermion] (a) -- [fermion] (f1),
    			(f3) -- [fermion] (a) -- [fermion] (f2),
    		};
        \end{feynman}
        \end{tikzpicture}
        \caption*{$C_{LL,1221}$, $C_{LL,2112}$}
    \end{subfigure}
    \caption{Muon decay diagrams with the effective operators contributing to $\var \hat G_F$, given in \eq{shift_GF}.}
    \label{fig:G_F diagrams}
\end{figure*}
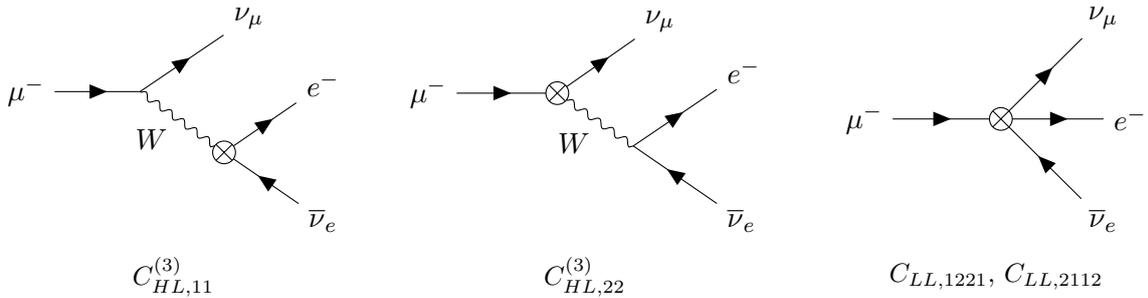

We denote with a hat, e.g.~$\hat{\alpha}$, a parameter directly measured, or derived from measured values using tree-level SM relations. For a parameter directly measured we have
\begin{equation}
    \hat{\alpha} = \bar{\alpha} + \var \hat{\alpha} = \alpha + \var \bar{\alpha} + \var \hat{\alpha}.
\end{equation}
Here there is an additional contribution, labelled $\var \hat{\alpha}$, which comes from other diagrams contributing to the measurement process, and which cannot be absorbed into a redefinition of the parameter as above. Such diagrams typically arise at the loop order or at $\order{(v/\Lambda)^4}$, which we neglect, therefore leaving $\var\hat\alpha = 0$. The sole exception to this is the Fermi constant, which acquires tree-level contributions to $\var\hat{G}_F$ from the diagrams in Figure~\ref{fig:G_F diagrams}.

For a parameter computed from the input parameters using the tree-level SM relations, $\hat{g} \equiv g(\hat{\alpha}_i)$ (one such example would be computing the elementary charge from the measured value of the fine structure constant via $\hat{e} = \sqrt{4\pi \hat{\alpha}}$), we instead have
\begin{equation}
    \hat{g} = g + \sum_i \pdv{g}{\alpha_i} \pqty{ \var \bar{\alpha}_i + \var \hat{\alpha}_i} \equiv g + \var \hat{g}. \label{eq:g hat}
\end{equation}
By combining Eqs.~\eqref{eq:g bar} and \eqref{eq:g hat}, we see that
\begin{eqns}
    \bar{g} &=& \hat{g} + \var \bar{g} - \var \hat{g} \\\nonumber
    &\equiv& \hat{g} + \var g.
\end{eqns}%
As will soon become apparent, we will be most interested in computing this total shift,
\begin{equation}\label{eq:var g}
    \var g \equiv \var \bar{g} - \sum_i \pdv{g}{\alpha_i} \pqty{ \var \bar{\alpha}_i + \var \hat{\alpha}_i }.
\end{equation}

\paragraph{Part 3: Shifts in Observables}

We now come to the climax of this exposition. Suppose now that SMEFT represents the ``true'' theory, and that we wish to compute the correction to a quantity such as a cross section which has been calculated under the mistaken assumption that the SM was the correct theory: $\Delta \sigma = \sigma_{\mathrm{SMEFT}} - \sigma_{\mathrm{SM}}$. In addition to direct contributions from new operators, there will be an indirect contribution from the shifts of the SM parameters.

To arrive at this conclusion carefully, convince yourself that the correct cross section should be written in terms of the barred parameters, $\sigma(\bar{g}_i)$, as---after all---these parameters are the ones extant in the ``true'' Lagrangian. In the SM, however, it's evidently the case that both $\var \bar{g}_i = 0$ and $\var \hat{g}_i = 0$,\footnote{Of course it's not strictly true that $\var \hat{g}_i = 0$, as there will be loop effects contributing to the screening of $g_i$ even in the SM. This is however a separate shift to the one under consideration and it can be independently dealt with, so we do not treat it here.} so one sees no issue with setting $g_i = \hat{g}_i$ and computing the cross section as $\sigma(\hat{g}_i)$. Explicitly,
\begin{equation}
    \sigma_{\mathrm{SMEFT}} = \sigma(\bar{g}_i) + \Delta \sigma_{\mathrm{Direct}} \qand \sigma_{\mathrm{SM}} = \sigma(\hat{g}_i),
\end{equation}
and so
\begin{align}
    \Delta \sigma &= \Delta \sigma_{\mathrm{Direct}} + \sum_i \pdv{\sigma}{g_i} \eval_{g_i = \hat{g}_i} \pqty{ \bar{g}_i - \hat{g}_i } \label{eq:general shift}
    \\
    &= \Delta \sigma_{\mathrm{Direct}} + \sum_i \pdv{\sigma}{g_i} \eval_{g_i = \hat{g}_i} \var g_i. \nonumber
\end{align}
It is worth emphasising that \eq{general shift} applies only when computing a correction to an SM prediction, which we do so in Sections~\ref{sec:higgsstrahlung} and \ref{sec:electroweak sector} of the main text.
However, as in Sections~\ref{sec:LFU} and \ref{sec:LFV}, it does \emph{not} apply when one is only interested in the SMEFT prediction, $\sigma_{\mathrm{SMEFT}}$.


\subsection*{An explicit example}

We illustrate the above procedure by working out the shift in $g_{ZZh}$, which appears in the SM Lagrangian as
\begin{equation}
    \mathcal{L} \supset \frac{1}{4} g_Z^2 v Z_\mu Z^\mu h \equiv \frac{1}{2} g_{ZZh} Z_\mu Z^\mu h. 
\end{equation}
In SMEFT this becomes---with additional complications arising from the fact that the Higgs field acquires a new normalisation, see Ref.~\cite{Alonso:2013hga}, and that $\mathcal{O}_{HD}$ contributes a $Z_\mu Z^\mu h$ term---
\begin{eqns}
    \mathcal{L} &\supset& \frac{1}{4} \bar{g}_Z^2 \vev Z_\mu Z^\mu \pqty{1 + \vev^2 C_{H\square} - \frac{1}{4} \vev^2C_{HD} } h + C_{HD} \frac{1}{2} g_Z^2 \vev^3 Z_\mu Z^\mu h
    \\\nonumber
    &=& \frac{1}{4} g_Z^2 v \pqty{ 1 + 2\frac{\var \bar{g}_Z}{g_Z} + \frac{\var\bar v}{v} + \vev^2 C_{H\square} + \frac{3}{4} \vev^2 C_{HD} } Z_\mu Z^\mu h
    \\\nonumber
    &\equiv& \frac{1}{2} \bar{g}_{ZZh} Z_\mu Z^\mu h,
\end{eqns}%
from which we read off
\begin{eqns}
    \frac{\var \bar{g}_{ZZh}}{g_{ZZh}} &=& 2\frac{\var \bar{g}_Z}{g_Z} + \frac{\var\bar v}{v} + \vev^2 C_{H\square} + \frac{3}{4} \vev^2 C_{HD} \\\nonumber
    &=& 2c_w s_w v_T^2 C_{HWB} + \vev^2 C_{H\square} + \frac{3}{4} \vev^2 C_{HD} + \frac{\var\bar v}{v}.
\end{eqns}%
The $\var \hat{g}_{ZZh}$ part of the full shift depends on one's choice of input parameters; in the $(\alpha,m_Z,G_F)$ scheme one has
\begin{equation}
    \hat{g}_{ZZh} = 2^{5/4} \hat{m}_Z^2 \sqrt{\hat{G}_F},
\end{equation}
which is a tree-level relation valid in the SM. In accordance with \eq{g hat}, we then have
\begin{equation}
    \frac{\var \hat{g}_{ZZh}}{g_{ZZh}} = \pqty{ \frac{\var \bar{m}_Z^2}{m_Z^2} + \frac{\var \hat{m}_Z^2}{m_Z^2} } + \frac{1}{2} \pqty{ \frac{\var \bar{G}_F}{G_F} + \frac{\var \hat{G}_F}{G_F} }.
\end{equation}
Plugging in the known shifts\footnote{One arrives at the shift $\var \bar{G}_F$ by writing $$\bar G_F = \frac{1}{\sqrt{2} \vev^2} = \frac{1}{\sqrt{2} v^2} \pqty{ 1 - 2\frac{\var \bar v}{v} }.$$ $\var\bar m_Z$ is more involved, as $\bar m_Z^2$ additionally receives a direct contribution from $\mathcal{O}_{HD}$ and an indirect contribution from $\mathcal{O}_{HWB}$ due to the rediagonalisation of $Z_\mu$ and $A_\mu$. Lastly, $\var\hat m_Z$ is zero as there are no tree-level diagrams at order $(v/\Lambda)^2$ which contribute to the $Z$-boson self energy.}~\cite{Alonso:2013hga}
\begin{equation}
    \frac{\var \bar{m}_Z^2}{m_Z^2} = \frac{1}{2} v_T^2 C_{HD} + 2c_w s_w v_T^2 C_{HWB} + 2\frac{\var\bar v}{v} \qcomma
    \frac{\var \hat{m}_Z^2}{m_Z^2} = 0 \qcomma
    \frac{\var \bar{G}_F}{G_F} = -2\frac{\var\bar v}{v},
\end{equation}
and leaving $\var \hat{G}_F / G_F$ symbolic as it has a cumbersome expression (see \eq{shift_GF}), this evaluates to
\begin{equation}
    \frac{\var \hat{g}_{ZZh}}{g_{ZZh}} = \frac{1}{2} v_T^2 C_{HD} + 2c_w s_w v_T^2 C_{HWB} + \frac{\var\bar v}{v} + \frac{1}{2} \frac{\var \hat{G}_F}{G_F}.
\end{equation}
Lastly, we bring it all together as per \eq{var g} to obtain
\begin{eqns}
    \frac{\var g_{ZZh}}{g_{ZZh}} &=& \frac{\var \bar{g}_{ZZh}}{g_{ZZh}} - \frac{\var \hat{g}_{ZZh}}{g_{ZZh}}
    \\\nonumber
    &=& \vev^2 \pqty{ C_{H\square} + \frac{1}{4} C_{HD} } - \frac{1}{2} \frac{\var \hat{G}_F}{G_F},
\end{eqns}%
which one sees matches \eq{Higgsstrahlung_ParameterShift1}. We note that the dependence on $\var\bar v$ completely cancels. This is a general trend for all shifts we consider -- thus, for our purposes, the replacement $v \to v_T$ is functionally unphysical.


\section{Approximate Matching Conditions}
\label{sec:approx_matching}

From \texttt{DsixTools} we numerically get the matching condition for the electromagnetic dipole operator, relevant to LFV decays:
\begin{eqns}\label{eq:dipole_matching}
    \frac{C_{e\gamma,ij}\,(\mu = \SI{5}{\GeV})}{\si{\GeV}} &\approx& 150.732\,C_{eB,ij} - 82.394\,C_{eW,ij}
    + 3.204\,C_{LeQu,ij33}^{(3)}  
    \\\nonumber && {} + A_{ij}\left(C_{HL,ij}^{(3)} - 0.27353\,C_{HL,ij}^{(1)}\right), 
\end{eqns}%
where the Wilson coefficients on the right-hand side are evaluated at the scale $\mu = m_Z$, repeated indices do not indicate summation, and
\begin{equation}
    A_{ij} = 10^{-3} \mqty( \textcolor{gray}{-} & 0.2661 & 4.4758 \\
    0.0013 & \textcolor{gray}{-} & \textcolor{gray}{-} \\
    0.0013 & \textcolor{gray}{-}& \textcolor{gray}{-} )_{ij},
\end{equation}
with the dashed entries being irrelevant for our purposes.

We are conscious that the scales associated with the decays of taus and muons are smaller than $\mu = \SI{5}{\GeV}$. The largest contributions from RG running at lower scales can be expected to originate from QCD. The only SMEFT operator which one may a priori expect to yield a sizeable contribution to LEFT operators which involve quark fields and mix into $\mathcal{O}_{e\gamma}$ is $\mathcal{O}_{LeQu}^{(3)} = (\overline L\sigma_{\mu\nu}e_R)\epsilon(\overline Q\sigma^{\mu\nu}u_R)$. Still, this operator is not induced at 1-loop level in the Seesaw models under consideration~\cite{Du:2022vso, Zhang:2021jdf, Coy:2021hyr}, and it only yields a tiny sub-percent contribution to $\mathcal{O}_{e\gamma}$. Therefore, we do not expect to have missed any sizeable effects from RG running below $\mu = \SI{5}{\GeV}$.\footnote{Paraphrasing this, even if the numerical factor multiplying $C_{LeQu}^{(3)}(m_Z)$ in Eq.~(\ref{eq:dipole_matching}) changes appreciably upon further lowering the scale on the left-hand side to, say, $\mu = m_\mu$, $C_{LeQu}^{(3)}(m_Z)$ itself is so small for the fermionic Seesaw models that we do not expect the resulting (relative) contribution to $C_{e\gamma}(m_\mu)$ to become sizeable in any way.}
The same holds for the (semi-)leptonic vector operators mediating trilepton decays and $\mu - e$ conversion; we have explicitly checked that the contribution from quark operators is at most 1\% in Eqs.~\eqref{eq:vector operators 1} and \eqref{eq:vector operators 2} below.

For all the matching conditions listed in the following, the respective LEFT Wilson coefficient on the left-hand side is given at $\mu = \SI{5}{\GeV}$, while the SMEFT Wilson coefficients entering on the right-hand side are evaluated at the scale $\mu = m_Z$. For the operators relevant to trilepton decays we find:
\begin{eqns}
    C_{ee,jijj}^{VLL} &\approx& -0.266\, C_{HL,ji}^{(1)} - 0.271\, C_{HL,ji}^{(3)} + 0.973\, C_{LL,jjji}, \IEEEyesnumber \label{eq:vector operators 1} \IEEEyessubnumber
    \\
    C_{ee,jijj}^{VRR} &\approx& 0.974\,C_{ee,jjji} + 0.235\,C_{He,ji} - 0.006\,C_{eu,ji33} + 0.003\,C_{Qe,33ji}, \IEEEyessubnumber
    \\
    C_{ee,jijj}^{VLR} &\approx& 0.4912\, C_{HL,ji}^{(1)} + 0.4909\, C_{HL,ji}^{(3)} + 1.018\,C_{Le,jijj} - 0.012\,C_{Lu,ji33} \qcand \IEEEyessubnumber
    \\
    C_{ee,jjji}^{VLR} &\approx& - 0.556\,C_{He,ji} + 1.018\,C_{Le,jjji} - 0.015\,C_{Qe,33ji} + 0.011\,C_{eu,ji33}. \IEEEyessubnumber
\end{eqns}%
For $\mu - e$ conversion the matching conditions for the vector operators are
\begin{eqns}
    C_{eu,1211}^{VLL} &\approx& 0.708\,C_{HL,12}^{(1)} + 0.734\,C_{HL,12}^{(3)} - 1.047\,C_{LQ,1211}^{(3)}, \IEEEyesnumber \label{eq:vector operators 2} \IEEEyessubnumber
    \\
    C_{eu,1211}^{VLR} &\approx& -0.3172 C_{HL,12}^{(1)} - 0.3170\, C_{HL,12}^{(3)} + 0.984\,C_{Lu,1211}, \IEEEyessubnumber
    \\
    C_{eu,1211}^{VRR} &\approx& - 0.321\,C_{He,12} + 0.008\,C_{eu,1233} - 0.005\, C_{Qe,3312}, \IEEEyessubnumber
    \\
    C_{ue,1112}^{VLR} &\approx& 0.696\,C_{He,12} + 0.017\,C_{Qe,3312} - 0.014\,C_{eu,1233}, \IEEEyessubnumber
    \\
    C_{ed,1211}^{VLL} &\approx& - 0.856\,C_{HL,12}^{(1)} - 0.864\,C_{HL,12}^{(3)} + 0.987\,C_{LQ,1211}^{(3)}, \IEEEyessubnumber
    \\
    C_{ed,1211}^{VLR} &\approx& 0.1617\, C_{HL,12}^{(1)} + 0.1615\, C_{HL,12}^{(3)} + 1.006\,C_{Ld,1211}, \IEEEyessubnumber
    \\
    C_{ed,1211}^{VRR} &\approx& 0.158\,C_{He,12} -0.004\,C_{eu,1233} + 0.002\, C_{Qe,3312} \qcand \IEEEyessubnumber
    \\
    C_{de,1112}^{VLR} &\approx& - 0.867\,C_{He,12} - 0.020\,C_{Qe,3312} + 0.018\,C_{eu,1233}. \IEEEyessubnumber
\end{eqns}%

\label{Bibliography}
\bibliography{literature}

\begin{thebibliography}{127}%
\makeatletter
\providecommand \@ifxundefined [1]{%
 \@ifx{#1\undefined}
}%
\providecommand \@ifnum [1]{%
 \ifnum #1\expandafter \@firstoftwo
 \else \expandafter \@secondoftwo
 \fi
}%
\providecommand \@ifx [1]{%
 \ifx #1\expandafter \@firstoftwo
 \else \expandafter \@secondoftwo
 \fi
}%
\providecommand \natexlab [1]{#1}%
\providecommand \enquote  [1]{``#1''}%
\providecommand \bibnamefont  [1]{#1}%
\providecommand \bibfnamefont [1]{#1}%
\providecommand \citenamefont [1]{#1}%
\providecommand \href@noop [0]{\@secondoftwo}%
\providecommand \href [0]{\begingroup \@sanitize@url \@href}%
\providecommand \@href[1]{\@@startlink{#1}\@@href}%
\providecommand \@@href[1]{\endgroup#1\@@endlink}%
\providecommand \@sanitize@url [0]{\catcode `\\12\catcode `\$12\catcode
  `\&12\catcode `\#12\catcode `\^12\catcode `\_12\catcode `\%12\relax}%
\providecommand \@@startlink[1]{}%
\providecommand \@@endlink[0]{}%
\providecommand \url  [0]{\begingroup\@sanitize@url \@url }%
\providecommand \@url [1]{\endgroup\@href {#1}{\urlprefix }}%
\providecommand \urlprefix  [0]{URL }%
\providecommand \Eprint [0]{\href }%
\providecommand \doibase [0]{https://doi.org/}%
\providecommand \selectlanguage [0]{\@gobble}%
\providecommand \bibinfo  [0]{\@secondoftwo}%
\providecommand \bibfield  [0]{\@secondoftwo}%
\providecommand \translation [1]{[#1]}%
\providecommand \BibitemOpen [0]{}%
\providecommand \bibitemStop [0]{}%
\providecommand \bibitemNoStop [0]{.\EOS\space}%
\providecommand \EOS [0]{\spacefactor3000\relax}%
\providecommand \BibitemShut  [1]{\csname bibitem#1\endcsname}%
\let\auto@bib@innerbib\@empty
\bibitem [{\citenamefont {Aad}\ \emph {et~al.}(2012)\citenamefont {Aad} \emph
  {et~al.}}]{ATLAS:2012yve}%
  \BibitemOpen
  \bibfield  {author} {\bibinfo {author} {\bibfnamefont {G.}~\bibnamefont
  {Aad}} \emph {et~al.} (\bibinfo {collaboration} {ATLAS}),\ }\bibfield
  {title} {\emph {\bibinfo {title} {{Observation of a new particle in the
  search for the Standard Model Higgs boson with the ATLAS detector at the
  LHC}}},\ }\href {https://doi.org/10.1016/j.physletb.2012.08.020} {\bibfield
  {journal} {\bibinfo  {journal} {Phys. Lett. B}\ }\textbf {\bibinfo {volume}
  {716}},\ \bibinfo {pages} {1} (\bibinfo {year} {2012})},\ \Eprint
  {https://arxiv.org/abs/1207.7214} {arXiv:1207.7214 [hep-ex]}\BibitemShut
  {NoStop}%
\bibitem [{\citenamefont {Chatrchyan}\ \emph {et~al.}(2012)\citenamefont
  {Chatrchyan} \emph {et~al.}}]{CMS:2012qbp}%
  \BibitemOpen
  \bibfield  {author} {\bibinfo {author} {\bibfnamefont {S.}~\bibnamefont
  {Chatrchyan}} \emph {et~al.} (\bibinfo {collaboration} {CMS}),\ }\bibfield
  {title} {\emph {\bibinfo {title} {{Observation of a New Boson at a Mass of
  125 GeV with the CMS Experiment at the LHC}}},\ }\href
  {https://doi.org/10.1016/j.physletb.2012.08.021} {\bibfield  {journal}
  {\bibinfo  {journal} {Phys. Lett. B}\ }\textbf {\bibinfo {volume} {716}},\
  \bibinfo {pages} {30} (\bibinfo {year} {2012})},\ \Eprint
  {https://arxiv.org/abs/1207.7235} {arXiv:1207.7235 [hep-ex]}\BibitemShut
  {NoStop}%
\bibitem [{\citenamefont {Dawson}\ \emph {et~al.}(2022)\citenamefont {Dawson}
  \emph {et~al.}}]{Dawson:2022zbb}%
  \BibitemOpen
  \bibfield  {author} {\bibinfo {author} {\bibfnamefont {S.}~\bibnamefont
  {Dawson}} \emph {et~al.},\ }\bibfield  {title} {\emph {\bibinfo {title}
  {{Report of the Topical Group on Higgs Physics for Snowmass 2021: The Case
  for Precision Higgs Physics}}},\ }(\bibinfo {year} {2022}),\ \Eprint
  {https://arxiv.org/abs/2209.07510} {arXiv:2209.07510 [hep-ph]}\BibitemShut
  {NoStop}%
\bibitem [{\citenamefont {Ahmad}\ \emph {et~al.}(2015)\citenamefont {Ahmad}
  \emph {et~al.}}]{CEPC-SPPCStudyGroup:2015csa}%
  \BibitemOpen
  \bibfield  {author} {\bibinfo {author} {\bibfnamefont {M.}~\bibnamefont
  {Ahmad}} \emph {et~al.},\ }\bibfield  {title} {\emph {\bibinfo {title}
  {{CEPC-SPPC Preliminary Conceptual Design Report. 1. Physics and
  Detector}}},\ }\href@noop {} {\  (\bibinfo {year} {2015})}\BibitemShut
  {NoStop}%
\bibitem [{\citenamefont {Guimar{\~a}es~da Costa}\ \emph
  {et~al.}(2018)\citenamefont {Guimar{\~a}es~da Costa} \emph
  {et~al.}}]{CEPCStudyGroup:2018ghi}%
  \BibitemOpen
  \bibfield  {author} {\bibinfo {author} {\bibfnamefont {J.~B.}\ \bibnamefont
  {Guimar{\~a}es~da Costa}} \emph {et~al.} (\bibinfo {collaboration} {CEPC
  Study Group}),\ }\bibfield  {title} {\emph {\bibinfo {title} {{CEPC
  Conceptual Design Report: Volume 2 - Physics \& Detector}}},\ }\href@noop {}
  {\  (\bibinfo {year} {2018})},\ \Eprint {https://arxiv.org/abs/1811.10545}
  {arXiv:1811.10545 [hep-ex]}\BibitemShut {NoStop}%
\bibitem [{\citenamefont {Cheng}\ \emph {et~al.}(2022)\citenamefont {Cheng}
  \emph {et~al.}}]{CEPCPhysicsStudyGroup:2022uwl}%
  \BibitemOpen
  \bibfield  {author} {\bibinfo {author} {\bibfnamefont {H.}~\bibnamefont
  {Cheng}} \emph {et~al.} (\bibinfo {collaboration} {CEPC Physics Study
  Group}),\ }\bibfield  {title} {\emph {\bibinfo {title} {{The Physics
  potential of the CEPC. Prepared for the US Snowmass Community Planning
  Exercise (Snowmass 2021)}}},\ }\bibfield  {booktitle} {\emph {\bibinfo
  {booktitle} {{2022 Snowmass Summer Study}}},\ }\href@noop {} {\  (\bibinfo
  {year} {2022})},\ \Eprint {https://arxiv.org/abs/2205.08553}
  {arXiv:2205.08553 [hep-ph]}\BibitemShut {NoStop}%
\bibitem [{\citenamefont {Baer}\ \emph {et~al.}(2013)\citenamefont {Baer} \emph
  {et~al.}}]{Baer:2013cma}%
  \BibitemOpen
  \bibfield  {author} {\bibinfo {author} {\bibfnamefont {H.}~\bibnamefont
  {Baer}} \emph {et~al.},\ }\bibfield  {title} {\emph {\bibinfo {title} {{The
  International Linear Collider Technical Design Report - Volume 2:
  Physics}}},\ }\href@noop {} {\  (\bibinfo {year} {2013})},\ \Eprint
  {https://arxiv.org/abs/1306.6352} {arXiv:1306.6352 [hep-ph]}\BibitemShut
  {NoStop}%
\bibitem [{\citenamefont {Asner}\ \emph {et~al.}(2013)\citenamefont {Asner}
  \emph {et~al.}}]{Asner:2013psa}%
  \BibitemOpen
  \bibfield  {author} {\bibinfo {author} {\bibfnamefont {D.~M.}\ \bibnamefont
  {Asner}} \emph {et~al.},\ }\bibfield  {title} {\emph {\bibinfo {title} {{ILC
  Higgs White Paper}}},\ }(\bibinfo {year} {2013}),\ \Eprint
  {https://arxiv.org/abs/1310.0763} {arXiv:1310.0763 [hep-ph]}\BibitemShut
  {NoStop}%
\bibitem [{\citenamefont {Bambade}\ \emph {et~al.}(2019)\citenamefont {Bambade}
  \emph {et~al.}}]{Bambade:2019fyw}%
  \BibitemOpen
  \bibfield  {author} {\bibinfo {author} {\bibfnamefont {P.}~\bibnamefont
  {Bambade}} \emph {et~al.},\ }\bibfield  {title} {\emph {\bibinfo {title}
  {{The International Linear Collider: A Global Project}}},\ }\href@noop {} {\
  (\bibinfo {year} {2019})},\ \Eprint {https://arxiv.org/abs/1903.01629}
  {arXiv:1903.01629 [hep-ex]}\BibitemShut {NoStop}%
\bibitem [{\citenamefont {Aryshev}\ \emph {et~al.}(2022)\citenamefont {Aryshev}
  \emph {et~al.}}]{ILCInternationalDevelopmentTeam:2022izu}%
  \BibitemOpen
  \bibfield  {author} {\bibinfo {author} {\bibfnamefont {A.}~\bibnamefont
  {Aryshev}} \emph {et~al.} (\bibinfo {collaboration} {ILC International
  Development Team}),\ }\bibfield  {title} {\emph {\bibinfo {title} {{The
  International Linear Collider: Report to Snowmass 2021}}},\ }\href@noop {} {\
   (\bibinfo {year} {2022})},\ \Eprint {https://arxiv.org/abs/2203.07622}
  {arXiv:2203.07622 [physics.acc-ph]}\BibitemShut {NoStop}%
\bibitem [{\citenamefont {Bicer}\ \emph {et~al.}(2014)\citenamefont {Bicer}
  \emph {et~al.}}]{TLEPDesignStudyWorkingGroup:2013myl}%
  \BibitemOpen
  \bibfield  {author} {\bibinfo {author} {\bibfnamefont {M.}~\bibnamefont
  {Bicer}} \emph {et~al.} (\bibinfo {collaboration} {TLEP Design Study Working
  Group}),\ }\bibfield  {title} {\emph {\bibinfo {title} {{First Look at the
  Physics Case of TLEP}}},\ }\href {https://doi.org/10.1007/JHEP01(2014)164}
  {\bibfield  {journal} {\bibinfo  {journal} {JHEP}\ }\textbf {\bibinfo
  {volume} {01}},\ \bibinfo {pages} {164}},\ \Eprint
  {https://arxiv.org/abs/1308.6176} {arXiv:1308.6176 [hep-ex]}\BibitemShut
  {NoStop}%
\bibitem [{\citenamefont {Pyarelal}\ \emph {et~al.}(2019)\citenamefont
  {Pyarelal}, \citenamefont {Song},\ and\ \citenamefont {Su}}]{10150/633402}%
  \BibitemOpen
  \bibfield  {author} {\bibinfo {author} {\bibfnamefont {A.}~\bibnamefont
  {Pyarelal}}, \bibinfo {author} {\bibfnamefont {H.}~\bibnamefont {Song}},\
  and\ \bibinfo {author} {\bibfnamefont {S.}~\bibnamefont {Su}},\ }\bibfield
  {title} {\emph {\bibinfo {title} {{FCC-ee: The Lepton Collider: Future
  Circular Collider Conceptual Design Report Volume 2}}},\ }\bibfield
  {journal} {\bibinfo  {journal} {EUROPEAN PHYSICAL JOURNAL-SPECIAL TOPICS}\
  }\href {https://doi.org/10.1140/epjst/e2019-900045-4}
  {10.1140/epjst/e2019-900045-4} (\bibinfo {year} {2019})\BibitemShut {NoStop}%
\bibitem [{\citenamefont {Agapov}\ \emph {et~al.}(2022)\citenamefont {Agapov}
  \emph {et~al.}}]{Agapov:2022bhm}%
  \BibitemOpen
  \bibfield  {author} {\bibinfo {author} {\bibfnamefont {I.}~\bibnamefont
  {Agapov}} \emph {et~al.},\ }\bibfield  {title} {\emph {\bibinfo {title}
  {{Future Circular Lepton Collider FCC-ee: Overview and Status}}},\ }(\bibinfo
  {year} {2022}),\ \Eprint {https://arxiv.org/abs/2203.08310} {arXiv:2203.08310
  [physics.acc-ph]}\BibitemShut {NoStop}%
\bibitem [{\citenamefont {Bernardi}\ \emph {et~al.}(2022)\citenamefont
  {Bernardi} \emph {et~al.}}]{Bernardi:2022hny}%
  \BibitemOpen
  \bibfield  {author} {\bibinfo {author} {\bibfnamefont {G.}~\bibnamefont
  {Bernardi}} \emph {et~al.},\ }\bibfield  {title} {\emph {\bibinfo {title}
  {{The Future Circular Collider: a Summary for the US 2021 Snowmass
  Process}}},\ }\href@noop {} {\  (\bibinfo {year} {2022})},\ \Eprint
  {https://arxiv.org/abs/2203.06520} {arXiv:2203.06520 [hep-ex]}\BibitemShut
  {NoStop}%
\bibitem [{\citenamefont {Lebrun}\ \emph {et~al.}(2012)\citenamefont {Lebrun},
  \citenamefont {Linssen}, \citenamefont {Lucaci-Timoce}, \citenamefont
  {Schulte}, \citenamefont {Simon}, \citenamefont {Stapnes}, \citenamefont
  {Toge}, \citenamefont {Weerts},\ and\ \citenamefont {Wells}}]{Lebrun:2012hj}%
  \BibitemOpen
  \bibfield  {author} {\bibinfo {author} {\bibfnamefont {P.}~\bibnamefont
  {Lebrun}}, \bibinfo {author} {\bibfnamefont {L.}~\bibnamefont {Linssen}},
  \bibinfo {author} {\bibfnamefont {A.}~\bibnamefont {Lucaci-Timoce}}, \bibinfo
  {author} {\bibfnamefont {D.}~\bibnamefont {Schulte}}, \bibinfo {author}
  {\bibfnamefont {F.}~\bibnamefont {Simon}}, \bibinfo {author} {\bibfnamefont
  {S.}~\bibnamefont {Stapnes}}, \bibinfo {author} {\bibfnamefont
  {N.}~\bibnamefont {Toge}}, \bibinfo {author} {\bibfnamefont {H.}~\bibnamefont
  {Weerts}},\ and\ \bibinfo {author} {\bibfnamefont {J.}~\bibnamefont
  {Wells}},\ }\bibfield  {title} {\emph {\bibinfo {title} {{The CLIC Programme:
  Towards a Staged $e^+e^-$ Linear Collider Exploring the Terascale : CLIC
  Conceptual Design Report}}}\ }\href {https://doi.org/10.5170/CERN-2012-005}
  {10.5170/CERN-2012-005} (\bibinfo {year} {2012}),\ \Eprint
  {https://arxiv.org/abs/1209.2543} {arXiv:1209.2543
  [physics.ins-det]}\BibitemShut {NoStop}%
\bibitem [{\citenamefont {de~Blas}\ \emph {et~al.}(2018)\citenamefont {de~Blas}
  \emph {et~al.}}]{CLIC:2018fvx}%
  \BibitemOpen
  \bibfield  {author} {\bibinfo {author} {\bibfnamefont {J.}~\bibnamefont
  {de~Blas}} \emph {et~al.} (\bibinfo {collaboration} {CLIC}),\ }\bibfield
  {title} {\emph {\bibinfo {title} {{The CLIC Potential for New Physics}}}\
  }\textbf {\bibinfo {volume} {3/2018}},\ \href
  {https://doi.org/10.23731/CYRM-2018-003} {10.23731/CYRM-2018-003} (\bibinfo
  {year} {2018}),\ \Eprint {https://arxiv.org/abs/1812.02093} {arXiv:1812.02093
  [hep-ph]}\BibitemShut {NoStop}%
\bibitem [{\citenamefont {Charles}\ \emph {et~al.}(2018)\citenamefont {Charles}
  \emph {et~al.}}]{CLICdp:2018cto}%
  \BibitemOpen
  \bibfield  {author} {\bibinfo {author} {\bibfnamefont {T.~K.}\ \bibnamefont
  {Charles}} \emph {et~al.} (\bibinfo {collaboration} {CLICdp, CLIC}),\
  }\bibfield  {title} {\emph {\bibinfo {title} {{The Compact Linear Collider
  (CLIC) - 2018 Summary Report}}}\ }\textbf {\bibinfo {volume} {2/2018}},\
  \href {https://doi.org/10.23731/CYRM-2018-002} {10.23731/CYRM-2018-002}
  (\bibinfo {year} {2018}),\ \Eprint {https://arxiv.org/abs/1812.06018}
  {arXiv:1812.06018 [physics.acc-ph]}\BibitemShut {NoStop}%
\bibitem [{\citenamefont {Bai}\ \emph {et~al.}(2021)\citenamefont {Bai} \emph
  {et~al.}}]{Bai:2021rdg}%
  \BibitemOpen
  \bibfield  {author} {\bibinfo {author} {\bibfnamefont {M.}~\bibnamefont
  {Bai}} \emph {et~al.},\ }\bibfield  {title} {\emph {\bibinfo {title} {{C$^3$:
  A ''Cool'' Route to the Higgs Boson and Beyond}}},\ }(\bibinfo {year}
  {2021}),\ \Eprint {https://arxiv.org/abs/2110.15800} {arXiv:2110.15800
  [hep-ex]}\BibitemShut {NoStop}%
\bibitem [{\citenamefont {Nanni}\ \emph {et~al.}(2022)\citenamefont {Nanni}
  \emph {et~al.}}]{Nanni:2022oha}%
  \BibitemOpen
  \bibfield  {author} {\bibinfo {author} {\bibfnamefont {E.~A.}\ \bibnamefont
  {Nanni}} \emph {et~al.},\ }\bibfield  {title} {\emph {\bibinfo {title}
  {{C$^3$ Demonstration Research and Development Plan}}},\ }(\bibinfo {year}
  {2022}),\ \Eprint {https://arxiv.org/abs/2203.09076} {arXiv:2203.09076
  [physics.acc-ph]}\BibitemShut {NoStop}%
\bibitem [{\citenamefont {Dasu}\ \emph {et~al.}(2022)\citenamefont {Dasu} \emph
  {et~al.}}]{Dasu:2022nux}%
  \BibitemOpen
  \bibfield  {author} {\bibinfo {author} {\bibfnamefont {S.}~\bibnamefont
  {Dasu}} \emph {et~al.},\ }\bibfield  {title} {\emph {\bibinfo {title}
  {{Strategy for Understanding the Higgs Physics: The Cool Copper Collider}}},\
  }(\bibinfo {year} {2022}),\ \Eprint {https://arxiv.org/abs/2203.07646}
  {arXiv:2203.07646 [hep-ex]}\BibitemShut {NoStop}%
\bibitem [{\citenamefont {Aslanides}\ \emph {et~al.}(2019)\citenamefont
  {Aslanides} \emph {et~al.}}]{Aslanides:2019gew}%
  \BibitemOpen
  \bibfield  {author} {\bibinfo {author} {\bibfnamefont {E.}~\bibnamefont
  {Aslanides}} \emph {et~al.},\ }\bibfield  {title} {\emph {\bibinfo {title}
  {{Charting the European Course to the High-Energy Frontier}}},\ }\href@noop
  {} {\  (\bibinfo {year} {2019})},\ \Eprint {https://arxiv.org/abs/1912.13466}
  {arXiv:1912.13466 [hep-ex]}\BibitemShut {NoStop}%
\bibitem [{\citenamefont {Fukuda}\ \emph {et~al.}(1998)\citenamefont {Fukuda}
  \emph {et~al.}}]{Super-Kamiokande:1998kpq}%
  \BibitemOpen
  \bibfield  {author} {\bibinfo {author} {\bibfnamefont {Y.}~\bibnamefont
  {Fukuda}} \emph {et~al.} (\bibinfo {collaboration} {Super-Kamiokande}),\
  }\bibfield  {title} {\emph {\bibinfo {title} {{Evidence for oscillation of
  atmospheric neutrinos}}},\ }\href
  {https://doi.org/10.1103/PhysRevLett.81.1562} {\bibfield  {journal} {\bibinfo
   {journal} {Phys. Rev. Lett.}\ }\textbf {\bibinfo {volume} {81}},\ \bibinfo
  {pages} {1562} (\bibinfo {year} {1998})},\ \Eprint
  {https://arxiv.org/abs/hep-ex/9807003} {arXiv:hep-ex/9807003}\BibitemShut
  {NoStop}%
\bibitem [{\citenamefont {Ahmad}\ \emph {et~al.}(2001)\citenamefont {Ahmad}
  \emph {et~al.}}]{SNO:2001kpb}%
  \BibitemOpen
  \bibfield  {author} {\bibinfo {author} {\bibfnamefont {Q.~R.}\ \bibnamefont
  {Ahmad}} \emph {et~al.} (\bibinfo {collaboration} {SNO}),\ }\bibfield
  {title} {\emph {\bibinfo {title} {{Measurement of the rate of $\nu_e+d \to
  p+p+e^-$ interactions produced by $^8$B solar neutrinos at the Sudbury
  Neutrino Observatory}}},\ }\href
  {https://doi.org/10.1103/PhysRevLett.87.071301} {\bibfield  {journal}
  {\bibinfo  {journal} {Phys. Rev. Lett.}\ }\textbf {\bibinfo {volume} {87}},\
  \bibinfo {pages} {071301} (\bibinfo {year} {2001})},\ \Eprint
  {https://arxiv.org/abs/nucl-ex/0106015} {arXiv:nucl-ex/0106015}\BibitemShut
  {NoStop}%
\bibitem [{\citenamefont {Ahmad}\ \emph {et~al.}(2002)\citenamefont {Ahmad}
  \emph {et~al.}}]{SNO:2002tuh}%
  \BibitemOpen
  \bibfield  {author} {\bibinfo {author} {\bibfnamefont {Q.~R.}\ \bibnamefont
  {Ahmad}} \emph {et~al.} (\bibinfo {collaboration} {SNO}),\ }\bibfield
  {title} {\emph {\bibinfo {title} {{Direct evidence for neutrino flavor
  transformation from neutral current interactions in the Sudbury Neutrino
  Observatory}}},\ }\href {https://doi.org/10.1103/PhysRevLett.89.011301}
  {\bibfield  {journal} {\bibinfo  {journal} {Phys. Rev. Lett.}\ }\textbf
  {\bibinfo {volume} {89}},\ \bibinfo {pages} {011301} (\bibinfo {year}
  {2002})},\ \Eprint {https://arxiv.org/abs/nucl-ex/0204008}
  {arXiv:nucl-ex/0204008}\BibitemShut {NoStop}%
\bibitem [{\citenamefont {del Aguila}\ \emph {et~al.}(2005)\citenamefont {del
  Aguila}, \citenamefont {Aguilar-Saavedra}, \citenamefont {Martinez de~la
  Ossa},\ and\ \citenamefont {Meloni}}]{delAguila:2005ssc}%
  \BibitemOpen
  \bibfield  {author} {\bibinfo {author} {\bibfnamefont {F.}~\bibnamefont {del
  Aguila}}, \bibinfo {author} {\bibfnamefont {J.~A.}\ \bibnamefont
  {Aguilar-Saavedra}}, \bibinfo {author} {\bibfnamefont {A.}~\bibnamefont
  {Martinez de~la Ossa}},\ and\ \bibinfo {author} {\bibfnamefont
  {D.}~\bibnamefont {Meloni}},\ }\bibfield  {title} {\emph {\bibinfo {title}
  {{Flavor and polarisation in heavy neutrino production at $e^+ e^-$
  colliders}}},\ }\href {https://doi.org/10.1016/j.j.physletb.2005.03.054}
  {\bibfield  {journal} {\bibinfo  {journal} {Phys. Lett. B}\ }\textbf
  {\bibinfo {volume} {613}},\ \bibinfo {pages} {170} (\bibinfo {year}
  {2005})},\ \Eprint {https://arxiv.org/abs/hep-ph/0502189}
  {arXiv:hep-ph/0502189}\BibitemShut {NoStop}%
\bibitem [{\citenamefont {del Aguila}\ and\ \citenamefont
  {Aguilar-Saavedra}(2005)}]{delAguila:2005pin}%
  \BibitemOpen
  \bibfield  {author} {\bibinfo {author} {\bibfnamefont {F.}~\bibnamefont {del
  Aguila}}\ and\ \bibinfo {author} {\bibfnamefont {J.~A.}\ \bibnamefont
  {Aguilar-Saavedra}},\ }\bibfield  {title} {\emph {\bibinfo {title} {{$\ell W
  \nu$ production at CLIC: A Window to TeV scale non-decoupled neutrinos}}},\
  }\href {https://doi.org/10.1088/1126-6708/2005/05/026} {\bibfield  {journal}
  {\bibinfo  {journal} {JHEP}\ }\textbf {\bibinfo {volume} {05}},\ \bibinfo
  {pages} {026}},\ \Eprint {https://arxiv.org/abs/hep-ph/0503026}
  {arXiv:hep-ph/0503026}\BibitemShut {NoStop}%
\bibitem [{\citenamefont {Antusch}\ and\ \citenamefont
  {Fischer}(2015)}]{Antusch:2015mia}%
  \BibitemOpen
  \bibfield  {author} {\bibinfo {author} {\bibfnamefont {S.}~\bibnamefont
  {Antusch}}\ and\ \bibinfo {author} {\bibfnamefont {O.}~\bibnamefont
  {Fischer}},\ }\bibfield  {title} {\emph {\bibinfo {title} {{Testing sterile
  neutrino extensions of the Standard Model at future lepton colliders}}},\
  }\href {https://doi.org/10.1007/JHEP05(2015)053} {\bibfield  {journal}
  {\bibinfo  {journal} {JHEP}\ }\textbf {\bibinfo {volume} {05}},\ \bibinfo
  {pages} {053}},\ \Eprint {https://arxiv.org/abs/1502.05915} {arXiv:1502.05915
  [hep-ph]}\BibitemShut {NoStop}%
\bibitem [{\citenamefont {Antusch}\ \emph
  {et~al.}(2016{\natexlab{a}})\citenamefont {Antusch}, \citenamefont
  {Cazzato},\ and\ \citenamefont {Fischer}}]{Antusch:2015gjw}%
  \BibitemOpen
  \bibfield  {author} {\bibinfo {author} {\bibfnamefont {S.}~\bibnamefont
  {Antusch}}, \bibinfo {author} {\bibfnamefont {E.}~\bibnamefont {Cazzato}},\
  and\ \bibinfo {author} {\bibfnamefont {O.}~\bibnamefont {Fischer}},\
  }\bibfield  {title} {\emph {\bibinfo {title} {{Higgs production from sterile
  neutrinos at future lepton colliders}}},\ }\href
  {https://doi.org/10.1007/JHEP04(2016)189} {\bibfield  {journal} {\bibinfo
  {journal} {JHEP}\ }\textbf {\bibinfo {volume} {04}},\ \bibinfo {pages}
  {189}},\ \Eprint {https://arxiv.org/abs/1512.06035} {arXiv:1512.06035
  [hep-ph]}\BibitemShut {NoStop}%
\bibitem [{\citenamefont {Antusch}\ \emph
  {et~al.}(2016{\natexlab{b}})\citenamefont {Antusch}, \citenamefont
  {Cazzato},\ and\ \citenamefont {Fischer}}]{Antusch:2016vyf}%
  \BibitemOpen
  \bibfield  {author} {\bibinfo {author} {\bibfnamefont {S.}~\bibnamefont
  {Antusch}}, \bibinfo {author} {\bibfnamefont {E.}~\bibnamefont {Cazzato}},\
  and\ \bibinfo {author} {\bibfnamefont {O.}~\bibnamefont {Fischer}},\
  }\bibfield  {title} {\emph {\bibinfo {title} {{Displaced vertex searches for
  sterile neutrinos at future lepton colliders}}},\ }\href
  {https://doi.org/10.1007/JHEP12(2016)007} {\bibfield  {journal} {\bibinfo
  {journal} {JHEP}\ }\textbf {\bibinfo {volume} {12}},\ \bibinfo {pages}
  {007}},\ \Eprint {https://arxiv.org/abs/1604.02420} {arXiv:1604.02420
  [hep-ph]}\BibitemShut {NoStop}%
\bibitem [{\citenamefont {Zhang}\ and\ \citenamefont
  {Zhang}(2019)}]{Zhang:2018rtr}%
  \BibitemOpen
  \bibfield  {author} {\bibinfo {author} {\bibfnamefont {Y.}~\bibnamefont
  {Zhang}}\ and\ \bibinfo {author} {\bibfnamefont {B.}~\bibnamefont {Zhang}},\
  }\bibfield  {title} {\emph {\bibinfo {title} {{A potential scenario for
  Majorana neutrino detection at future lepton colliders}}},\ }\href
  {https://doi.org/10.1007/JHEP02(2019)175} {\bibfield  {journal} {\bibinfo
  {journal} {JHEP}\ }\textbf {\bibinfo {volume} {02}},\ \bibinfo {pages}
  {175}},\ \Eprint {https://arxiv.org/abs/1805.09520} {arXiv:1805.09520
  [hep-ph]}\BibitemShut {NoStop}%
\bibitem [{\citenamefont {Das}\ \emph {et~al.}(2019)\citenamefont {Das},
  \citenamefont {Jana}, \citenamefont {Mandal},\ and\ \citenamefont
  {Nandi}}]{Das:2018usr}%
  \BibitemOpen
  \bibfield  {author} {\bibinfo {author} {\bibfnamefont {A.}~\bibnamefont
  {Das}}, \bibinfo {author} {\bibfnamefont {S.}~\bibnamefont {Jana}}, \bibinfo
  {author} {\bibfnamefont {S.}~\bibnamefont {Mandal}},\ and\ \bibinfo {author}
  {\bibfnamefont {S.}~\bibnamefont {Nandi}},\ }\bibfield  {title} {\emph
  {\bibinfo {title} {{Probing right handed neutrinos at the LHeC and lepton
  colliders using fat jet signatures}}},\ }\href
  {https://doi.org/10.1103/PhysRevD.99.055030} {\bibfield  {journal} {\bibinfo
  {journal} {Phys. Rev. D}\ }\textbf {\bibinfo {volume} {99}},\ \bibinfo
  {pages} {055030} (\bibinfo {year} {2019})},\ \Eprint
  {https://arxiv.org/abs/1811.04291} {arXiv:1811.04291 [hep-ph]}\BibitemShut
  {NoStop}%
\bibitem [{\citenamefont {Barducci}\ \emph {et~al.}(2021)\citenamefont
  {Barducci}, \citenamefont {Bertuzzo}, \citenamefont {Caputo}, \citenamefont
  {Hernandez},\ and\ \citenamefont {Mele}}]{Barducci:2020icf}%
  \BibitemOpen
  \bibfield  {author} {\bibinfo {author} {\bibfnamefont {D.}~\bibnamefont
  {Barducci}}, \bibinfo {author} {\bibfnamefont {E.}~\bibnamefont {Bertuzzo}},
  \bibinfo {author} {\bibfnamefont {A.}~\bibnamefont {Caputo}}, \bibinfo
  {author} {\bibfnamefont {P.}~\bibnamefont {Hernandez}},\ and\ \bibinfo
  {author} {\bibfnamefont {B.}~\bibnamefont {Mele}},\ }\bibfield  {title}
  {\emph {\bibinfo {title} {{The see-saw portal at future Higgs Factories}}},\
  }\href {https://doi.org/10.1007/JHEP03(2021)117} {\bibfield  {journal}
  {\bibinfo  {journal} {JHEP}\ }\textbf {\bibinfo {volume} {03}},\ \bibinfo
  {pages} {117}},\ \Eprint {https://arxiv.org/abs/2011.04725} {arXiv:2011.04725
  [hep-ph]}\BibitemShut {NoStop}%
\bibitem [{\citenamefont {Gao}\ and\ \citenamefont {Wang}(2022)}]{Gao:2021one}%
  \BibitemOpen
  \bibfield  {author} {\bibinfo {author} {\bibfnamefont {Y.}~\bibnamefont
  {Gao}}\ and\ \bibinfo {author} {\bibfnamefont {K.}~\bibnamefont {Wang}},\
  }\bibfield  {title} {\emph {\bibinfo {title} {{Heavy neutrino searches via
  same-sign lepton pairs at a Higgs boson factory}}},\ }\href
  {https://doi.org/10.1103/PhysRevD.105.076005} {\bibfield  {journal} {\bibinfo
   {journal} {Phys. Rev. D}\ }\textbf {\bibinfo {volume} {105}},\ \bibinfo
  {pages} {076005} (\bibinfo {year} {2022})},\ \Eprint
  {https://arxiv.org/abs/2102.12826} {arXiv:2102.12826 [hep-ph]}\BibitemShut
  {NoStop}%
\bibitem [{\citenamefont {Deppisch}\ \emph {et~al.}(2015)\citenamefont
  {Deppisch}, \citenamefont {Bhupal~Dev},\ and\ \citenamefont
  {Pilaftsis}}]{Deppisch:2015qwa}%
  \BibitemOpen
  \bibfield  {author} {\bibinfo {author} {\bibfnamefont {F.~F.}\ \bibnamefont
  {Deppisch}}, \bibinfo {author} {\bibfnamefont {P.~S.}\ \bibnamefont
  {Bhupal~Dev}},\ and\ \bibinfo {author} {\bibfnamefont {A.}~\bibnamefont
  {Pilaftsis}},\ }\bibfield  {title} {\emph {\bibinfo {title} {{Neutrinos and
  Collider Physics}}},\ }\href {https://doi.org/10.1088/1367-2630/17/7/075019}
  {\bibfield  {journal} {\bibinfo  {journal} {New J. Phys.}\ }\textbf {\bibinfo
  {volume} {17}},\ \bibinfo {pages} {075019} (\bibinfo {year} {2015})},\
  \Eprint {https://arxiv.org/abs/1502.06541} {arXiv:1502.06541
  [hep-ph]}\BibitemShut {NoStop}%
\bibitem [{\citenamefont {Cai}\ \emph {et~al.}(2018)\citenamefont {Cai},
  \citenamefont {Han}, \citenamefont {Li},\ and\ \citenamefont
  {Ruiz}}]{Cai:2017mow}%
  \BibitemOpen
  \bibfield  {author} {\bibinfo {author} {\bibfnamefont {Y.}~\bibnamefont
  {Cai}}, \bibinfo {author} {\bibfnamefont {T.}~\bibnamefont {Han}}, \bibinfo
  {author} {\bibfnamefont {T.}~\bibnamefont {Li}},\ and\ \bibinfo {author}
  {\bibfnamefont {R.}~\bibnamefont {Ruiz}},\ }\bibfield  {title} {\emph
  {\bibinfo {title} {{Lepton Number Violation: Seesaw Models and Their Collider
  Tests}}},\ }\href {https://doi.org/10.3389/fphy.2018.00040} {\bibfield
  {journal} {\bibinfo  {journal} {Front. in Phys.}\ }\textbf {\bibinfo {volume}
  {6}},\ \bibinfo {pages} {40} (\bibinfo {year} {2018})},\ \Eprint
  {https://arxiv.org/abs/1711.02180} {arXiv:1711.02180 [hep-ph]}\BibitemShut
  {NoStop}%
\bibitem [{\citenamefont {Das}(2018)}]{Das:2018hph}%
  \BibitemOpen
  \bibfield  {author} {\bibinfo {author} {\bibfnamefont {A.}~\bibnamefont
  {Das}},\ }\bibfield  {title} {\emph {\bibinfo {title} {{Searching for the
  minimal Seesaw models at the LHC and beyond}}},\ }\href
  {https://doi.org/10.1155/2018/9785318} {\bibfield  {journal} {\bibinfo
  {journal} {Adv. High Energy Phys.}\ }\textbf {\bibinfo {volume} {2018}},\
  \bibinfo {pages} {9785318} (\bibinfo {year} {2018})},\ \Eprint
  {https://arxiv.org/abs/1803.10940} {arXiv:1803.10940 [hep-ph]}\BibitemShut
  {NoStop}%
\bibitem [{\citenamefont {Abdullahi}\ \emph {et~al.}(2023)\citenamefont
  {Abdullahi} \emph {et~al.}}]{Abdullahi:2022jlv}%
  \BibitemOpen
  \bibfield  {author} {\bibinfo {author} {\bibfnamefont {A.~M.}\ \bibnamefont
  {Abdullahi}} \emph {et~al.},\ }\bibfield  {title} {\emph {\bibinfo {title}
  {{The present and future status of heavy neutral leptons}}},\ }\href
  {https://doi.org/10.1088/1361-6471/ac98f9} {\bibfield  {journal} {\bibinfo
  {journal} {J. Phys. G}\ }\textbf {\bibinfo {volume} {50}},\ \bibinfo {pages}
  {020501} (\bibinfo {year} {2023})},\ \Eprint
  {https://arxiv.org/abs/2203.08039} {arXiv:2203.08039 [hep-ph]}\BibitemShut
  {NoStop}%
\bibitem [{\citenamefont {Das}\ and\ \citenamefont
  {Mandal}(2021)}]{Das:2020uer}%
  \BibitemOpen
  \bibfield  {author} {\bibinfo {author} {\bibfnamefont {A.}~\bibnamefont
  {Das}}\ and\ \bibinfo {author} {\bibfnamefont {S.}~\bibnamefont {Mandal}},\
  }\bibfield  {title} {\emph {\bibinfo {title} {{Bounds on the triplet fermions
  in type-III seesaw and implications for collider searches}}},\ }\href
  {https://doi.org/10.1016/j.nuclphysb.2021.115374} {\bibfield  {journal}
  {\bibinfo  {journal} {Nucl. Phys. B}\ }\textbf {\bibinfo {volume} {966}},\
  \bibinfo {pages} {115374} (\bibinfo {year} {2021})},\ \Eprint
  {https://arxiv.org/abs/2006.04123} {arXiv:2006.04123 [hep-ph]}\BibitemShut
  {NoStop}%
\bibitem [{\citenamefont {Arg\"uelles}\ \emph {et~al.}(2023)\citenamefont
  {Arg\"uelles} \emph {et~al.}}]{Arguelles:2022tki}%
  \BibitemOpen
  \bibfield  {author} {\bibinfo {author} {\bibfnamefont {C.~A.}\ \bibnamefont
  {Arg\"uelles}} \emph {et~al.},\ }\bibfield  {title} {\emph {\bibinfo {title}
  {{Snowmass White Paper: Beyond the Standard Model effects on Neutrino
  Flavor}}},\ }\href {https://doi.org/10.1140/epjc/s10052-022-11049-7}
  {\bibfield  {journal} {\bibinfo  {journal} {Eur. Phys. J. C}\ }\textbf
  {\bibinfo {volume} {83}},\ \bibinfo {pages} {15} (\bibinfo {year} {2023})},\
  \Eprint {https://arxiv.org/abs/2203.10811} {arXiv:2203.10811
  [hep-ph]}\BibitemShut {NoStop}%
\bibitem [{\citenamefont {Ge}\ \emph {et~al.}(2016)\citenamefont {Ge},
  \citenamefont {He},\ and\ \citenamefont {Xiao}}]{Ge:2016zro}%
  \BibitemOpen
  \bibfield  {author} {\bibinfo {author} {\bibfnamefont {S.-F.}\ \bibnamefont
  {Ge}}, \bibinfo {author} {\bibfnamefont {H.-J.}\ \bibnamefont {He}},\ and\
  \bibinfo {author} {\bibfnamefont {R.-Q.}\ \bibnamefont {Xiao}},\ }\bibfield
  {title} {\emph {\bibinfo {title} {{Probing new physics scales from Higgs and
  electroweak observables at $e^+e^-$ Higgs factory}}},\ }\href
  {https://doi.org/10.1007/JHEP10(2016)007} {\bibfield  {journal} {\bibinfo
  {journal} {JHEP}\ }\textbf {\bibinfo {volume} {10}},\ \bibinfo {pages}
  {007}},\ \Eprint {https://arxiv.org/abs/1603.03385} {arXiv:1603.03385
  [hep-ph]}\BibitemShut {NoStop}%
\bibitem [{\citenamefont {Baglio}\ and\ \citenamefont
  {Weiland}(2017)}]{Baglio:2016bop}%
  \BibitemOpen
  \bibfield  {author} {\bibinfo {author} {\bibfnamefont {J.}~\bibnamefont
  {Baglio}}\ and\ \bibinfo {author} {\bibfnamefont {C.}~\bibnamefont
  {Weiland}},\ }\bibfield  {title} {\emph {\bibinfo {title} {{The triple Higgs
  coupling: A new probe of low-scale seesaw models}}},\ }\href
  {https://doi.org/10.1007/JHEP04(2017)038} {\bibfield  {journal} {\bibinfo
  {journal} {JHEP}\ }\textbf {\bibinfo {volume} {04}},\ \bibinfo {pages}
  {038}},\ \Eprint {https://arxiv.org/abs/1612.06403} {arXiv:1612.06403
  [hep-ph]}\BibitemShut {NoStop}%
\bibitem [{\citenamefont {Baglio}\ and\ \citenamefont
  {Weiland}(2016)}]{Baglio:2016ijw}%
  \BibitemOpen
  \bibfield  {author} {\bibinfo {author} {\bibfnamefont {J.}~\bibnamefont
  {Baglio}}\ and\ \bibinfo {author} {\bibfnamefont {C.}~\bibnamefont
  {Weiland}},\ }\bibfield  {title} {\emph {\bibinfo {title} {{Heavy neutrino
  impact on the triple Higgs coupling}}},\ }\href
  {https://doi.org/10.1103/PhysRevD.94.013002} {\bibfield  {journal} {\bibinfo
  {journal} {Phys. Rev. D}\ }\textbf {\bibinfo {volume} {94}},\ \bibinfo
  {pages} {013002} (\bibinfo {year} {2016})},\ \Eprint
  {https://arxiv.org/abs/1603.00879} {arXiv:1603.00879 [hep-ph]}\BibitemShut
  {NoStop}%
\bibitem [{\citenamefont {Minkowski}(1977)}]{Minkowski:1977sc}%
  \BibitemOpen
  \bibfield  {author} {\bibinfo {author} {\bibfnamefont {P.}~\bibnamefont
  {Minkowski}},\ }\bibfield  {title} {\emph {\bibinfo {title} {{$\mu \to
  e\gamma$ at a Rate of One Out of $10^{9}$ Muon Decays?}}},\ }\href
  {https://doi.org/10.1016/0370-2693(77)90435-X} {\bibfield  {journal}
  {\bibinfo  {journal} {Phys. Lett.}\ }\textbf {\bibinfo {volume} {67B}},\
  \bibinfo {pages} {421} (\bibinfo {year} {1977})}\BibitemShut {NoStop}%
\bibitem [{\citenamefont {Yanagida}(1979)}]{Yanagida:1979as}%
  \BibitemOpen
  \bibfield  {author} {\bibinfo {author} {\bibfnamefont {T.}~\bibnamefont
  {Yanagida}},\ }\bibfield  {title} {\emph {\bibinfo {title} {{Horizontal gauge
  symmetry and masses of neutrinos}}},\ }\bibfield  {booktitle} {\emph
  {\bibinfo {booktitle} {{Proceedings: Workshop on the Unified Theories and the
  Baryon Number in the Universe: Tsukuba, Japan, February 13-14, 1979}}},\
  }\href@noop {} {\bibfield  {journal} {\bibinfo  {journal} {Conf. Proc.}\
  }\textbf {\bibinfo {volume} {C7902131}},\ \bibinfo {pages} {95} (\bibinfo
  {year} {1979})}\BibitemShut {NoStop}%
\bibitem [{\citenamefont {Gell-Mann}\ \emph {et~al.}(1979)\citenamefont
  {Gell-Mann}, \citenamefont {Ramond},\ and\ \citenamefont
  {Slansky}}]{GellMann:1980vs}%
  \BibitemOpen
  \bibfield  {author} {\bibinfo {author} {\bibfnamefont {M.}~\bibnamefont
  {Gell-Mann}}, \bibinfo {author} {\bibfnamefont {P.}~\bibnamefont {Ramond}},\
  and\ \bibinfo {author} {\bibfnamefont {R.}~\bibnamefont {Slansky}},\
  }\bibfield  {title} {\emph {\bibinfo {title} {{Complex Spinors and Unified
  Theories}}},\ }\bibfield  {booktitle} {\emph {\bibinfo {booktitle}
  {{Supergravity Workshop Stony Brook, New York, September 27-28, 1979}}},\
  }\href@noop {} {\bibfield  {journal} {\bibinfo  {journal} {Conf. Proc.}\
  }\textbf {\bibinfo {volume} {C790927}},\ \bibinfo {pages} {315} (\bibinfo
  {year} {1979})},\ \Eprint {https://arxiv.org/abs/1306.4669} {arXiv:1306.4669
  [hep-th]}\BibitemShut {NoStop}%
\bibitem [{\citenamefont {Glashow}(1980)}]{Glashow1979}%
  \BibitemOpen
  \bibfield  {author} {\bibinfo {author} {\bibfnamefont {S.~L.}\ \bibnamefont
  {Glashow}},\ }\href@noop {} {\emph {\bibinfo {title} {{Quarks and Leptons,
  Carg\`ese 1979}}}},\ edited by\ \bibinfo {editor} {\bibfnamefont
  {M.}~\bibnamefont {L\'evy}}\ (\bibinfo  {publisher} {Plenum Press},\ \bibinfo
  {address} {New York},\ \bibinfo {year} {1980}),\ p.\ \bibinfo {pages}
  {720}\BibitemShut {NoStop}%
\bibitem [{\citenamefont {Mohapatra}\ and\ \citenamefont
  {Senjanovic}(1980)}]{Mohapatra:1979ia}%
  \BibitemOpen
  \bibfield  {author} {\bibinfo {author} {\bibfnamefont {R.~N.}\ \bibnamefont
  {Mohapatra}}\ and\ \bibinfo {author} {\bibfnamefont {G.}~\bibnamefont
  {Senjanovic}},\ }\bibfield  {title} {\emph {\bibinfo {title} {{Neutrino Mass
  and Spontaneous Parity Nonconservation}}},\ }\href
  {https://doi.org/10.1103/PhysRevLett.44.912} {\bibfield  {journal} {\bibinfo
  {journal} {Phys. Rev. Lett.}\ }\textbf {\bibinfo {volume} {44}},\ \bibinfo
  {pages} {912} (\bibinfo {year} {1980})}\BibitemShut {NoStop}%
\bibitem [{\citenamefont {Foot}\ \emph {et~al.}(1989)\citenamefont {Foot},
  \citenamefont {Lew}, \citenamefont {He},\ and\ \citenamefont
  {Joshi}}]{Foot:1988aq}%
  \BibitemOpen
  \bibfield  {author} {\bibinfo {author} {\bibfnamefont {R.}~\bibnamefont
  {Foot}}, \bibinfo {author} {\bibfnamefont {H.}~\bibnamefont {Lew}}, \bibinfo
  {author} {\bibfnamefont {X.~G.}\ \bibnamefont {He}},\ and\ \bibinfo {author}
  {\bibfnamefont {G.~C.}\ \bibnamefont {Joshi}},\ }\bibfield  {title} {\emph
  {\bibinfo {title} {{Seesaw Neutrino Masses Induced by a Triplet of
  Leptons}}},\ }\href {https://doi.org/10.1007/BF01415558} {\bibfield
  {journal} {\bibinfo  {journal} {Z. Phys.}\ }\textbf {\bibinfo {volume}
  {C44}},\ \bibinfo {pages} {441} (\bibinfo {year} {1989})}\BibitemShut
  {NoStop}%
\bibitem [{\citenamefont {Freitas}\ and\ \citenamefont
  {Song}(2023)}]{Freitas:2022hyp}%
  \BibitemOpen
  \bibfield  {author} {\bibinfo {author} {\bibfnamefont {A.}~\bibnamefont
  {Freitas}}\ and\ \bibinfo {author} {\bibfnamefont {Q.}~\bibnamefont {Song}},\
  }\bibfield  {title} {\emph {\bibinfo {title} {{Two-Loop Electroweak
  Corrections with Fermion Loops to $e^+e^-\to ZH$}}},\ }\href
  {https://doi.org/10.1103/PhysRevLett.130.031801} {\bibfield  {journal}
  {\bibinfo  {journal} {Phys. Rev. Lett.}\ }\textbf {\bibinfo {volume} {130}},\
  \bibinfo {pages} {031801} (\bibinfo {year} {2023})},\ \Eprint
  {https://arxiv.org/abs/2209.07612} {arXiv:2209.07612 [hep-ph]}\BibitemShut
  {NoStop}%
\bibitem [{\citenamefont {Chen}\ \emph {et~al.}(2022)\citenamefont {Chen},
  \citenamefont {Guan}, \citenamefont {He}, \citenamefont {Li}, \citenamefont
  {Liu},\ and\ \citenamefont {Ma}}]{Chen:2022mre}%
  \BibitemOpen
  \bibfield  {author} {\bibinfo {author} {\bibfnamefont {X.}~\bibnamefont
  {Chen}}, \bibinfo {author} {\bibfnamefont {X.}~\bibnamefont {Guan}}, \bibinfo
  {author} {\bibfnamefont {C.-Q.}\ \bibnamefont {He}}, \bibinfo {author}
  {\bibfnamefont {Z.}~\bibnamefont {Li}}, \bibinfo {author} {\bibfnamefont
  {X.}~\bibnamefont {Liu}},\ and\ \bibinfo {author} {\bibfnamefont {Y.-Q.}\
  \bibnamefont {Ma}},\ }\bibfield  {title} {\emph {\bibinfo {title} {{Complete
  two-loop electroweak corrections to $e^+e^-\rightarrow HZ$}}},\ }\href@noop
  {} {\  (\bibinfo {year} {2022})},\ \Eprint {https://arxiv.org/abs/2209.14953}
  {arXiv:2209.14953 [hep-ph]}\BibitemShut {NoStop}%
\bibitem [{\citenamefont {Fleischer}\ and\ \citenamefont
  {Jegerlehner}(1983)}]{FLEISCHER1983469}%
  \BibitemOpen
  \bibfield  {author} {\bibinfo {author} {\bibfnamefont {J.}~\bibnamefont
  {Fleischer}}\ and\ \bibinfo {author} {\bibfnamefont {F.}~\bibnamefont
  {Jegerlehner}},\ }\bibfield  {title} {\emph {\bibinfo {title} {{Radiative
  corrections to Higgs production by $e^+e^-\to Zh$ in the Weinberg-Salam
  model}}},\ }\href
  {https://doi.org/https://doi.org/10.1016/0550-3213(83)90296-1} {\bibfield
  {journal} {\bibinfo  {journal} {Nuclear Physics B}\ }\textbf {\bibinfo
  {volume} {216}},\ \bibinfo {pages} {469} (\bibinfo {year}
  {1983})}\BibitemShut {NoStop}%
\bibitem [{\citenamefont {Kniehl}(1991)}]{Kniehl1991RadiativeCF}%
  \BibitemOpen
  \bibfield  {author} {\bibinfo {author} {\bibfnamefont {B.~A.}\ \bibnamefont
  {Kniehl}},\ }\bibfield  {title} {\emph {\bibinfo {title} {{Radiative
  corrections for associated ZH production at future $e^+e^-$ colliders}}},\
  }\href@noop {} {\bibfield  {journal} {\bibinfo  {journal} {Zeitschrift
  f{\"u}r Physik C Particles and Fields}\ }\textbf {\bibinfo {volume} {55}},\
  \bibinfo {pages} {605} (\bibinfo {year} {1991})}\BibitemShut {NoStop}%
\bibitem [{\citenamefont {Denner}\ \emph {et~al.}(1992)\citenamefont {Denner},
  \citenamefont {Kublbeck}, \citenamefont {Mertig},\ and\ \citenamefont
  {Bohm}}]{Denner:1992bc}%
  \BibitemOpen
  \bibfield  {author} {\bibinfo {author} {\bibfnamefont {A.}~\bibnamefont
  {Denner}}, \bibinfo {author} {\bibfnamefont {J.}~\bibnamefont {Kublbeck}},
  \bibinfo {author} {\bibfnamefont {R.}~\bibnamefont {Mertig}},\ and\ \bibinfo
  {author} {\bibfnamefont {M.}~\bibnamefont {Bohm}},\ }\bibfield  {title}
  {\emph {\bibinfo {title} {{Electroweak radiative corrections to $e^+e^-\to
  Zh$}}},\ }\href {https://doi.org/10.1007/BF01555523} {\bibfield  {journal}
  {\bibinfo  {journal} {Z. Phys. C}\ }\textbf {\bibinfo {volume} {56}},\
  \bibinfo {pages} {261} (\bibinfo {year} {1992})}\BibitemShut {NoStop}%
\bibitem [{\citenamefont {Bondarenko}\ \emph {et~al.}(2019)\citenamefont
  {Bondarenko}, \citenamefont {Dydyshka}, \citenamefont {Kalinovskaya},
  \citenamefont {Rumyantsev}, \citenamefont {Sadykov},\ and\ \citenamefont
  {Yermolchyk}}]{Bondarenko:2018sgg}%
  \BibitemOpen
  \bibfield  {author} {\bibinfo {author} {\bibfnamefont {S.}~\bibnamefont
  {Bondarenko}}, \bibinfo {author} {\bibfnamefont {Y.}~\bibnamefont
  {Dydyshka}}, \bibinfo {author} {\bibfnamefont {L.}~\bibnamefont
  {Kalinovskaya}}, \bibinfo {author} {\bibfnamefont {L.}~\bibnamefont
  {Rumyantsev}}, \bibinfo {author} {\bibfnamefont {R.}~\bibnamefont
  {Sadykov}},\ and\ \bibinfo {author} {\bibfnamefont {V.}~\bibnamefont
  {Yermolchyk}},\ }\bibfield  {title} {\emph {\bibinfo {title} {{One-loop
  electroweak radiative corrections to polarized $e^+e^- \to ZH$}}},\ }\href
  {https://doi.org/10.1103/PhysRevD.100.073002} {\bibfield  {journal} {\bibinfo
   {journal} {Phys. Rev. D}\ }\textbf {\bibinfo {volume} {100}},\ \bibinfo
  {pages} {073002} (\bibinfo {year} {2019})},\ \Eprint
  {https://arxiv.org/abs/1812.10965} {arXiv:1812.10965 [hep-ph]}\BibitemShut
  {NoStop}%
\bibitem [{\citenamefont {Gong}\ \emph {et~al.}(2017)\citenamefont {Gong},
  \citenamefont {Li}, \citenamefont {Xu}, \citenamefont {Yang},\ and\
  \citenamefont {Zhao}}]{Gong:2016jys}%
  \BibitemOpen
  \bibfield  {author} {\bibinfo {author} {\bibfnamefont {Y.}~\bibnamefont
  {Gong}}, \bibinfo {author} {\bibfnamefont {Z.}~\bibnamefont {Li}}, \bibinfo
  {author} {\bibfnamefont {X.}~\bibnamefont {Xu}}, \bibinfo {author}
  {\bibfnamefont {L.~L.}\ \bibnamefont {Yang}},\ and\ \bibinfo {author}
  {\bibfnamefont {X.}~\bibnamefont {Zhao}},\ }\bibfield  {title} {\emph
  {\bibinfo {title} {{Mixed QCD-EW corrections for Higgs boson production at
  $e^+e^-$ colliders}}},\ }\href {https://doi.org/10.1103/PhysRevD.95.093003}
  {\bibfield  {journal} {\bibinfo  {journal} {Phys. Rev. D}\ }\textbf {\bibinfo
  {volume} {95}},\ \bibinfo {pages} {093003} (\bibinfo {year} {2017})},\
  \Eprint {https://arxiv.org/abs/1609.03955} {arXiv:1609.03955
  [hep-ph]}\BibitemShut {NoStop}%
\bibitem [{\citenamefont {Sun}\ \emph {et~al.}(2017)\citenamefont {Sun},
  \citenamefont {Feng}, \citenamefont {Jia},\ and\ \citenamefont
  {Sang}}]{Sun:2016bel}%
  \BibitemOpen
  \bibfield  {author} {\bibinfo {author} {\bibfnamefont {Q.-F.}\ \bibnamefont
  {Sun}}, \bibinfo {author} {\bibfnamefont {F.}~\bibnamefont {Feng}}, \bibinfo
  {author} {\bibfnamefont {Y.}~\bibnamefont {Jia}},\ and\ \bibinfo {author}
  {\bibfnamefont {W.-L.}\ \bibnamefont {Sang}},\ }\bibfield  {title} {\emph
  {\bibinfo {title} {{Mixed electroweak-QCD corrections to $e^+e^-\to ZH$ at
  Higgs factories}}},\ }\href {https://doi.org/10.1103/PhysRevD.96.051301}
  {\bibfield  {journal} {\bibinfo  {journal} {Phys. Rev. D}\ }\textbf {\bibinfo
  {volume} {96}},\ \bibinfo {pages} {051301} (\bibinfo {year} {2017})},\
  \Eprint {https://arxiv.org/abs/1609.03995} {arXiv:1609.03995
  [hep-ph]}\BibitemShut {NoStop}%
\bibitem [{\citenamefont {Chen}\ \emph {et~al.}(2019)\citenamefont {Chen},
  \citenamefont {Feng}, \citenamefont {Jia},\ and\ \citenamefont
  {Sang}}]{Chen:2018xau}%
  \BibitemOpen
  \bibfield  {author} {\bibinfo {author} {\bibfnamefont {W.}~\bibnamefont
  {Chen}}, \bibinfo {author} {\bibfnamefont {F.}~\bibnamefont {Feng}}, \bibinfo
  {author} {\bibfnamefont {Y.}~\bibnamefont {Jia}},\ and\ \bibinfo {author}
  {\bibfnamefont {W.-L.}\ \bibnamefont {Sang}},\ }\bibfield  {title} {\emph
  {\bibinfo {title} {{Mixed electroweak-QCD corrections to $e^+e^-\to
  \mu^+\mu^- H$ at CEPC with finite-width effect}}},\ }\href
  {https://doi.org/10.1088/1674-1137/43/1/013108} {\bibfield  {journal}
  {\bibinfo  {journal} {Chin. Phys. C}\ }\textbf {\bibinfo {volume} {43}},\
  \bibinfo {pages} {013108} (\bibinfo {year} {2019})},\ \Eprint
  {https://arxiv.org/abs/1811.05453} {arXiv:1811.05453 [hep-ph]}\BibitemShut
  {NoStop}%
\bibitem [{\citenamefont {Abramowicz}\ \emph {et~al.}(2017)\citenamefont
  {Abramowicz} \emph {et~al.}}]{Abramowicz:2016zbo}%
  \BibitemOpen
  \bibfield  {author} {\bibinfo {author} {\bibfnamefont {H.}~\bibnamefont
  {Abramowicz}} \emph {et~al.},\ }\bibfield  {title} {\emph {\bibinfo {title}
  {{Higgs physics at the CLIC electron\textendash{}positron linear
  collider}}},\ }\href {https://doi.org/10.1140/epjc/s10052-017-4968-5}
  {\bibfield  {journal} {\bibinfo  {journal} {Eur. Phys. J. C}\ }\textbf
  {\bibinfo {volume} {77}},\ \bibinfo {pages} {475} (\bibinfo {year} {2017})},\
  \Eprint {https://arxiv.org/abs/1608.07538} {arXiv:1608.07538
  [hep-ex]}\BibitemShut {NoStop}%
\bibitem [{\citenamefont {Kersten}\ and\ \citenamefont
  {Smirnov}(2007)}]{Kersten:2007vk}%
  \BibitemOpen
  \bibfield  {author} {\bibinfo {author} {\bibfnamefont {J.}~\bibnamefont
  {Kersten}}\ and\ \bibinfo {author} {\bibfnamefont {A.~Y.}\ \bibnamefont
  {Smirnov}},\ }\bibfield  {title} {\emph {\bibinfo {title} {{Right-Handed
  Neutrinos at CERN LHC and the Mechanism of Neutrino Mass Generation}}},\
  }\href {https://doi.org/10.1103/PhysRevD.76.073005} {\bibfield  {journal}
  {\bibinfo  {journal} {Phys. Rev. D}\ }\textbf {\bibinfo {volume} {76}},\
  \bibinfo {pages} {073005} (\bibinfo {year} {2007})},\ \Eprint
  {https://arxiv.org/abs/0705.3221} {arXiv:0705.3221 [hep-ph]}\BibitemShut
  {NoStop}%
\bibitem [{\citenamefont {Drewes}\ \emph {et~al.}(2019)\citenamefont {Drewes},
  \citenamefont {Klari\'c},\ and\ \citenamefont {Klose}}]{Drewes:2019byd}%
  \BibitemOpen
  \bibfield  {author} {\bibinfo {author} {\bibfnamefont {M.}~\bibnamefont
  {Drewes}}, \bibinfo {author} {\bibfnamefont {J.}~\bibnamefont {Klari\'c}},\
  and\ \bibinfo {author} {\bibfnamefont {P.}~\bibnamefont {Klose}},\ }\bibfield
   {title} {\emph {\bibinfo {title} {{On lepton number violation in heavy
  neutrino decays at colliders}}},\ }\href
  {https://doi.org/10.1007/JHEP11(2019)032} {\bibfield  {journal} {\bibinfo
  {journal} {JHEP}\ }\textbf {\bibinfo {volume} {11}},\ \bibinfo {pages}
  {032}},\ \Eprint {https://arxiv.org/abs/1907.13034} {arXiv:1907.13034
  [hep-ph]}\BibitemShut {NoStop}%
\bibitem [{\citenamefont {Abada}\ \emph {et~al.}(2022)\citenamefont {Abada},
  \citenamefont {Escribano}, \citenamefont {Marcano},\ and\ \citenamefont
  {Piazza}}]{Abada:2022wvh}%
  \BibitemOpen
  \bibfield  {author} {\bibinfo {author} {\bibfnamefont {A.}~\bibnamefont
  {Abada}}, \bibinfo {author} {\bibfnamefont {P.}~\bibnamefont {Escribano}},
  \bibinfo {author} {\bibfnamefont {X.}~\bibnamefont {Marcano}},\ and\ \bibinfo
  {author} {\bibfnamefont {G.}~\bibnamefont {Piazza}},\ }\bibfield  {title}
  {\emph {\bibinfo {title} {{Collider searches for heavy neutral leptons:
  beyond simplified scenarios}}},\ }\href
  {https://doi.org/10.1140/epjc/s10052-022-11011-7} {\bibfield  {journal}
  {\bibinfo  {journal} {Eur. Phys. J. C}\ }\textbf {\bibinfo {volume} {82}},\
  \bibinfo {pages} {1030} (\bibinfo {year} {2022})},\ \Eprint
  {https://arxiv.org/abs/2208.13882} {arXiv:2208.13882 [hep-ph]}\BibitemShut
  {NoStop}%
\bibitem [{\citenamefont {Wyler}\ and\ \citenamefont
  {Wolfenstein}(1983)}]{Wyler:1982dd}%
  \BibitemOpen
  \bibfield  {author} {\bibinfo {author} {\bibfnamefont {D.}~\bibnamefont
  {Wyler}}\ and\ \bibinfo {author} {\bibfnamefont {L.}~\bibnamefont
  {Wolfenstein}},\ }\bibfield  {title} {\emph {\bibinfo {title} {{Massless
  Neutrinos in Left-Right Symmetric Models}}},\ }\href
  {https://doi.org/10.1016/0550-3213(83)90482-0} {\bibfield  {journal}
  {\bibinfo  {journal} {Nucl. Phys. B}\ }\textbf {\bibinfo {volume} {218}},\
  \bibinfo {pages} {205} (\bibinfo {year} {1983})}\BibitemShut {NoStop}%
\bibitem [{\citenamefont {Bernabeu}\ \emph {et~al.}(1987)\citenamefont
  {Bernabeu}, \citenamefont {Santamaria}, \citenamefont {Vidal}, \citenamefont
  {Mendez},\ and\ \citenamefont {Valle}}]{Bernabeu:1987gr}%
  \BibitemOpen
  \bibfield  {author} {\bibinfo {author} {\bibfnamefont {J.}~\bibnamefont
  {Bernabeu}}, \bibinfo {author} {\bibfnamefont {A.}~\bibnamefont
  {Santamaria}}, \bibinfo {author} {\bibfnamefont {J.}~\bibnamefont {Vidal}},
  \bibinfo {author} {\bibfnamefont {A.}~\bibnamefont {Mendez}},\ and\ \bibinfo
  {author} {\bibfnamefont {J.~W.~F.}\ \bibnamefont {Valle}},\ }\bibfield
  {title} {\emph {\bibinfo {title} {{Lepton Flavor Nonconservation at
  High-Energies in a Superstring Inspired Standard Model}}},\ }\href
  {https://doi.org/10.1016/0370-2693(87)91100-2} {\bibfield  {journal}
  {\bibinfo  {journal} {Phys. Lett. B}\ }\textbf {\bibinfo {volume} {187}},\
  \bibinfo {pages} {303} (\bibinfo {year} {1987})}\BibitemShut {NoStop}%
\bibitem [{\citenamefont {Branco}\ \emph {et~al.}(1989)\citenamefont {Branco},
  \citenamefont {Grimus},\ and\ \citenamefont {Lavoura}}]{Branco:1988ex}%
  \BibitemOpen
  \bibfield  {author} {\bibinfo {author} {\bibfnamefont {G.~C.}\ \bibnamefont
  {Branco}}, \bibinfo {author} {\bibfnamefont {W.}~\bibnamefont {Grimus}},\
  and\ \bibinfo {author} {\bibfnamefont {L.}~\bibnamefont {Lavoura}},\
  }\bibfield  {title} {\emph {\bibinfo {title} {{The Seesaw Mechanism in the
  Presence of a Conserved Lepton Number}}},\ }\href
  {https://doi.org/10.1016/0550-3213(89)90304-0} {\bibfield  {journal}
  {\bibinfo  {journal} {Nucl. Phys. B}\ }\textbf {\bibinfo {volume} {312}},\
  \bibinfo {pages} {492} (\bibinfo {year} {1989})}\BibitemShut {NoStop}%
\bibitem [{\citenamefont {Tommasini}\ \emph {et~al.}(1995)\citenamefont
  {Tommasini}, \citenamefont {Barenboim}, \citenamefont {Bernabeu},\ and\
  \citenamefont {Jarlskog}}]{Tommasini:1995ii}%
  \BibitemOpen
  \bibfield  {author} {\bibinfo {author} {\bibfnamefont {D.}~\bibnamefont
  {Tommasini}}, \bibinfo {author} {\bibfnamefont {G.}~\bibnamefont
  {Barenboim}}, \bibinfo {author} {\bibfnamefont {J.}~\bibnamefont
  {Bernabeu}},\ and\ \bibinfo {author} {\bibfnamefont {C.}~\bibnamefont
  {Jarlskog}},\ }\bibfield  {title} {\emph {\bibinfo {title} {{Nondecoupling of
  heavy neutrinos and lepton flavor violation}}},\ }\href
  {https://doi.org/10.1016/0550-3213(95)00201-3} {\bibfield  {journal}
  {\bibinfo  {journal} {Nucl. Phys. B}\ }\textbf {\bibinfo {volume} {444}},\
  \bibinfo {pages} {451} (\bibinfo {year} {1995})},\ \Eprint
  {https://arxiv.org/abs/hep-ph/9503228} {arXiv:hep-ph/9503228}\BibitemShut
  {NoStop}%
\bibitem [{\citenamefont {Pilaftsis}(2005)}]{Pilaftsis:2004xx}%
  \BibitemOpen
  \bibfield  {author} {\bibinfo {author} {\bibfnamefont {A.}~\bibnamefont
  {Pilaftsis}},\ }\bibfield  {title} {\emph {\bibinfo {title} {{Resonant
  tau-leptogenesis with observable lepton number violation}}},\ }\href
  {https://doi.org/10.1103/PhysRevLett.95.081602} {\bibfield  {journal}
  {\bibinfo  {journal} {Phys. Rev. Lett.}\ }\textbf {\bibinfo {volume} {95}},\
  \bibinfo {pages} {081602} (\bibinfo {year} {2005})},\ \Eprint
  {https://arxiv.org/abs/hep-ph/0408103} {arXiv:hep-ph/0408103}\BibitemShut
  {NoStop}%
\bibitem [{\citenamefont {Pilaftsis}\ and\ \citenamefont
  {Underwood}(2005)}]{Pilaftsis:2005rv}%
  \BibitemOpen
  \bibfield  {author} {\bibinfo {author} {\bibfnamefont {A.}~\bibnamefont
  {Pilaftsis}}\ and\ \bibinfo {author} {\bibfnamefont {T.~E.~J.}\ \bibnamefont
  {Underwood}},\ }\bibfield  {title} {\emph {\bibinfo {title}
  {{Electroweak-scale resonant leptogenesis}}},\ }\href
  {https://doi.org/10.1103/PhysRevD.72.113001} {\bibfield  {journal} {\bibinfo
  {journal} {Phys. Rev. D}\ }\textbf {\bibinfo {volume} {72}},\ \bibinfo
  {pages} {113001} (\bibinfo {year} {2005})},\ \Eprint
  {https://arxiv.org/abs/hep-ph/0506107} {arXiv:hep-ph/0506107}\BibitemShut
  {NoStop}%
\bibitem [{\citenamefont {Shaposhnikov}(2007)}]{Shaposhnikov:2006nn}%
  \BibitemOpen
  \bibfield  {author} {\bibinfo {author} {\bibfnamefont {M.}~\bibnamefont
  {Shaposhnikov}},\ }\bibfield  {title} {\emph {\bibinfo {title} {{A Possible
  symmetry of the $\nu$MSM}}},\ }\href
  {https://doi.org/10.1016/j.nuclphysb.2006.11.003} {\bibfield  {journal}
  {\bibinfo  {journal} {Nucl. Phys. B}\ }\textbf {\bibinfo {volume} {763}},\
  \bibinfo {pages} {49} (\bibinfo {year} {2007})},\ \Eprint
  {https://arxiv.org/abs/hep-ph/0605047} {arXiv:hep-ph/0605047}\BibitemShut
  {NoStop}%
\bibitem [{\citenamefont {Abada}\ \emph {et~al.}(2007)\citenamefont {Abada},
  \citenamefont {Biggio}, \citenamefont {Bonnet}, \citenamefont {Gavela},\ and\
  \citenamefont {Hambye}}]{Abada:2007ux}%
  \BibitemOpen
  \bibfield  {author} {\bibinfo {author} {\bibfnamefont {A.}~\bibnamefont
  {Abada}}, \bibinfo {author} {\bibfnamefont {C.}~\bibnamefont {Biggio}},
  \bibinfo {author} {\bibfnamefont {F.}~\bibnamefont {Bonnet}}, \bibinfo
  {author} {\bibfnamefont {M.~B.}\ \bibnamefont {Gavela}},\ and\ \bibinfo
  {author} {\bibfnamefont {T.}~\bibnamefont {Hambye}},\ }\bibfield  {title}
  {\emph {\bibinfo {title} {{Low energy effects of neutrino masses}}},\ }\href
  {https://doi.org/10.1088/1126-6708/2007/12/061} {\bibfield  {journal}
  {\bibinfo  {journal} {JHEP}\ }\textbf {\bibinfo {volume} {12}},\ \bibinfo
  {pages} {061}},\ \Eprint {https://arxiv.org/abs/0707.4058} {arXiv:0707.4058
  [hep-ph]}\BibitemShut {NoStop}%
\bibitem [{\citenamefont {Gavela}\ \emph {et~al.}(2009)\citenamefont {Gavela},
  \citenamefont {Hambye}, \citenamefont {Hernandez},\ and\ \citenamefont
  {Hernandez}}]{Gavela:2009cd}%
  \BibitemOpen
  \bibfield  {author} {\bibinfo {author} {\bibfnamefont {M.~B.}\ \bibnamefont
  {Gavela}}, \bibinfo {author} {\bibfnamefont {T.}~\bibnamefont {Hambye}},
  \bibinfo {author} {\bibfnamefont {D.}~\bibnamefont {Hernandez}},\ and\
  \bibinfo {author} {\bibfnamefont {P.}~\bibnamefont {Hernandez}},\ }\bibfield
  {title} {\emph {\bibinfo {title} {{Minimal Flavour Seesaw Models}}},\ }\href
  {https://doi.org/10.1088/1126-6708/2009/09/038} {\bibfield  {journal}
  {\bibinfo  {journal} {JHEP}\ }\textbf {\bibinfo {volume} {09}},\ \bibinfo
  {pages} {038}},\ \Eprint {https://arxiv.org/abs/0906.1461} {arXiv:0906.1461
  [hep-ph]}\BibitemShut {NoStop}%
\bibitem [{\citenamefont {Eboli}\ \emph {et~al.}(2011)\citenamefont {Eboli},
  \citenamefont {Gonzalez-Fraile},\ and\ \citenamefont
  {Gonzalez-Garcia}}]{Eboli:2011ia}%
  \BibitemOpen
  \bibfield  {author} {\bibinfo {author} {\bibfnamefont {O.~J.~P.}\
  \bibnamefont {Eboli}}, \bibinfo {author} {\bibfnamefont {J.}~\bibnamefont
  {Gonzalez-Fraile}},\ and\ \bibinfo {author} {\bibfnamefont {M.~C.}\
  \bibnamefont {Gonzalez-Garcia}},\ }\bibfield  {title} {\emph {\bibinfo
  {title} {{Neutrino Masses at LHC: Minimal Lepton Flavour Violation in
  Type-III See-saw}}},\ }\href {https://doi.org/10.1007/JHEP12(2011)009}
  {\bibfield  {journal} {\bibinfo  {journal} {JHEP}\ }\textbf {\bibinfo
  {volume} {12}},\ \bibinfo {pages} {009}},\ \Eprint
  {https://arxiv.org/abs/1108.0661} {arXiv:1108.0661 [hep-ph]}\BibitemShut
  {NoStop}%
\bibitem [{\citenamefont {Fernandez-Martinez}\ \emph
  {et~al.}(2015)\citenamefont {Fernandez-Martinez}, \citenamefont
  {Hernandez-Garcia}, \citenamefont {Lopez-Pavon},\ and\ \citenamefont
  {Lucente}}]{Fernandez-Martinez:2015hxa}%
  \BibitemOpen
  \bibfield  {author} {\bibinfo {author} {\bibfnamefont {E.}~\bibnamefont
  {Fernandez-Martinez}}, \bibinfo {author} {\bibfnamefont {J.}~\bibnamefont
  {Hernandez-Garcia}}, \bibinfo {author} {\bibfnamefont {J.}~\bibnamefont
  {Lopez-Pavon}},\ and\ \bibinfo {author} {\bibfnamefont {M.}~\bibnamefont
  {Lucente}},\ }\bibfield  {title} {\emph {\bibinfo {title} {{Loop level
  constraints on Seesaw neutrino mixing}}},\ }\href
  {https://doi.org/10.1007/JHEP10(2015)130} {\bibfield  {journal} {\bibinfo
  {journal} {JHEP}\ }\textbf {\bibinfo {volume} {10}},\ \bibinfo {pages}
  {130}},\ \Eprint {https://arxiv.org/abs/1508.03051} {arXiv:1508.03051
  [hep-ph]}\BibitemShut {NoStop}%
\bibitem [{\citenamefont {Grzadkowski}\ \emph {et~al.}(2010)\citenamefont
  {Grzadkowski}, \citenamefont {Iskrzynski}, \citenamefont {Misiak},\ and\
  \citenamefont {Rosiek}}]{Grzadkowski:2010es}%
  \BibitemOpen
  \bibfield  {author} {\bibinfo {author} {\bibfnamefont {B.}~\bibnamefont
  {Grzadkowski}}, \bibinfo {author} {\bibfnamefont {M.}~\bibnamefont
  {Iskrzynski}}, \bibinfo {author} {\bibfnamefont {M.}~\bibnamefont {Misiak}},\
  and\ \bibinfo {author} {\bibfnamefont {J.}~\bibnamefont {Rosiek}},\
  }\bibfield  {title} {\emph {\bibinfo {title} {{Dimension-Six Terms in the
  Standard Model Lagrangian}}},\ }\href
  {https://doi.org/10.1007/JHEP10(2010)085} {\bibfield  {journal} {\bibinfo
  {journal} {JHEP}\ }\textbf {\bibinfo {volume} {10}},\ \bibinfo {pages}
  {085}},\ \Eprint {https://arxiv.org/abs/1008.4884} {arXiv:1008.4884
  [hep-ph]}\BibitemShut {NoStop}%
\bibitem [{\citenamefont {Berthier}\ and\ \citenamefont
  {Trott}(2015)}]{Berthier:2015oma}%
  \BibitemOpen
  \bibfield  {author} {\bibinfo {author} {\bibfnamefont {L.}~\bibnamefont
  {Berthier}}\ and\ \bibinfo {author} {\bibfnamefont {M.}~\bibnamefont
  {Trott}},\ }\bibfield  {title} {\emph {\bibinfo {title} {{Towards consistent
  Electroweak Precision Data constraints in the SMEFT}}},\ }\href
  {https://doi.org/10.1007/JHEP05(2015)024} {\bibfield  {journal} {\bibinfo
  {journal} {JHEP}\ }\textbf {\bibinfo {volume} {05}},\ \bibinfo {pages}
  {024}},\ \Eprint {https://arxiv.org/abs/1502.02570} {arXiv:1502.02570
  [hep-ph]}\BibitemShut {NoStop}%
\bibitem [{\citenamefont {Alonso}\ \emph {et~al.}(2014)\citenamefont {Alonso},
  \citenamefont {Jenkins}, \citenamefont {Manohar},\ and\ \citenamefont
  {Trott}}]{Alonso:2013hga}%
  \BibitemOpen
  \bibfield  {author} {\bibinfo {author} {\bibfnamefont {R.}~\bibnamefont
  {Alonso}}, \bibinfo {author} {\bibfnamefont {E.~E.}\ \bibnamefont {Jenkins}},
  \bibinfo {author} {\bibfnamefont {A.~V.}\ \bibnamefont {Manohar}},\ and\
  \bibinfo {author} {\bibfnamefont {M.}~\bibnamefont {Trott}},\ }\bibfield
  {title} {\emph {\bibinfo {title} {{Renormalization Group Evolution of the
  Standard Model Dimension Six Operators III: Gauge Coupling Dependence and
  Phenomenology}}},\ }\href {https://doi.org/10.1007/JHEP04(2014)159}
  {\bibfield  {journal} {\bibinfo  {journal} {JHEP}\ }\textbf {\bibinfo
  {volume} {04}},\ \bibinfo {pages} {159}},\ \Eprint
  {https://arxiv.org/abs/1312.2014} {arXiv:1312.2014 [hep-ph]}\BibitemShut
  {NoStop}%
\bibitem [{\citenamefont {Jenkins}\ \emph {et~al.}(2018)\citenamefont
  {Jenkins}, \citenamefont {Manohar},\ and\ \citenamefont
  {Stoffer}}]{Jenkins:2017jig}%
  \BibitemOpen
  \bibfield  {author} {\bibinfo {author} {\bibfnamefont {E.~E.}\ \bibnamefont
  {Jenkins}}, \bibinfo {author} {\bibfnamefont {A.~V.}\ \bibnamefont
  {Manohar}},\ and\ \bibinfo {author} {\bibfnamefont {P.}~\bibnamefont
  {Stoffer}},\ }\bibfield  {title} {\emph {\bibinfo {title} {{Low-Energy
  Effective Field Theory below the Electroweak Scale: Operators and
  Matching}}},\ }\href {https://doi.org/10.1007/JHEP03(2018)016} {\bibfield
  {journal} {\bibinfo  {journal} {JHEP}\ }\textbf {\bibinfo {volume} {03}},\
  \bibinfo {pages} {016}},\ \Eprint {https://arxiv.org/abs/1709.04486}
  {arXiv:1709.04486 [hep-ph]}\BibitemShut {NoStop}%
\bibitem [{\citenamefont {Celis}\ \emph {et~al.}(2017)\citenamefont {Celis},
  \citenamefont {Fuentes-Martin}, \citenamefont {Vicente},\ and\ \citenamefont
  {Virto}}]{Celis:2017hod}%
  \BibitemOpen
  \bibfield  {author} {\bibinfo {author} {\bibfnamefont {A.}~\bibnamefont
  {Celis}}, \bibinfo {author} {\bibfnamefont {J.}~\bibnamefont
  {Fuentes-Martin}}, \bibinfo {author} {\bibfnamefont {A.}~\bibnamefont
  {Vicente}},\ and\ \bibinfo {author} {\bibfnamefont {J.}~\bibnamefont
  {Virto}},\ }\bibfield  {title} {\emph {\bibinfo {title} {{DsixTools: The
  Standard Model Effective Field Theory Toolkit}}},\ }\href
  {https://doi.org/10.1140/epjc/s10052-017-4967-6} {\bibfield  {journal}
  {\bibinfo  {journal} {Eur. Phys. J. C}\ }\textbf {\bibinfo {volume} {77}},\
  \bibinfo {pages} {405} (\bibinfo {year} {2017})},\ \Eprint
  {https://arxiv.org/abs/1704.04504} {arXiv:1704.04504 [hep-ph]}\BibitemShut
  {NoStop}%
\bibitem [{\citenamefont {Du}\ \emph {et~al.}(2022)\citenamefont {Du},
  \citenamefont {Li},\ and\ \citenamefont {Yu}}]{Du:2022vso}%
  \BibitemOpen
  \bibfield  {author} {\bibinfo {author} {\bibfnamefont {Y.}~\bibnamefont
  {Du}}, \bibinfo {author} {\bibfnamefont {X.-X.}\ \bibnamefont {Li}},\ and\
  \bibinfo {author} {\bibfnamefont {J.-H.}\ \bibnamefont {Yu}},\ }\bibfield
  {title} {\emph {\bibinfo {title} {{Neutrino seesaw models at one-loop
  matching: discrimination by effective operators}}},\ }\href
  {https://doi.org/10.1007/JHEP09(2022)207} {\bibfield  {journal} {\bibinfo
  {journal} {JHEP}\ }\textbf {\bibinfo {volume} {09}},\ \bibinfo {pages}
  {207}},\ \Eprint {https://arxiv.org/abs/2201.04646} {arXiv:2201.04646
  [hep-ph]}\BibitemShut {NoStop}%
\bibitem [{\citenamefont {Zhang}\ and\ \citenamefont
  {Zhou}(2021)}]{Zhang:2021jdf}%
  \BibitemOpen
  \bibfield  {author} {\bibinfo {author} {\bibfnamefont {D.}~\bibnamefont
  {Zhang}}\ and\ \bibinfo {author} {\bibfnamefont {S.}~\bibnamefont {Zhou}},\
  }\bibfield  {title} {\emph {\bibinfo {title} {{Complete one-loop matching of
  the type-I seesaw model onto the Standard Model effective field theory}}},\
  }\href {https://doi.org/10.1007/JHEP09(2021)163} {\bibfield  {journal}
  {\bibinfo  {journal} {JHEP}\ }\textbf {\bibinfo {volume} {09}},\ \bibinfo
  {pages} {163}},\ \Eprint {https://arxiv.org/abs/2107.12133} {arXiv:2107.12133
  [hep-ph]}\BibitemShut {NoStop}%
\bibitem [{\citenamefont {Coy}\ and\ \citenamefont
  {Frigerio}(2022)}]{Coy:2021hyr}%
  \BibitemOpen
  \bibfield  {author} {\bibinfo {author} {\bibfnamefont {R.}~\bibnamefont
  {Coy}}\ and\ \bibinfo {author} {\bibfnamefont {M.}~\bibnamefont {Frigerio}},\
  }\bibfield  {title} {\emph {\bibinfo {title} {{Effective comparison of
  neutrino-mass models}}},\ }\href
  {https://doi.org/10.1103/PhysRevD.105.115041} {\bibfield  {journal} {\bibinfo
   {journal} {Phys. Rev. D}\ }\textbf {\bibinfo {volume} {105}},\ \bibinfo
  {pages} {115041} (\bibinfo {year} {2022})},\ \Eprint
  {https://arxiv.org/abs/2110.09126} {arXiv:2110.09126 [hep-ph]}\BibitemShut
  {NoStop}%
\bibitem [{\citenamefont {Pal}(2011)}]{Pal2010}%
  \BibitemOpen
  \bibfield  {author} {\bibinfo {author} {\bibfnamefont {P.~B.}\ \bibnamefont
  {Pal}},\ }\bibfield  {title} {\emph {\bibinfo {title} {{Dirac, Majorana, and
  Weyl fermions}}},\ }\href {https://doi.org/10.1119/1.3549729} {\bibfield
  {journal} {\bibinfo  {journal} {Am. J. Phys.}\ }\textbf {\bibinfo {volume}
  {79}},\ \bibinfo {pages} {485} (\bibinfo {year} {2011})},\ \Eprint
  {https://arxiv.org/abs/1006.1718} {arXiv:1006.1718 [hep-ph]}\BibitemShut
  {NoStop}%
\bibitem [{\citenamefont {Biggio}\ \emph {et~al.}(2020)\citenamefont {Biggio},
  \citenamefont {Fernandez-Martinez}, \citenamefont {Filaci}, \citenamefont
  {Hernandez-Garcia},\ and\ \citenamefont {Lopez-Pavon}}]{Biggio:2019eeo}%
  \BibitemOpen
  \bibfield  {author} {\bibinfo {author} {\bibfnamefont {C.}~\bibnamefont
  {Biggio}}, \bibinfo {author} {\bibfnamefont {E.}~\bibnamefont
  {Fernandez-Martinez}}, \bibinfo {author} {\bibfnamefont {M.}~\bibnamefont
  {Filaci}}, \bibinfo {author} {\bibfnamefont {J.}~\bibnamefont
  {Hernandez-Garcia}},\ and\ \bibinfo {author} {\bibfnamefont {J.}~\bibnamefont
  {Lopez-Pavon}},\ }\bibfield  {title} {\emph {\bibinfo {title} {{Global Bounds
  on the Type-III Seesaw}}},\ }\href {https://doi.org/10.1007/JHEP05(2020)022}
  {\bibfield  {journal} {\bibinfo  {journal} {JHEP}\ }\textbf {\bibinfo
  {volume} {05}},\ \bibinfo {pages} {022}},\ \Eprint
  {https://arxiv.org/abs/1911.11790} {arXiv:1911.11790 [hep-ph]}\BibitemShut
  {NoStop}%
\bibitem [{\citenamefont {Mohapatra}(1986)}]{Mohapatra:1986aw}%
  \BibitemOpen
  \bibfield  {author} {\bibinfo {author} {\bibfnamefont {R.~N.}\ \bibnamefont
  {Mohapatra}},\ }\bibfield  {title} {\emph {\bibinfo {title} {{Mechanism for
  Understanding Small Neutrino Mass in Superstring Theories}}},\ }\href
  {https://doi.org/10.1103/PhysRevLett.56.561} {\bibfield  {journal} {\bibinfo
  {journal} {Phys. Rev. Lett.}\ }\textbf {\bibinfo {volume} {56}},\ \bibinfo
  {pages} {561} (\bibinfo {year} {1986})}\BibitemShut {NoStop}%
\bibitem [{\citenamefont {Mohapatra}\ and\ \citenamefont
  {Valle}(1986)}]{Mohapatra:1986bd}%
  \BibitemOpen
  \bibfield  {author} {\bibinfo {author} {\bibfnamefont {R.~N.}\ \bibnamefont
  {Mohapatra}}\ and\ \bibinfo {author} {\bibfnamefont {J.~W.~F.}\ \bibnamefont
  {Valle}},\ }\bibfield  {title} {\emph {\bibinfo {title} {{Neutrino Mass and
  Baryon Number Nonconservation in Superstring Models}}},\ }\href
  {https://doi.org/10.1103/PhysRevD.34.1642} {\bibfield  {journal} {\bibinfo
  {journal} {Phys. Rev. D}\ }\textbf {\bibinfo {volume} {34}},\ \bibinfo
  {pages} {1642} (\bibinfo {year} {1986})}\BibitemShut {NoStop}%
\bibitem [{\citenamefont {Akhmedov}\ \emph
  {et~al.}(1996{\natexlab{a}})\citenamefont {Akhmedov}, \citenamefont
  {Lindner}, \citenamefont {Schnapka},\ and\ \citenamefont
  {Valle}}]{Akhmedov:1995ip}%
  \BibitemOpen
  \bibfield  {author} {\bibinfo {author} {\bibfnamefont {E.~K.}\ \bibnamefont
  {Akhmedov}}, \bibinfo {author} {\bibfnamefont {M.}~\bibnamefont {Lindner}},
  \bibinfo {author} {\bibfnamefont {E.}~\bibnamefont {Schnapka}},\ and\
  \bibinfo {author} {\bibfnamefont {J.~W.~F.}\ \bibnamefont {Valle}},\
  }\bibfield  {title} {\emph {\bibinfo {title} {{Left-right symmetry breaking
  in NJL approach}}},\ }\href {https://doi.org/10.1016/0370-2693(95)01504-3}
  {\bibfield  {journal} {\bibinfo  {journal} {Phys. Lett. B}\ }\textbf
  {\bibinfo {volume} {368}},\ \bibinfo {pages} {270} (\bibinfo {year}
  {1996}{\natexlab{a}})},\ \Eprint {https://arxiv.org/abs/hep-ph/9507275}
  {arXiv:hep-ph/9507275}\BibitemShut {NoStop}%
\bibitem [{\citenamefont {Akhmedov}\ \emph
  {et~al.}(1996{\natexlab{b}})\citenamefont {Akhmedov}, \citenamefont
  {Lindner}, \citenamefont {Schnapka},\ and\ \citenamefont
  {Valle}}]{Akhmedov:1995vm}%
  \BibitemOpen
  \bibfield  {author} {\bibinfo {author} {\bibfnamefont {E.~K.}\ \bibnamefont
  {Akhmedov}}, \bibinfo {author} {\bibfnamefont {M.}~\bibnamefont {Lindner}},
  \bibinfo {author} {\bibfnamefont {E.}~\bibnamefont {Schnapka}},\ and\
  \bibinfo {author} {\bibfnamefont {J.~W.~F.}\ \bibnamefont {Valle}},\
  }\bibfield  {title} {\emph {\bibinfo {title} {{Dynamical left-right symmetry
  breaking}}},\ }\href {https://doi.org/10.1103/PhysRevD.53.2752} {\bibfield
  {journal} {\bibinfo  {journal} {Phys. Rev. D}\ }\textbf {\bibinfo {volume}
  {53}},\ \bibinfo {pages} {2752} (\bibinfo {year} {1996}{\natexlab{b}})},\
  \Eprint {https://arxiv.org/abs/hep-ph/9509255}
  {arXiv:hep-ph/9509255}\BibitemShut {NoStop}%
\bibitem [{\citenamefont {Donini}\ \emph {et~al.}(2012)\citenamefont {Donini},
  \citenamefont {Hernandez}, \citenamefont {Lopez-Pavon}, \citenamefont
  {Maltoni},\ and\ \citenamefont {Schwetz}}]{Donini:2012tt}%
  \BibitemOpen
  \bibfield  {author} {\bibinfo {author} {\bibfnamefont {A.}~\bibnamefont
  {Donini}}, \bibinfo {author} {\bibfnamefont {P.}~\bibnamefont {Hernandez}},
  \bibinfo {author} {\bibfnamefont {J.}~\bibnamefont {Lopez-Pavon}}, \bibinfo
  {author} {\bibfnamefont {M.}~\bibnamefont {Maltoni}},\ and\ \bibinfo {author}
  {\bibfnamefont {T.}~\bibnamefont {Schwetz}},\ }\bibfield  {title} {\emph
  {\bibinfo {title} {{The minimal 3+2 neutrino model versus oscillation
  anomalies}}},\ }\href {https://doi.org/10.1007/JHEP07(2012)161} {\bibfield
  {journal} {\bibinfo  {journal} {JHEP}\ }\textbf {\bibinfo {volume} {07}},\
  \bibinfo {pages} {161}},\ \Eprint {https://arxiv.org/abs/1205.5230}
  {arXiv:1205.5230 [hep-ph]}\BibitemShut {NoStop}%
\bibitem [{\citenamefont {Sirunyan}\ \emph {et~al.}(2018)\citenamefont
  {Sirunyan} \emph {et~al.}}]{CMS:2018iaf}%
  \BibitemOpen
  \bibfield  {author} {\bibinfo {author} {\bibfnamefont {A.~M.}\ \bibnamefont
  {Sirunyan}} \emph {et~al.} (\bibinfo {collaboration} {CMS}),\ }\bibfield
  {title} {\emph {\bibinfo {title} {{Search for heavy neutral leptons in events
  with three charged leptons in proton-proton collisions at $\sqrt{s} =$ 13
  TeV}}},\ }\href {https://doi.org/10.1103/PhysRevLett.120.221801} {\bibfield
  {journal} {\bibinfo  {journal} {Phys. Rev. Lett.}\ }\textbf {\bibinfo
  {volume} {120}},\ \bibinfo {pages} {221801} (\bibinfo {year} {2018})},\
  \Eprint {https://arxiv.org/abs/1802.02965} {arXiv:1802.02965
  [hep-ex]}\BibitemShut {NoStop}%
\bibitem [{\citenamefont {{CMS Collaboration}}(2022)}]{CMS:2022rqc}%
  \BibitemOpen
  \bibfield  {author} {\bibinfo {author} {\bibnamefont {{CMS Collaboration}}}
  (\bibinfo {collaboration} {CMS}),\ }\bibfield  {title} {\emph {\bibinfo
  {title} {{Probing heavy Majorana neutrinos and the Weinberg operator through
  vector boson fusion processes in proton-proton collisions at $\sqrt{s}$ = 13
  TeV}}},\ }\href@noop {} {\  (\bibinfo {year} {2022})},\ \Eprint
  {https://arxiv.org/abs/2206.08956} {arXiv:2206.08956 [hep-ex]}\BibitemShut
  {NoStop}%
\bibitem [{\citenamefont {Aad}\ \emph {et~al.}(2022)\citenamefont {Aad} \emph
  {et~al.}}]{ATLAS:2022yhd}%
  \BibitemOpen
  \bibfield  {author} {\bibinfo {author} {\bibfnamefont {G.}~\bibnamefont
  {Aad}} \emph {et~al.} (\bibinfo {collaboration} {ATLAS}),\ }\bibfield
  {title} {\emph {\bibinfo {title} {{Search for type-III seesaw heavy leptons
  in leptonic final states in pp collisions at $\sqrt{s} = 13~\text {TeV}$ with
  the ATLAS detector}}},\ }\href
  {https://doi.org/10.1140/epjc/s10052-022-10785-0} {\bibfield  {journal}
  {\bibinfo  {journal} {Eur. Phys. J. C}\ }\textbf {\bibinfo {volume} {82}},\
  \bibinfo {pages} {988} (\bibinfo {year} {2022})},\ \Eprint
  {https://arxiv.org/abs/2202.02039} {arXiv:2202.02039 [hep-ex]}\BibitemShut
  {NoStop}%
\bibitem [{\citenamefont {Kniehl}\ and\ \citenamefont
  {Pilaftsis}(1994)}]{Kniehl:1994vk}%
  \BibitemOpen
  \bibfield  {author} {\bibinfo {author} {\bibfnamefont {B.}~\bibnamefont
  {Kniehl}}\ and\ \bibinfo {author} {\bibfnamefont {A.}~\bibnamefont
  {Pilaftsis}},\ }\bibfield  {title} {\emph {\bibinfo {title} {{Quantum effects
  on Higgs-boson production and decay due to Majorana neutrinos}}},\ }\href
  {https://doi.org/10.1016/0550-3213(94)90086-8} {\bibfield  {journal}
  {\bibinfo  {journal} {Nucl. Phys. B}\ }\textbf {\bibinfo {volume} {424}},\
  \bibinfo {pages} {18} (\bibinfo {year} {1994})},\ \Eprint
  {https://arxiv.org/abs/hep-ph/9402314} {arXiv:hep-ph/9402314
  [hep-ph]}\BibitemShut {NoStop}%
\bibitem [{\citenamefont {Craig}\ \emph {et~al.}(2015)\citenamefont {Craig},
  \citenamefont {Farina}, \citenamefont {McCullough},\ and\ \citenamefont
  {Perelstein}}]{Craig:2014una}%
  \BibitemOpen
  \bibfield  {author} {\bibinfo {author} {\bibfnamefont {N.}~\bibnamefont
  {Craig}}, \bibinfo {author} {\bibfnamefont {M.}~\bibnamefont {Farina}},
  \bibinfo {author} {\bibfnamefont {M.}~\bibnamefont {McCullough}},\ and\
  \bibinfo {author} {\bibfnamefont {M.}~\bibnamefont {Perelstein}},\ }\bibfield
   {title} {\emph {\bibinfo {title} {{Precision Higgsstrahlung as a Probe of
  New Physics}}},\ }\href {https://doi.org/10.1007/JHEP03(2015)146} {\bibfield
  {journal} {\bibinfo  {journal} {JHEP}\ }\textbf {\bibinfo {volume} {03}},\
  \bibinfo {pages} {146}},\ \Eprint {https://arxiv.org/abs/1411.0676}
  {arXiv:1411.0676 [hep-ph]}\BibitemShut {NoStop}%
\bibitem [{\citenamefont {Yan}\ \emph {et~al.}(2016)\citenamefont {Yan},
  \citenamefont {Watanuki}, \citenamefont {Fujii}, \citenamefont {Ishikawa},
  \citenamefont {Jeans}, \citenamefont {Strube}, \citenamefont {Tian},\ and\
  \citenamefont {Yamamoto}}]{Yan:2016xyx}%
  \BibitemOpen
  \bibfield  {author} {\bibinfo {author} {\bibfnamefont {J.}~\bibnamefont
  {Yan}}, \bibinfo {author} {\bibfnamefont {S.}~\bibnamefont {Watanuki}},
  \bibinfo {author} {\bibfnamefont {K.}~\bibnamefont {Fujii}}, \bibinfo
  {author} {\bibfnamefont {A.}~\bibnamefont {Ishikawa}}, \bibinfo {author}
  {\bibfnamefont {D.}~\bibnamefont {Jeans}}, \bibinfo {author} {\bibfnamefont
  {J.}~\bibnamefont {Strube}}, \bibinfo {author} {\bibfnamefont
  {J.}~\bibnamefont {Tian}},\ and\ \bibinfo {author} {\bibfnamefont
  {H.}~\bibnamefont {Yamamoto}},\ }\bibfield  {title} {\emph {\bibinfo {title}
  {{Measurement of the Higgs boson mass and $e^+e^- \to ZH$ cross section using
  $Z \to \mu^+\mu^-$ and $Z \to e^+ e^-$ at the ILC}}},\ }\href
  {https://doi.org/10.1103/PhysRevD.94.113002} {\bibfield  {journal} {\bibinfo
  {journal} {Phys. Rev. D}\ }\textbf {\bibinfo {volume} {94}},\ \bibinfo
  {pages} {113002} (\bibinfo {year} {2016})},\ \bibinfo {note} {[Erratum:
  Phys.Rev.D 103, 099903 (2021)]},\ \Eprint {https://arxiv.org/abs/1604.07524}
  {arXiv:1604.07524 [hep-ex]}\BibitemShut {NoStop}%
\bibitem [{\citenamefont {Thomson}(2016)}]{Thomson:2015jda}%
  \BibitemOpen
  \bibfield  {author} {\bibinfo {author} {\bibfnamefont {M.}~\bibnamefont
  {Thomson}},\ }\bibfield  {title} {\emph {\bibinfo {title} {{Model-independent
  measurement of the $e^+e^- \to\text{HZ}$ cross section at a future $e^+e^-$
  linear collider using hadronic Z decays}}},\ }\href
  {https://doi.org/10.1140/epjc/s10052-016-3911-5} {\bibfield  {journal}
  {\bibinfo  {journal} {Eur. Phys. J. C}\ }\textbf {\bibinfo {volume} {76}},\
  \bibinfo {pages} {72} (\bibinfo {year} {2016})},\ \Eprint
  {https://arxiv.org/abs/1509.02853} {arXiv:1509.02853 [hep-ex]}\BibitemShut
  {NoStop}%
\bibitem [{\citenamefont {Miyamoto}(2013)}]{Miyamoto:2013zva}%
  \BibitemOpen
  \bibfield  {author} {\bibinfo {author} {\bibfnamefont {A.}~\bibnamefont
  {Miyamoto}},\ }\bibfield  {title} {\emph {\bibinfo {title} {{A measurement of
  the total cross section of $\sigma_{Zh}$ at a future $e^{+}e^{-}$ collider
  using the hadronic decay mode of $Z$}}},\ }\href@noop {} {\  (\bibinfo {year}
  {2013})},\ \Eprint {https://arxiv.org/abs/1311.2248} {arXiv:1311.2248
  [hep-ex]}\BibitemShut {NoStop}%
\bibitem [{\citenamefont {de~Blas}\ \emph {et~al.}(2022)\citenamefont
  {de~Blas}, \citenamefont {Ciuchini}, \citenamefont {Franco}, \citenamefont
  {Goncalves}, \citenamefont {Mishima}, \citenamefont {Pierini}, \citenamefont
  {Reina},\ and\ \citenamefont {Silvestrini}}]{deBlas:2021wap}%
  \BibitemOpen
  \bibfield  {author} {\bibinfo {author} {\bibfnamefont {J.}~\bibnamefont
  {de~Blas}}, \bibinfo {author} {\bibfnamefont {M.}~\bibnamefont {Ciuchini}},
  \bibinfo {author} {\bibfnamefont {E.}~\bibnamefont {Franco}}, \bibinfo
  {author} {\bibfnamefont {A.}~\bibnamefont {Goncalves}}, \bibinfo {author}
  {\bibfnamefont {S.}~\bibnamefont {Mishima}}, \bibinfo {author} {\bibfnamefont
  {M.}~\bibnamefont {Pierini}}, \bibinfo {author} {\bibfnamefont
  {L.}~\bibnamefont {Reina}},\ and\ \bibinfo {author} {\bibfnamefont
  {L.}~\bibnamefont {Silvestrini}},\ }\bibfield  {title} {\emph {\bibinfo
  {title} {{Global analysis of electroweak data in the Standard Model}}},\
  }\href {https://doi.org/10.1103/PhysRevD.106.033003} {\bibfield  {journal}
  {\bibinfo  {journal} {Phys. Rev. D}\ }\textbf {\bibinfo {volume} {106}},\
  \bibinfo {pages} {033003} (\bibinfo {year} {2022})},\ \Eprint
  {https://arxiv.org/abs/2112.07274} {arXiv:2112.07274 [hep-ph]}\BibitemShut
  {NoStop}%
\bibitem [{\citenamefont {Schael}\ \emph {et~al.}(2006)\citenamefont {Schael}
  \emph {et~al.}}]{ALEPH:2005ab}%
  \BibitemOpen
  \bibfield  {author} {\bibinfo {author} {\bibfnamefont {S.}~\bibnamefont
  {Schael}} \emph {et~al.} (\bibinfo {collaboration} {ALEPH, DELPHI, L3, OPAL,
  SLD, LEP Electroweak Working Group, SLD Electroweak Group, SLD Heavy Flavour
  Group}),\ }\bibfield  {title} {\emph {\bibinfo {title} {{Precision
  electroweak measurements on the $Z$ resonance}}},\ }\href
  {https://doi.org/10.1016/j.physrep.2005.12.006} {\bibfield  {journal}
  {\bibinfo  {journal} {Phys. Rept.}\ }\textbf {\bibinfo {volume} {427}},\
  \bibinfo {pages} {257} (\bibinfo {year} {2006})},\ \Eprint
  {https://arxiv.org/abs/hep-ex/0509008} {arXiv:hep-ex/0509008
  [hep-ex]}\BibitemShut {NoStop}%
\bibitem [{\citenamefont {Workman}(2022)}]{Workman:2022ynf}%
  \BibitemOpen
  \bibfield  {author} {\bibinfo {author} {\bibfnamefont {R.~L.}\ \bibnamefont
  {Workman}} (\bibinfo {collaboration} {Particle Data Group}),\ }\bibfield
  {title} {\emph {\bibinfo {title} {{Review of Particle Physics}}},\ }\href
  {https://doi.org/10.1093/ptep/ptac097} {\bibfield  {journal} {\bibinfo
  {journal} {PTEP}\ }\textbf {\bibinfo {volume} {2022}},\ \bibinfo {pages}
  {083C01} (\bibinfo {year} {2022})}\BibitemShut {NoStop}%
\bibitem [{\citenamefont {Aaltonen}\ \emph {et~al.}(2022)\citenamefont
  {Aaltonen} \emph {et~al.}}]{CDF:2022hxs}%
  \BibitemOpen
  \bibfield  {author} {\bibinfo {author} {\bibfnamefont {T.}~\bibnamefont
  {Aaltonen}} \emph {et~al.} (\bibinfo {collaboration} {CDF}),\ }\bibfield
  {title} {\emph {\bibinfo {title} {{High-precision measurement of the W boson
  mass with the CDF II detector}}},\ }\href
  {https://doi.org/10.1126/science.abk1781} {\bibfield  {journal} {\bibinfo
  {journal} {Science}\ }\textbf {\bibinfo {volume} {376}},\ \bibinfo {pages}
  {170} (\bibinfo {year} {2022})}\BibitemShut {NoStop}%
\bibitem [{\citenamefont {Pich}(2021)}]{Pich:2020qna}%
  \BibitemOpen
  \bibfield  {author} {\bibinfo {author} {\bibfnamefont {A.}~\bibnamefont
  {Pich}},\ }\bibfield  {title} {\emph {\bibinfo {title} {{Challenges for tau
  physics at the TeraZ}}},\ }\href
  {https://doi.org/10.1140/epjp/s13360-021-02077-5} {\bibfield  {journal}
  {\bibinfo  {journal} {Eur. Phys. J. Plus}\ }\textbf {\bibinfo {volume}
  {136}},\ \bibinfo {pages} {1117} (\bibinfo {year} {2021})},\ \Eprint
  {https://arxiv.org/abs/2012.07099} {arXiv:2012.07099 [hep-ph]}\BibitemShut
  {NoStop}%
\bibitem [{\citenamefont {Seng}\ \emph {et~al.}(2022)\citenamefont {Seng},
  \citenamefont {Galviz}, \citenamefont {Gorchtein},\ and\ \citenamefont
  {Mei\ss{}ner}}]{Seng:2022wcw}%
  \BibitemOpen
  \bibfield  {author} {\bibinfo {author} {\bibfnamefont {C.-Y.}\ \bibnamefont
  {Seng}}, \bibinfo {author} {\bibfnamefont {D.}~\bibnamefont {Galviz}},
  \bibinfo {author} {\bibfnamefont {M.}~\bibnamefont {Gorchtein}},\ and\
  \bibinfo {author} {\bibfnamefont {U.-G.}\ \bibnamefont {Mei\ss{}ner}},\
  }\bibfield  {title} {\emph {\bibinfo {title} {{Complete theory of radiative
  corrections to K$_{\ell 3}$ decays and the V$_{us}$ update}}},\ }\href
  {https://doi.org/10.1007/JHEP07(2022)071} {\bibfield  {journal} {\bibinfo
  {journal} {JHEP}\ }\textbf {\bibinfo {volume} {07}},\ \bibinfo {pages}
  {071}},\ \Eprint {https://arxiv.org/abs/2203.05217} {arXiv:2203.05217
  [hep-ph]}\BibitemShut {NoStop}%
\bibitem [{\citenamefont {Crivellin}\ and\ \citenamefont
  {Hoferichter}(2020)}]{Crivellin:2020lzu}%
  \BibitemOpen
  \bibfield  {author} {\bibinfo {author} {\bibfnamefont {A.}~\bibnamefont
  {Crivellin}}\ and\ \bibinfo {author} {\bibfnamefont {M.}~\bibnamefont
  {Hoferichter}},\ }\bibfield  {title} {\emph {\bibinfo {title}
  {{\ensuremath{\beta} Decays as Sensitive Probes of Lepton Flavor
  Universality}}},\ }\href {https://doi.org/10.1103/PhysRevLett.125.111801}
  {\bibfield  {journal} {\bibinfo  {journal} {Phys. Rev. Lett.}\ }\textbf
  {\bibinfo {volume} {125}},\ \bibinfo {pages} {111801} (\bibinfo {year}
  {2020})},\ \Eprint {https://arxiv.org/abs/2002.07184} {arXiv:2002.07184
  [hep-ph]}\BibitemShut {NoStop}%
\bibitem [{\citenamefont {Bryman}\ \emph {et~al.}(2021)\citenamefont {Bryman},
  \citenamefont {Cirigliano}, \citenamefont {Crivellin},\ and\ \citenamefont
  {Inguglia}}]{Bryman:2021teu}%
  \BibitemOpen
  \bibfield  {author} {\bibinfo {author} {\bibfnamefont {D.}~\bibnamefont
  {Bryman}}, \bibinfo {author} {\bibfnamefont {V.}~\bibnamefont {Cirigliano}},
  \bibinfo {author} {\bibfnamefont {A.}~\bibnamefont {Crivellin}},\ and\
  \bibinfo {author} {\bibfnamefont {G.}~\bibnamefont {Inguglia}},\ }\bibfield
  {title} {\emph {\bibinfo {title} {{Testing Lepton Flavor Universality with
  Pion, Kaon, Tau, and Beta Decays}}}\ }\href
  {https://doi.org/10.1146/annurev-nucl-110121-051223}
  {10.1146/annurev-nucl-110121-051223} (\bibinfo {year} {2021}),\ \Eprint
  {https://arxiv.org/abs/2111.05338} {arXiv:2111.05338 [hep-ph]}\BibitemShut
  {NoStop}%
\bibitem [{\citenamefont {Baldini}\ \emph {et~al.}(2016)\citenamefont {Baldini}
  \emph {et~al.}}]{MEG:2016leq}%
  \BibitemOpen
  \bibfield  {author} {\bibinfo {author} {\bibfnamefont {A.~M.}\ \bibnamefont
  {Baldini}} \emph {et~al.} (\bibinfo {collaboration} {MEG}),\ }\bibfield
  {title} {\emph {\bibinfo {title} {{Search for the lepton flavour violating
  decay $\mu ^+ \rightarrow \mathrm {e}^+ \gamma $ with the full dataset of the
  MEG experiment}}},\ }\href {https://doi.org/10.1140/epjc/s10052-016-4271-x}
  {\bibfield  {journal} {\bibinfo  {journal} {Eur. Phys. J. C}\ }\textbf
  {\bibinfo {volume} {76}},\ \bibinfo {pages} {434} (\bibinfo {year} {2016})},\
  \Eprint {https://arxiv.org/abs/1605.05081} {arXiv:1605.05081
  [hep-ex]}\BibitemShut {NoStop}%
\bibitem [{\citenamefont {Baldini}\ \emph {et~al.}(2021)\citenamefont {Baldini}
  \emph {et~al.}}]{MEGII:2021fah}%
  \BibitemOpen
  \bibfield  {author} {\bibinfo {author} {\bibfnamefont {A.~M.}\ \bibnamefont
  {Baldini}} \emph {et~al.} (\bibinfo {collaboration} {MEG II}),\ }\bibfield
  {title} {\emph {\bibinfo {title} {{The Search for $\mu^+\to e^+ \gamma$ with
  10$^{-14}$ Sensitivity: the Upgrade of the MEG Experiment}}},\ }\href
  {https://doi.org/10.3390/sym13091591} {\bibfield  {journal} {\bibinfo
  {journal} {Symmetry}\ }\textbf {\bibinfo {volume} {13}},\ \bibinfo {pages}
  {1591} (\bibinfo {year} {2021})},\ \Eprint {https://arxiv.org/abs/2107.10767}
  {arXiv:2107.10767 [hep-ex]}\BibitemShut {NoStop}%
\bibitem [{\citenamefont {Bellgardt}\ \emph {et~al.}(1988)\citenamefont
  {Bellgardt} \emph {et~al.}}]{SINDRUM:1987nra}%
  \BibitemOpen
  \bibfield  {author} {\bibinfo {author} {\bibfnamefont {U.}~\bibnamefont
  {Bellgardt}} \emph {et~al.} (\bibinfo {collaboration} {SINDRUM}),\ }\bibfield
   {title} {\emph {\bibinfo {title} {{Search for the Decay $\mu^+ \to e^+ e^+
  e^-$}}},\ }\href {https://doi.org/10.1016/0550-3213(88)90462-2} {\bibfield
  {journal} {\bibinfo  {journal} {Nucl. Phys. B}\ }\textbf {\bibinfo {volume}
  {299}},\ \bibinfo {pages} {1} (\bibinfo {year} {1988})}\BibitemShut {NoStop}%
\bibitem [{\citenamefont {Blondel}\ \emph {et~al.}(2013)\citenamefont {Blondel}
  \emph {et~al.}}]{Blondel:2013ia}%
  \BibitemOpen
  \bibfield  {author} {\bibinfo {author} {\bibfnamefont {A.}~\bibnamefont
  {Blondel}} \emph {et~al.},\ }\bibfield  {title} {\emph {\bibinfo {title}
  {{Research Proposal for an Experiment to Search for the Decay $\mu \to
  eee$}}},\ }\href@noop {} {\  (\bibinfo {year} {2013})},\ \Eprint
  {https://arxiv.org/abs/1301.6113} {arXiv:1301.6113
  [physics.ins-det]}\BibitemShut {NoStop}%
\bibitem [{\citenamefont {Bertl}\ \emph {et~al.}(2006)\citenamefont {Bertl}
  \emph {et~al.}}]{SINDRUMII:2006dvw}%
  \BibitemOpen
  \bibfield  {author} {\bibinfo {author} {\bibfnamefont {W.~H.}\ \bibnamefont
  {Bertl}} \emph {et~al.} (\bibinfo {collaboration} {SINDRUM II}),\ }\bibfield
  {title} {\emph {\bibinfo {title} {{A Search for muon to electron conversion
  in muonic gold}}},\ }\href {https://doi.org/10.1140/epjc/s2006-02582-x}
  {\bibfield  {journal} {\bibinfo  {journal} {Eur. Phys. J. C}\ }\textbf
  {\bibinfo {volume} {47}},\ \bibinfo {pages} {337} (\bibinfo {year}
  {2006})}\BibitemShut {NoStop}%
\bibitem [{\citenamefont {Abramishvili}\ \emph {et~al.}(2020)\citenamefont
  {Abramishvili} \emph {et~al.}}]{COMET:2018auw}%
  \BibitemOpen
  \bibfield  {author} {\bibinfo {author} {\bibfnamefont {R.}~\bibnamefont
  {Abramishvili}} \emph {et~al.} (\bibinfo {collaboration} {COMET}),\
  }\bibfield  {title} {\emph {\bibinfo {title} {{COMET Phase-I Technical Design
  Report}}},\ }\href {https://doi.org/10.1093/ptep/ptz125} {\bibfield
  {journal} {\bibinfo  {journal} {PTEP}\ }\textbf {\bibinfo {volume} {2020}},\
  \bibinfo {pages} {033C01} (\bibinfo {year} {2020})},\ \Eprint
  {https://arxiv.org/abs/1812.09018} {arXiv:1812.09018
  [physics.ins-det]}\BibitemShut {NoStop}%
\bibitem [{\citenamefont {Hedges}(2023)}]{Hedges:2022tnh}%
  \BibitemOpen
  \bibfield  {author} {\bibinfo {author} {\bibfnamefont {M.~T.}\ \bibnamefont
  {Hedges}} (\bibinfo {collaboration} {Mu2e}),\ }\bibfield  {title} {\emph
  {\bibinfo {title} {{The Mu2e experiment \textemdash{} Searching for charged
  lepton flavor violation}}},\ }\href
  {https://doi.org/10.1016/j.nima.2022.167589} {\bibfield  {journal} {\bibinfo
  {journal} {Nucl. Instrum. Meth. A}\ }\textbf {\bibinfo {volume} {1045}},\
  \bibinfo {pages} {167589} (\bibinfo {year} {2023})},\ \Eprint
  {https://arxiv.org/abs/2210.14317} {arXiv:2210.14317 [hep-ex]}\BibitemShut
  {NoStop}%
\bibitem [{\citenamefont {Wintz}(1998)}]{Wintz:1998rp}%
  \BibitemOpen
  \bibfield  {author} {\bibinfo {author} {\bibfnamefont {P.}~\bibnamefont
  {Wintz}},\ }\bibfield  {title} {\emph {\bibinfo {title} {{Results of the
  SINDRUM-II experiment}}},\ }\href@noop {} {\bibfield  {journal} {\bibinfo
  {journal} {Conf. Proc. C}\ }\textbf {\bibinfo {volume} {980420}},\ \bibinfo
  {pages} {534} (\bibinfo {year} {1998})}\BibitemShut {NoStop}%
\bibitem [{\citenamefont {Kuno}(2005)}]{Kuno:2005mm}%
  \BibitemOpen
  \bibfield  {author} {\bibinfo {author} {\bibfnamefont {Y.}~\bibnamefont
  {Kuno}},\ }\bibfield  {title} {\emph {\bibinfo {title} {{PRISM/PRIME}}},\
  }\href {https://doi.org/10.1016/j.nuclphysbps.2005.05.073} {\bibfield
  {journal} {\bibinfo  {journal} {Nucl. Phys. B Proc. Suppl.}\ }\textbf
  {\bibinfo {volume} {149}},\ \bibinfo {pages} {376} (\bibinfo {year}
  {2005})}\BibitemShut {NoStop}%
\bibitem [{\citenamefont {Honecker}\ \emph {et~al.}(1996)\citenamefont
  {Honecker} \emph {et~al.}}]{SINDRUMII:1996fti}%
  \BibitemOpen
  \bibfield  {author} {\bibinfo {author} {\bibfnamefont {W.}~\bibnamefont
  {Honecker}} \emph {et~al.} (\bibinfo {collaboration} {SINDRUM II}),\
  }\bibfield  {title} {\emph {\bibinfo {title} {{Improved limit on the
  branching ratio of $\mu\to e$ conversion on lead}}},\ }\href
  {https://doi.org/10.1103/PhysRevLett.76.200} {\bibfield  {journal} {\bibinfo
  {journal} {Phys. Rev. Lett.}\ }\textbf {\bibinfo {volume} {76}},\ \bibinfo
  {pages} {200} (\bibinfo {year} {1996})}\BibitemShut {NoStop}%
\bibitem [{\citenamefont {Badertscher}\ \emph {et~al.}(1980)\citenamefont
  {Badertscher} \emph {et~al.}}]{Badertscher:1980bt}%
  \BibitemOpen
  \bibfield  {author} {\bibinfo {author} {\bibfnamefont {A.}~\bibnamefont
  {Badertscher}} \emph {et~al.},\ }\bibfield  {title} {\emph {\bibinfo {title}
  {{New Upper Limits for Muon - Electron Conversion in Sulfur}}},\ }\href
  {https://doi.org/10.1007/BF02776193} {\bibfield  {journal} {\bibinfo
  {journal} {Lett. Nuovo Cim.}\ }\textbf {\bibinfo {volume} {28}},\ \bibinfo
  {pages} {401} (\bibinfo {year} {1980})}\BibitemShut {NoStop}%
\bibitem [{\citenamefont {Aubert}\ \emph {et~al.}(2010)\citenamefont {Aubert}
  \emph {et~al.}}]{BaBar:2009hkt}%
  \BibitemOpen
  \bibfield  {author} {\bibinfo {author} {\bibfnamefont {B.}~\bibnamefont
  {Aubert}} \emph {et~al.} (\bibinfo {collaboration} {BaBar}),\ }\bibfield
  {title} {\emph {\bibinfo {title} {{Searches for Lepton Flavor Violation in
  the Decays $\tau^\pm\to e^\pm\gamma$ and $\tau^\pm\to \mu^\pm\gamma$}}},\
  }\href {https://doi.org/10.1103/PhysRevLett.104.021802} {\bibfield  {journal}
  {\bibinfo  {journal} {Phys. Rev. Lett.}\ }\textbf {\bibinfo {volume} {104}},\
  \bibinfo {pages} {021802} (\bibinfo {year} {2010})},\ \Eprint
  {https://arxiv.org/abs/0908.2381} {arXiv:0908.2381 [hep-ex]}\BibitemShut
  {NoStop}%
\bibitem [{\citenamefont {Banerjee}\ \emph {et~al.}(2022)\citenamefont
  {Banerjee} \emph {et~al.}}]{Banerjee:2022xuw}%
  \BibitemOpen
  \bibfield  {author} {\bibinfo {author} {\bibfnamefont {S.}~\bibnamefont
  {Banerjee}} \emph {et~al.},\ }\bibfield  {title} {\emph {\bibinfo {title}
  {{Snowmass 2021 White Paper: Charged lepton flavor violation in the tau
  sector}}},\ }(\bibinfo {year} {2022}),\ \Eprint
  {https://arxiv.org/abs/2203.14919} {arXiv:2203.14919 [hep-ph]}\BibitemShut
  {NoStop}%
\bibitem [{\citenamefont {Hayasaka}\ \emph {et~al.}(2010)\citenamefont
  {Hayasaka} \emph {et~al.}}]{Hayasaka:2010np}%
  \BibitemOpen
  \bibfield  {author} {\bibinfo {author} {\bibfnamefont {K.}~\bibnamefont
  {Hayasaka}} \emph {et~al.},\ }\bibfield  {title} {\emph {\bibinfo {title}
  {{Search for Lepton Flavor Violating Tau Decays into Three Leptons with 719
  Million Produced $\tau^+\tau^-$ Pairs}}},\ }\href
  {https://doi.org/10.1016/j.physletb.2010.03.037} {\bibfield  {journal}
  {\bibinfo  {journal} {Phys. Lett. B}\ }\textbf {\bibinfo {volume} {687}},\
  \bibinfo {pages} {139} (\bibinfo {year} {2010})},\ \Eprint
  {https://arxiv.org/abs/1001.3221} {arXiv:1001.3221 [hep-ex]}\BibitemShut
  {NoStop}%
\bibitem [{\citenamefont {Crivellin}\ \emph {et~al.}(2022)\citenamefont
  {Crivellin}, \citenamefont {Kirk},\ and\ \citenamefont
  {Manzari}}]{Crivellin:2022cve}%
  \BibitemOpen
  \bibfield  {author} {\bibinfo {author} {\bibfnamefont {A.}~\bibnamefont
  {Crivellin}}, \bibinfo {author} {\bibfnamefont {F.}~\bibnamefont {Kirk}},\
  and\ \bibinfo {author} {\bibfnamefont {C.~A.}\ \bibnamefont {Manzari}},\
  }\bibfield  {title} {\emph {\bibinfo {title} {{Comprehensive analysis of
  charged lepton flavour violation in the symmetry protected type-I seesaw}}},\
  }\href {https://doi.org/10.1007/JHEP12(2022)031} {\bibfield  {journal}
  {\bibinfo  {journal} {JHEP}\ }\textbf {\bibinfo {volume} {12}},\ \bibinfo
  {pages} {031}},\ \Eprint {https://arxiv.org/abs/2208.00020} {arXiv:2208.00020
  [hep-ph]}\BibitemShut {NoStop}%
\bibitem [{\citenamefont {Kuno}\ and\ \citenamefont
  {Okada}(2001)}]{Kuno:1999jp}%
  \BibitemOpen
  \bibfield  {author} {\bibinfo {author} {\bibfnamefont {Y.}~\bibnamefont
  {Kuno}}\ and\ \bibinfo {author} {\bibfnamefont {Y.}~\bibnamefont {Okada}},\
  }\bibfield  {title} {\emph {\bibinfo {title} {{Muon decay and physics beyond
  the standard model}}},\ }\href {https://doi.org/10.1103/RevModPhys.73.151}
  {\bibfield  {journal} {\bibinfo  {journal} {Rev. Mod. Phys.}\ }\textbf
  {\bibinfo {volume} {73}},\ \bibinfo {pages} {151} (\bibinfo {year} {2001})},\
  \Eprint {https://arxiv.org/abs/hep-ph/9909265}
  {arXiv:hep-ph/9909265}\BibitemShut {NoStop}%
\bibitem [{\citenamefont {Calibbi}\ \emph {et~al.}(2021)\citenamefont
  {Calibbi}, \citenamefont {Marcano},\ and\ \citenamefont
  {Roy}}]{Calibbi:2021pyh}%
  \BibitemOpen
  \bibfield  {author} {\bibinfo {author} {\bibfnamefont {L.}~\bibnamefont
  {Calibbi}}, \bibinfo {author} {\bibfnamefont {X.}~\bibnamefont {Marcano}},\
  and\ \bibinfo {author} {\bibfnamefont {J.}~\bibnamefont {Roy}},\ }\bibfield
  {title} {\emph {\bibinfo {title} {{Z lepton flavour violation as a probe for
  new physics at future $e^+e^-$ colliders}}},\ }\href
  {https://doi.org/10.1140/epjc/s10052-021-09777-3} {\bibfield  {journal}
  {\bibinfo  {journal} {Eur. Phys. J. C}\ }\textbf {\bibinfo {volume} {81}},\
  \bibinfo {pages} {1054} (\bibinfo {year} {2021})},\ \Eprint
  {https://arxiv.org/abs/2107.10273} {arXiv:2107.10273 [hep-ph]}\BibitemShut
  {NoStop}%
\bibitem [{\citenamefont {Brignole}\ and\ \citenamefont
  {Rossi}(2004)}]{Brignole:2004ah}%
  \BibitemOpen
  \bibfield  {author} {\bibinfo {author} {\bibfnamefont {A.}~\bibnamefont
  {Brignole}}\ and\ \bibinfo {author} {\bibfnamefont {A.}~\bibnamefont
  {Rossi}},\ }\bibfield  {title} {\emph {\bibinfo {title} {{Anatomy and
  phenomenology of mu-tau lepton flavor violation in the MSSM}}},\ }\href
  {https://doi.org/10.1016/j.nuclphysb.2004.08.037} {\bibfield  {journal}
  {\bibinfo  {journal} {Nucl. Phys. B}\ }\textbf {\bibinfo {volume} {701}},\
  \bibinfo {pages} {3} (\bibinfo {year} {2004})},\ \Eprint
  {https://arxiv.org/abs/hep-ph/0404211} {arXiv:hep-ph/0404211}\BibitemShut
  {NoStop}%
\bibitem [{\citenamefont {Kitano}\ \emph {et~al.}(2002)\citenamefont {Kitano},
  \citenamefont {Koike},\ and\ \citenamefont {Okada}}]{Kitano:2002mt}%
  \BibitemOpen
  \bibfield  {author} {\bibinfo {author} {\bibfnamefont {R.}~\bibnamefont
  {Kitano}}, \bibinfo {author} {\bibfnamefont {M.}~\bibnamefont {Koike}},\ and\
  \bibinfo {author} {\bibfnamefont {Y.}~\bibnamefont {Okada}},\ }\bibfield
  {title} {\emph {\bibinfo {title} {{Detailed calculation of lepton flavor
  violating muon electron conversion rate for various nuclei}}},\ }\href
  {https://doi.org/10.1103/PhysRevD.76.059902} {\bibfield  {journal} {\bibinfo
  {journal} {Phys. Rev. D}\ }\textbf {\bibinfo {volume} {66}},\ \bibinfo
  {pages} {096002} (\bibinfo {year} {2002})},\ \bibinfo {note} {[Erratum:
  Phys.Rev.D 76, 059902 (2007)]},\ \Eprint
  {https://arxiv.org/abs/hep-ph/0203110} {arXiv:hep-ph/0203110}\BibitemShut
  {NoStop}%
\bibitem [{\citenamefont {Cirigliano}\ \emph {et~al.}(2009)\citenamefont
  {Cirigliano}, \citenamefont {Kitano}, \citenamefont {Okada},\ and\
  \citenamefont {Tuzon}}]{Cirigliano:2009bz}%
  \BibitemOpen
  \bibfield  {author} {\bibinfo {author} {\bibfnamefont {V.}~\bibnamefont
  {Cirigliano}}, \bibinfo {author} {\bibfnamefont {R.}~\bibnamefont {Kitano}},
  \bibinfo {author} {\bibfnamefont {Y.}~\bibnamefont {Okada}},\ and\ \bibinfo
  {author} {\bibfnamefont {P.}~\bibnamefont {Tuzon}},\ }\bibfield  {title}
  {\emph {\bibinfo {title} {{On the model discriminating power of $\mu\to e$
  conversion in nuclei}}},\ }\href {https://doi.org/10.1103/PhysRevD.80.013002}
  {\bibfield  {journal} {\bibinfo  {journal} {Phys. Rev. D}\ }\textbf {\bibinfo
  {volume} {80}},\ \bibinfo {pages} {013002} (\bibinfo {year} {2009})},\
  \Eprint {https://arxiv.org/abs/0904.0957} {arXiv:0904.0957
  [hep-ph]}\BibitemShut {NoStop}%
\bibitem [{\citenamefont {Suzuki}\ \emph {et~al.}(1987)\citenamefont {Suzuki},
  \citenamefont {Measday},\ and\ \citenamefont {Roalsvig}}]{PhysRevC.35.2212}%
  \BibitemOpen
  \bibfield  {author} {\bibinfo {author} {\bibfnamefont {T.}~\bibnamefont
  {Suzuki}}, \bibinfo {author} {\bibfnamefont {D.~F.}\ \bibnamefont
  {Measday}},\ and\ \bibinfo {author} {\bibfnamefont {J.~P.}\ \bibnamefont
  {Roalsvig}},\ }\bibfield  {title} {\emph {\bibinfo {title} {Total nuclear
  capture rates for negative muons}},\ }\href
  {https://doi.org/10.1103/PhysRevC.35.2212} {\bibfield  {journal} {\bibinfo
  {journal} {Phys. Rev. C}\ }\textbf {\bibinfo {volume} {35}},\ \bibinfo
  {pages} {2212} (\bibinfo {year} {1987})}\BibitemShut {NoStop}%
\bibitem [{\citenamefont {Alonso}\ \emph {et~al.}(2013)\citenamefont {Alonso},
  \citenamefont {Dhen}, \citenamefont {Gavela},\ and\ \citenamefont
  {Hambye}}]{Alonso:2012ji}%
  \BibitemOpen
  \bibfield  {author} {\bibinfo {author} {\bibfnamefont {R.}~\bibnamefont
  {Alonso}}, \bibinfo {author} {\bibfnamefont {M.}~\bibnamefont {Dhen}},
  \bibinfo {author} {\bibfnamefont {M.~B.}\ \bibnamefont {Gavela}},\ and\
  \bibinfo {author} {\bibfnamefont {T.}~\bibnamefont {Hambye}},\ }\bibfield
  {title} {\emph {\bibinfo {title} {{Muon conversion to electron in nuclei in
  type-I seesaw models}}},\ }\href {https://doi.org/10.1007/JHEP01(2013)118}
  {\bibfield  {journal} {\bibinfo  {journal} {JHEP}\ }\textbf {\bibinfo
  {volume} {01}},\ \bibinfo {pages} {118}},\ \Eprint
  {https://arxiv.org/abs/1209.2679} {arXiv:1209.2679 [hep-ph]}\BibitemShut
  {NoStop}%
\bibitem [{\citenamefont {Craig}\ \emph {et~al.}(2016)\citenamefont {Craig},
  \citenamefont {Gu}, \citenamefont {Liu},\ and\ \citenamefont
  {Wang}}]{Craig:2015wwr}%
  \BibitemOpen
  \bibfield  {author} {\bibinfo {author} {\bibfnamefont {N.}~\bibnamefont
  {Craig}}, \bibinfo {author} {\bibfnamefont {J.}~\bibnamefont {Gu}}, \bibinfo
  {author} {\bibfnamefont {Z.}~\bibnamefont {Liu}},\ and\ \bibinfo {author}
  {\bibfnamefont {K.}~\bibnamefont {Wang}},\ }\bibfield  {title} {\emph
  {\bibinfo {title} {{Beyond Higgs Couplings: Probing the Higgs with Angular
  Observables at Future $e^+e^-$ Colliders}}},\ }\href
  {https://doi.org/10.1007/JHEP03(2016)050} {\bibfield  {journal} {\bibinfo
  {journal} {JHEP}\ }\textbf {\bibinfo {volume} {03}},\ \bibinfo {pages}
  {050}},\ \Eprint {https://arxiv.org/abs/1512.06877} {arXiv:1512.06877
  [hep-ph]}\BibitemShut {NoStop}%
\bibitem [{\citenamefont {Ellis}(2017)}]{Ellis:2016jkw}%
  \BibitemOpen
  \bibfield  {author} {\bibinfo {author} {\bibfnamefont {J.}~\bibnamefont
  {Ellis}},\ }\bibfield  {title} {\emph {\bibinfo {title} {{TikZ-Feynman:
  Feynman diagrams with TikZ}}},\ }\href
  {https://doi.org/10.1016/j.cpc.2016.08.019} {\bibfield  {journal} {\bibinfo
  {journal} {Comput. Phys. Commun.}\ }\textbf {\bibinfo {volume} {210}},\
  \bibinfo {pages} {103} (\bibinfo {year} {2017})},\ \Eprint
  {https://arxiv.org/abs/1601.05437} {arXiv:1601.05437 [hep-ph]}\BibitemShut
  {NoStop}%
\end{thebibliography}%

\end{document}